\documentclass[%
 reprint,
 superscriptaddress,
amsmath,amssymb,
aps,
]{revtex4-2}
\usepackage{graphicx}
\usepackage{dcolumn}
\usepackage{bm}
\usepackage{capt-of}
\usepackage[colorlinks,linkcolor=blue, urlcolor=blue,anchorcolor=blue, citecolor=blue]{hyperref}
\begin{document}
\maxdeadcycles=1000
\newlength{\originaltextfloatsep}
\setlength{\originaltextfloatsep}{\textfloatsep}
\preprint{APS}

\title{Stable charged energy and phase transitions in an anisotropic Dicke quantum battery with Kerr nonlinearity}

\author{Shao-Fan He}
\affiliation{College of Physics and Electronic Engineering, Northwest Normal University, Lanzhou, 730070, China}
\author{Lv-Ting Gong}
\affiliation{College of Physics and Electronic Engineering, Northwest Normal University, Lanzhou, 730070, China}
\author{Fu-Quan Dou}
\email{doufq@nwnu.edu.cn}
\affiliation{College of Physics and Electronic Engineering, Northwest Normal University, Lanzhou, 730070, China}
\affiliation{Gansu Provincial Research Center for Basic Disciplines of Quantum Physics, Lanzhou, 730000, China}

\begin{abstract}
Phase transitions (PTs) in quantum many-body systems offer a powerful route to manipulate the performance of quantum batteries (QBs). Here, we investigate an anisotropic Dicke QB with Kerr nonlinearity, focusing on the role of PTs in the charging process of the QB system. By analyzing how the Kerr nonlinearity reshapes the cavity level-spacing structure, we find that multiple near-degeneracies emerge at specific nonlinearity strengths. Away from the region of multiple near-degeneracies, the stable stored energy is governed primarily by the ergodic-nonergodic transition (ENET), whereas within this region it is controlled by first- and second-order dissipative phase transitions (DPTs). We employ Liouvillian spectral theory to quantify the critical slowing down associated with the DPTs, and use the Wigner function to characterize the properties of the cavity field in the vicinity of criticality. We optimize the charging process via reinforcement learning (RL) and reveal the physical mechanism underlying performance enhancement from the perspective of ergodic and nonergodic phase dynamics. Our results offer a feasible pathway for manipulating and optimizing quantum battery performance based on intrinsic PT behaviors.

\end{abstract}

\maketitle


\section{INTRODUCTION}
Quantum technologies, ranging from quantum computing \cite{Acharya2025} and quantum simulation \cite{Daley2022} to quantum communication \cite{Chen2021} and quantum sensing \cite{RevModPhys.89.035002}, have achieved remarkable progress. Motivated by these advances, questions arose regarding quantum-scale energy storage: how would a battery operate at the quantum level? The concept of quantum batteries (QBs) was proposed more than a decade ago and has since developed rapidly \cite{PhysRevE.87.042123,Binder_2015,RevModPhys.96.031001,Ferraro2026}. QBs are devices that store and transfer energy by harnessing quantum-mechanical effects and deliver more efficient energy extraction and faster charging than classical counterparts \cite{PhysRevLett.118.150601,PhysRevLett.122.047702,PhysRevB.99.205437,Ferraro2026}. Substantial effort is directed toward developing high-performance QBs  \cite{Binder_2015,PhysRevLett.120.117702,PhysRevB.99.035421,PhysRevE.99.052106,PhysRevLett.118.150601,PhysRevLett.122.047702,PhysRevApplied.14.024092,PhysRevResearch.2.023113,PhysRevLett.128.140501,PhysRevA.106.032212,PhysRevA.107.023725,PhysRevLett.132.090401,kzvn-dj7v,vqnk-kzqg}. Ongoing studies of QB model construction \cite{PhysRevLett.120.117702,PhysRevB.99.035421,PhysRevApplied.14.024092,PhysRevLett.124.130601,PhysRevB.102.245407,PhysRevLett.125.236402,PhysRevA.104.043706,Dou2021,PhysRevLett.131.240401,PhysRevApplied.19.064069,10.1063/5.0156618,PhysRevLett.132.210402,PhysRevLett.134.180401,PhysRevLett.134.130401,3tm5-vsqw}, the role of quantum resources \cite{PhysRevLett.118.150601,PhysRevLett.122.047702,PhysRevB.99.205437,PhysRevLett.125.180603,PhysRevE.102.052109,PhysRevLett.129.130602,PhysRevA.106.012425,PhysRevLett.131.030402,PhysRevA.109.042424,PhysRevE.109.064103,2026-0106}, and many-body effects \cite{PhysRevA.97.022106,PhysRevB.99.205437,PhysRevB.100.115142,PhysRevLett.125.236402,PhysRevA.103.052220,PRXQuantum.5.030319,PhysRevA.109.032201,PhysRevApplied.23.024010,2z1g-rgr9} are vital for improving charging performance. Furthermore, experimental investigations have yielded encouraging advances in exploring QBs \cite{doi:10.1126/sciadv.abk3160,PhysRevA.106.042601,PhysRevLett.131.260401,PhysRevA.107.L030201,PhysRevLett.133.180401,PhysRevResearch.6.023091,g45c-ssfx,bhyh-53np,y3qx-cs3r,sp5l-c6m8,Hymas2026}.

Phase transitions (PTs) provide a fundamental framework for understanding and controlling QB performance. Cavity quantum electrodynamics (cavity QED) and spin chains---two classic platforms for QBs---both exhibit rich PT phenomena. Cavity-QED systems display distinct transitions \cite{PhysRevLett.90.044101,Baumann2010,PhysRevLett.117.173601,PhysRevLett.131.113601}, including normal-to-superradiant phase transition, excited-state quantum phase transition, and thermal phase transition, while spin chains are a canonical arena for quantum phase transitions \cite{PhysRevLett.110.135704,PhysRevLett.119.080501,Keesling2019,Feng2023}. Hallmark features near critical boundaries \cite{PhysRevLett.99.095701,PhysRevLett.124.120504,Ding2022}---energy-gap closing, divergence of the correlation length, and enhanced sensitivity to external control---further promote energy redistribution and efficient storage \cite{PhysRevA.101.032115,PhysRevResearch.4.L022017}.

A series of results has demonstrated that charging performance depends strongly on the PT dynamics of QB systems. In cavity-QED and topological systems, phase boundaries shape the maximum stored energy \cite{PhysRevLett.134.180401,PhysRevB.105.115405,PhysRevResearch.4.013172,PhysRevB.109.235432}, and in dissipative settings a quadratically driven battery yields abundant work near dissipative phase transition (DPT) critical points \cite{Downing2023}. In spin chains, extracted energy and thermodynamic efficiency improve near criticality; for quenches far from the critical lines, both stored energy and power exhibit pronounced enhancements at the critical times where dynamical quantum phase transitions occur \cite{42rw-l31q}. Criticality also confers robustness: in a dimerized XY spin chain, stored energy becomes insensitive to charging parameters when the quench crosses a quantum critical line \cite{PhysRevLett.133.197001}, and battery capacity itself serves as a sensitive indicator of PTs in anisotropic XY spin chains \cite{Hou2026}. Machine learning has recently been applied to enhance the charging performance of QBs \cite{PhysRevLett.133.243602,m1gm-f9zy,Sun_2025}. Despite these advances, how PTs, including DPTs in open systems and the dynamical features exploited by machine-learning protocols, govern battery performance remains incompletely understood.

The anisotropic Dicke model---characterized by distinct coupling strengths for rotating and counter-rotating wave interactions---serves as a natural framework for this purpose \cite{PhysRevA.93.043818,PhysRevA.110.042419,ymzt-2mq4}. It hosts a rich phase diagram encompassing normal-to-superradiant quantum phase transition, thermal phase transition, and ergodic-nonergodic transition (ENET) \cite{PhysRevLett.118.080601,PhysRevA.107.043706,PhysRevA.108.063716,e26070574,
das2026detailedstudyvariousphases,PhysRevA.110.063722}, and has been realized in circuit-QED platforms capable of ultrastrong and deep-strong light-matter coupling \cite{PhysRevLett.112.173601,RevModPhys.91.025005}. The Josephson nonlinearity underlying such couplings can introduce a Kerr nonlinearity analogously engineered in other systems \cite{Kirchmair2013,Tusnin2023}; driven-dissipative Kerr resonators, in particular, exhibit both first- and second-order DPTs \cite{PhysRevA.95.012128,PhysRevA.98.042118,Fink2018,Chen2023,Minganti_2023,PhysRevX.10.021022,PRXQuantum.5.010327}, recently observed in superconducting Kerr resonators \cite{Beaulieu2025}. Reinforcement learning (RL) has been used to optimize the charging process in a Dicke QB and a cavity-Heisenberg spin chain QB \cite{PhysRevLett.133.243602,Sun_2025}.
These features motivate introducing Kerr nonlinearity into an anisotropic Dicke QB with photon loss, and raise the question of whether anisotropy and Kerr nonlinearity can enhance QB performance, whether DPTs emerge in this setting, whether RL-optimized protocols can exploit PT dynamics to improve charging, and---most importantly---whether PTs and battery performance are fundamentally connected.
\begin{figure}[htbp]
\centering
\includegraphics[width=0.48\textwidth]{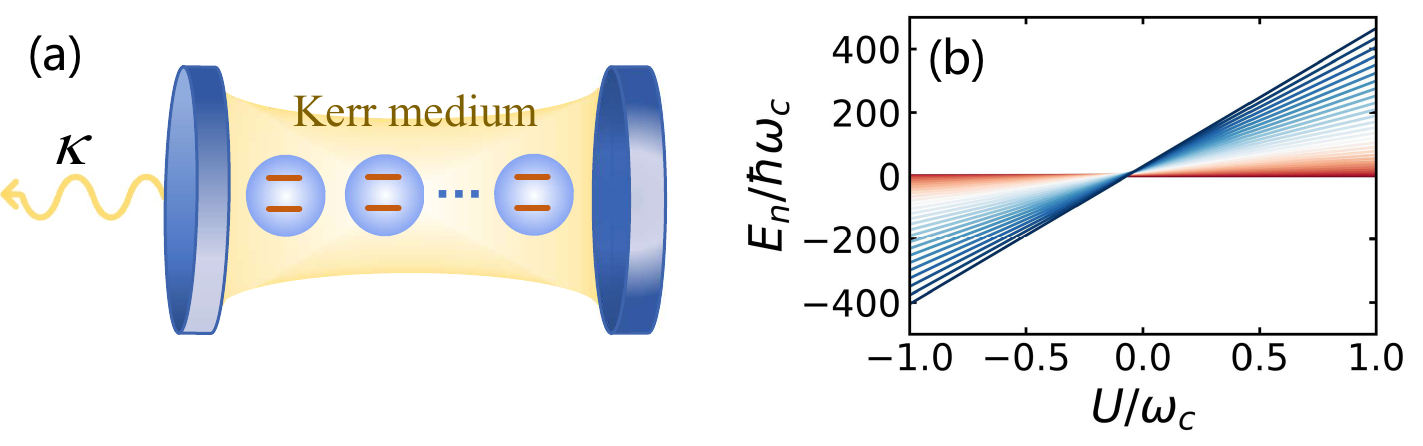}
\caption{\label{fig.1}(a) Sketch of an anisotropic Dicke battery composed by $N$ identical and independent TLSs coupled a single cavity mode with a Kerr nonlinearity. (b) The cavity spectrum as a function of Kerr nonlinearity strength for $N_{\rm ph}=30$. The color scale from red to blue denotes increasing photon number $n$.}
\end{figure}

In this work, we consider an anisotropic Dicke QB with Kerr nonlinearity and investigate the intrinsic relation between battery charging performance and PTs. We focus on how anisotropy and Kerr nonlinearity influence charging performance and PTs. Furthermore, we employ Liouvillian spectral theory to quantify the critical slowing down associated with DPT, and use the Wigner function to characterize the cavity properties near critical points. Finally, we optimize the charging protocols via RL and explain the enhancement from the perspective of the ergodic and nonergodic phase dynamics.

The rest of this paper is organized as follows. In Sec. \ref{section2}, we consider the anisotropic Dicke QB with Kerr nonlinearity and analyze the influence of the Kerr nonlinearity on the cavity energy spectrum. In Sec. \ref{section3}, we investigate the battery performance and PTs for further cavity-atoms couplings and Kerr nonlinearity strengths. We optimize the charging process and interpret the enhancement achieved by RL from the dynamics of the ergodic and nonergodic phases. Finally, we give a summary in Sec. \ref{section4}.

\section{MODEL} \label{section2}
We consider a QB consisting of $N$ identical and independent two-level systems (TLSs) coupled to a single-mode Kerr cavity via independent rotating and counter-rotating light-matter interactions, i.e., an anisotropic Dicke QB with Kerr nonlinearity, as shown in Fig.~\ref{fig.1}(a). The QB system can be described by the following Hamiltonian ($\hbar = 1$ throughout this paper):
\begin{eqnarray}\label{H(t)}
H(t)=H_{A}+H_{B}+\lambda(t)H_{I},
\end{eqnarray}
where the time-dependent parameter $\lambda(t)$ describes the charging time interval, which we assume to be given by a step function equal to $1$ for $t\in[0,\tau]$ and zero elsewhere. $H_{A,B,I}$ are the Hamiltonians of the charger, battery and interaction terms, respectively, with the following forms:

\begin{eqnarray}\label{H(A)}
H_{A}=\omega_{c}\hat{a}^{\dagger} \hat{a}+\frac{U}{2} \hat{a}^{\dagger 2} \hat{a}^2,
\end{eqnarray}
\begin{eqnarray}\label{H(B)}
H_{B}=\omega_{0}(\hat{J}_{z}+\frac{N}{2}),
\end{eqnarray}
\begin{eqnarray}\label{H(I)}
H_{I}=\frac{g_1}{\sqrt{N}} (\hat{a}^{\dagger} \hat{J}_- + \hat{a}\hat{J}_+) + \frac{g_2}{\sqrt{N}} (\hat{a}^{\dagger} \hat{J}_+ + \hat{a} \hat{J}_-).
\end{eqnarray}
Here $\hat{a}$ ($\hat{a}^\dagger$) is the annihilation (creation) operator of the cavity mode with frequency $\omega_c$. $\hat{J}_\alpha = \frac{1}{2} \sum_{i=1}^N \hat{\sigma}_i^\alpha$ with $\alpha = x, y, z$, indicate the components of a collective spin operator expressed in terms of the Pauli operators $\hat{\sigma}_i^\alpha$ of the $i$th TLS. $\hat{J}_\pm = (\hat{J}_x \pm i\hat{J}_y)$ are the up and down operators of the TLSs with the energy level spacing $\omega_0$. $U$ is Kerr nonlinearity strength which can be realized in a cavity filled with a nonlinear medium \cite{PhysRevLett.79.1467,Zhu:10}, and $g_1$, $g_{2}$ are the coupling strengths of rotating and counter-rotating wave terms, respectively.

We discuss the charging process of $N$ TLSs prepared in their ground state $|g\rangle$, via coupling to a single cavity mode residing in the $N$ photon Fock state. Thus, the initial state of the total system is
\begin{equation}\label{psi0}
\begin{split}
\rho(0) &= |\varPsi(0)\rangle\langle\varPsi(0)| \\
&= |N\rangle\otimes|\underbrace{g,g,...,g}_{N}\rangle\langle N|\otimes\langle \underbrace{g,g,...,g}_{N}|.
\end{split}
\end{equation}

Realistically, photon leakage from the cavity is inevitable, resulting in the nonunitary time evolution of the system described by the Lindblad master equation \cite{10.1093/acprof:oso/9780199213900.001.0001},
\begin{eqnarray}\label{Liouvillian}
\begin{split}
\frac{d\rho(t)}{dt} &= \hat{\mathcal{L}}[\rho(t)] \\
&= -i[\hat{H},\rho(t)] + \kappa[2\hat{a}\rho(t)\hat{a}^\dagger - \{\hat{a}^\dagger\hat{a},\rho(t)\}],
\end{split}
\end{eqnarray}
where $\kappa$ sets the photon loss rate and $\hat{\mathcal{L}}$ is the Liouvillian superoperator.
The system possesses a $\mathbb{Z}_2$ symmetry $[\hat{H}, \hat{\Pi}] = 0$, where the operator $\hat{\Pi} = \exp[i\pi(\hat{a}^\dagger \hat{a} + \hat{J}_z + N/2)]$ measures the parity of the total number of excitations. The Liouvillian superoperator $\hat{\mathcal{L}}$ inherits a weak $\mathbb{Z}_2$ symmetry, i.e., $[\hat{\mathcal{L}}, \hat{\mathit{\Pi}}] = 0$ \cite{PhysRevA.105.L050201,PhysRevLett.125.240405}.
The superoperator $\hat{\mathit{\Pi}}$ acts on the basis states in Liouville space as
\begin{eqnarray}\label{symmetry_operator}
\hat{\mathit{\Pi}}|\alpha\rangle\rangle = \hat{\Pi}|n_l, m_l\rangle\langle n_r, m_r|\hat{\Pi}^{\dagger}=\zeta |\alpha\rangle\rangle,
\end{eqnarray}
where $n$ indicates the number of photons and $m$ denotes the eigenvalue of $\hat{J}_z$. If the eigenvalue $\zeta=1$ ($\zeta=-1$), $|\alpha\rangle\rangle$ belongs to the even-parity subspace $\mathcal{L}_1$ (odd-parity subspace $\mathcal{L}_2$). This quantity characterizes the parity of the difference in the number of excitations between the left and right indices of operators.

The Kerr nonlinearity introduces an anharmonicity into the cavity spectrum, causing the adjacent level spacings to become photon-number dependent and evolve with the Kerr strength, as shown in Fig.~\ref{fig.1}(b). For $U>0$, the level spacing increases with excitation number, progressively suppressing multiphoton resonances and driving the cavity toward an effectively two-level behavior. 
For $U < 0$, the energy level spacings initially decrease as $ U $ becomes more negative. As a result, several pairs of levels become nearly degenerate, producing multiple near-degeneracies in the cavity spectrum rather than a single near-resonant pair. Beyond the region of multiple near-degeneracies, the magnitude of the adjacent level spacings then increases and the ordering of the energy levels is reversed, i.e., higher photon number Fock states become lower in energy. Since the photon number is neither conserved by the system Hamiltonian nor bounded from above, a sufficiently large truncation $N_{\mathrm{ph}} > 4N$ is required to ensure numerical convergence in the region of multiple near-degeneracies (see Appendix \ref{appendix1} for details).

The stored energy of QB at time $t$ is given by
\begin{eqnarray}\label{E(t)}
E(t)=\mathrm{Tr}[H_{B}\rho_{B}(t)],
\end{eqnarray}
where $\rho_{B}(t) = \mathrm{Tr_{A}}[\rho(t)]$ is the reduced density matrix of the system. We are interested in exploring the stable performance of the QB, since the QB is charged in an open quantum system. Therefore the stable stored energy of QB can be described as
\begin{eqnarray}\label{E_{ss}}
E_{s}=E(\infty)=\mathrm{Tr}[H_{B}\rho_{s}^{B}],
\end{eqnarray}
where $\rho_{s}^{B}$ is the reduced density matrix of the system steady-state formally defined by $\partial_t \rho_{s} = 0$.

In all subsequent calculations, we adopt the resonant condition $ \omega_0 = \omega_{\text{c}} $, where the charger and the battery share the same resonant frequency, and treat $ \omega_{\text{c}} $ as a dimensionless parameter. Numerical work has been performed by using the PYTHON toolbox QuTiP2 \cite{JOHANSSON20131234}.

\section{THE STORED ENERGY AND PHASE TRANSITIONS} \label{section3}
In this section, we focus on the stable stored energy of the QB and the PTs of the system. In general, a long relaxation time required to reach stable energy storage limits the effective charging power. We then employ RL to optimize the charging protocol, enabling the QB to attain a high stored energy within a substantially short time, and discuss the physical origin of the performance improvement by comparing the dynamical characteristics of the PT before and after optimization.

\vspace{0.5\baselineskip}
\subsection{Stored energy and ergodic-nonergodic transition}\label{subsection1}
We first investigate how ENET affects the stable energy storage of the QB. Various quantitative indicators have been proposed to 
characterize ENET in quantum systems, including the consecutive level-spacing ratio \cite{PhysRevLett.110.084101}, participation ratio (PR) \cite{PhysRevA.107.043706,das2026detailedstudyvariousphases}, and multifractal dimension \cite{PhysRevLett.123.180601}. Here, we employ the PR to examine the degree of localization in the steady state of the QB system. The PR for a steady-state $\rho_{s} = \sum_{\alpha}^{N_1} \psi_{\alpha} \, |\alpha\rangle\rangle$ is defined as
\begin{equation}\label{PR}
\text{PR}_{s} = \frac{\left( \sum_{\alpha}^{N_1} \left| \tilde{\psi}_{\alpha} \right| \right)^2}{\sum_{\alpha}^{N_1} \left| \tilde{\psi}_{\alpha} \right|^2},
\end{equation}
where $\tilde{\psi}_{\alpha}=\psi_{\alpha}^{*}\psi_{\alpha}$ and $N_1$ represents the dimension of the even-parity subspace $\mathcal{L}_1$. In the nonergodic phase the $\text{PR}_{s}$ value is low, indicating localization in the chosen basis, whereas in the ergodic phase its value is relatively higher, reflecting enhanced delocalization.

\begin{figure*}
\centering
\includegraphics[width=1.0\textwidth]{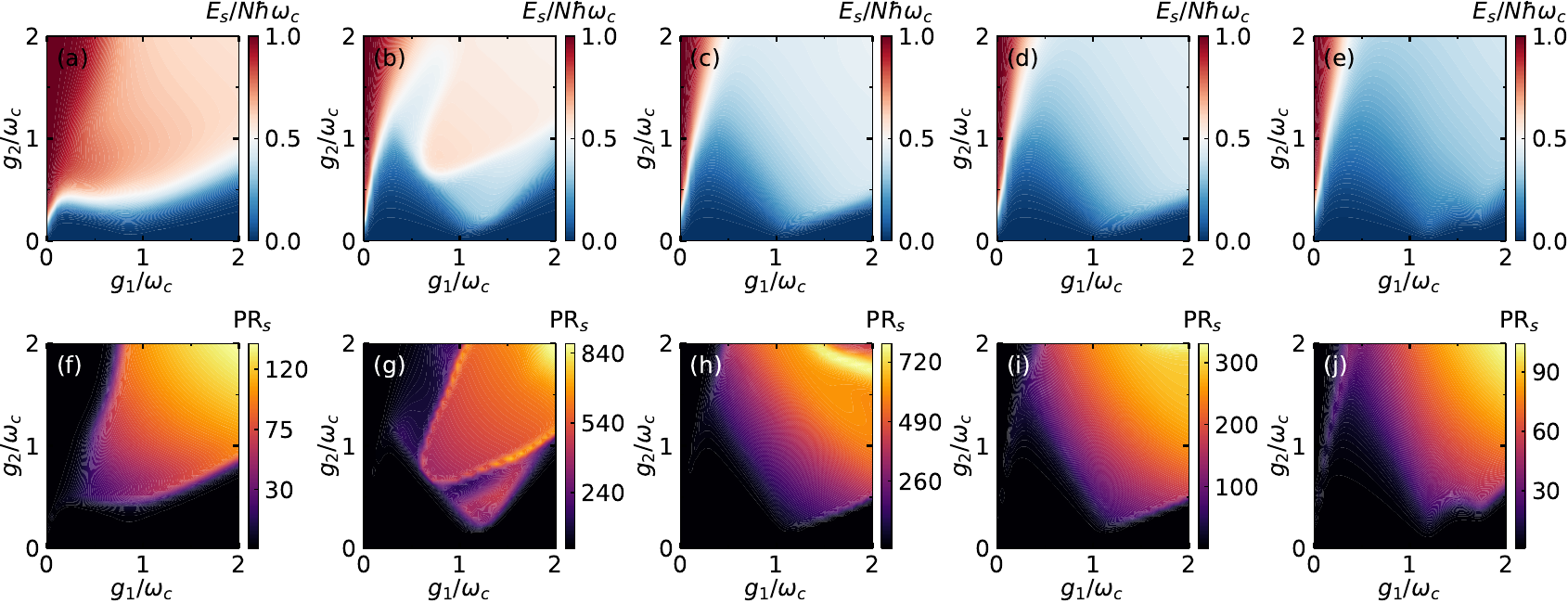}
\caption{\label{fig.2}The stable stored energy $  E_{s}  $ (in units of $N\hbar\omega_{c}$) (a)-(e) and the participation ratio $  \text{PR}_{s}  $ (f)-(j) as functions of the coupling strengths $  g_1 $ and $  g_2 $ (in units of $\omega_c$), for different Kerr nonlinearity strengths $  U  $ (in units of $\omega_{c}$). The values of the Kerr nonlinearity strength $  U  $ are as follows: (a) and (f) $U=-1$, (b) and (g) $U=-0.1$, (c) and (h) $U=0$, (d) and (i) $U=0.1$, (e) and (j) $U=1$, respectively. In all calculations, we use the same settings of $N=4$, $N_{\mathrm{ph}}=30$, $\kappa=0.2$ (in units of $\omega_c$) unless otherwise specified.}
\end{figure*}

Figure~\ref{fig.2} shows the stable stored energy and $\rm{PR}_s$ as functions of the coupling strengths $g_1$ and $g_2$ for different Kerr nonlinearity strengths. The stable stored energy exhibits a stark asymmetry governed by the competition between the two coupling strengths. In the regime $  g_1\gg g_2  $, $  E_{s}  $ approaches zero, corresponding to a low-energy storage regime. Conversely, for $  g_1\ll g_2  $, $  E_{s}  $ approaches its upper bound, signaling a high-energy storage regime where the QB is almost fully charged. When $ g_1 $ and $g_2$ are comparable, the battery exhibits intermediate-energy storage, with the normalized $E_{s}$ lying approximately between 0.4 and 0.6. Moreover, a comparison of Figs.~\ref{fig.2}(a)-(e) shows that the boundaries between these energy-storage regimes shift appreciably as \(U\) is varied, indicating that the Kerr nonlinearity provides an effective means of tuning the crossover among the low-, intermediate-, and high-energy storage regimes.
We next examine how these distinct energy-storage regimes are related to the ENET, as revealed by $\text{PR}_s$ in Figs.~\ref{fig.2}(f)–(j). A close correspondence emerges between the ENET phase diagram and the distribution of the stable stored energy. Specifically, extremal stored energy (either near-maximal or near-zero) is associated with the nonergodic phase, while intermediate-energy storage corresponds to the ergodic phase. The ENET boundary closely coincides with the crossover between extremal stored energy and intermediate-energy storage.
\begin{widetext}
\centering
\includegraphics[width=0.9\textwidth]{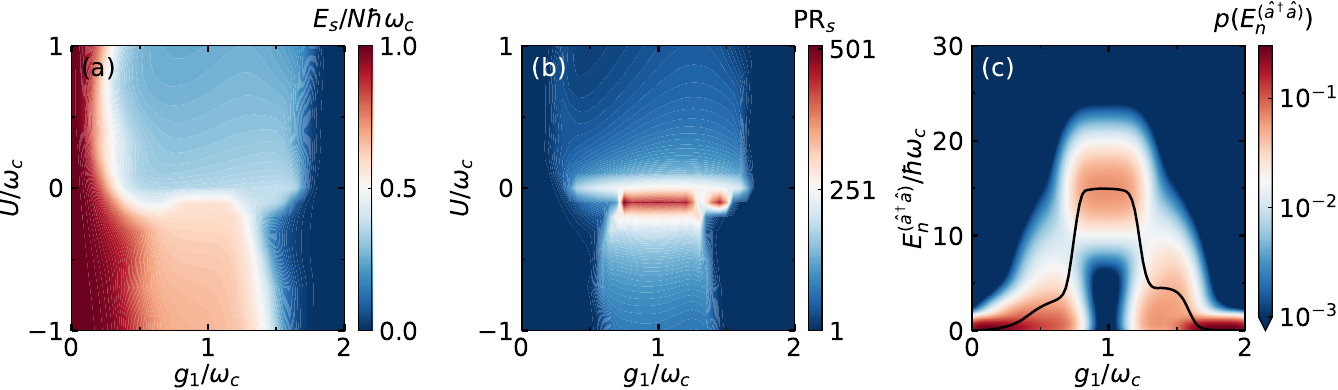}
\captionof{figure}{\label{fig.3}(a) The stable stored energy $E_{s}$ (in units of $N\hbar\omega_{c}$) as functions of the Kerr nonlinearity strength $U$ (in units of $\omega_{c}$) and the coupling strength $g_1$ (in units of $\omega_{c}$). (b) The participation ratio $\text{PR}_{s}$ as  functions of the Kerr nonlinearity strength $U$ and the coupling strength $g_1$. (c) The population of the steady-state onto the spectrum of $\hat{a}^\dagger \hat{a}$ at $U=-0.1$ as function of the coupling strength $g_1$. The mean photon energy is shown as a black line. We choose coupling strength $g_2=-g_1+2$ in this and subsequent calculations, unless otherwise stated.}
\end{widetext}

To examine the robustness of the correspondence between the stable stored energy and the ENET against variations in the Kerr nonlinearity and to identify possible deviations, we further consider a continuous range of $U$ and choose the line ($g_1+g_2 = 2$) to keep the overall coupling scale fixed while continuously tuning the competition between the rotating and counter-rotating channels. Along this line, we calculate the dependences of stable stored energy and $\text{PR}_{s}$ on $g_1$ and $U$, as well as the probability distribution of the steady-state over the Fock states, as shown in Fig.~\ref{fig.3}. The correspondence uncovered in Fig.~\ref{fig.2} persists across a broad range of Kerr nonlinearities, demonstrating the robustness of this relation [see Figs.~\ref{fig.3}(a) and (b)]. However, a noticeable deviation from this correspondence emerges for slightly negative $U$, where $\text{PR}_{s}$ is strongly enhanced once the two coupling channels contribute comparably. This parameter regime coincides with the occurrence of multiple near-degeneracies in the cavity spectrum, as shown in Fig.~\ref{fig.1}(b). To elucidate the origin of this deviation, we examine the probability distribution of cavity steady-state over Fock states at fixed $  U = -0.1  $ in Fig.~\ref{fig.3}(c). The results indicate that when a single coupling channel dominates, i.e., $  g_1\ll g_2  $ or $  g_1\gg g_2  $, the distribution is strongly localized in low-photon number states. Conversely, when $  g_1  $ and $  g_2  $ become comparable, the distribution spreads broadly across Fock states, exhibiting pronounced delocalization. In the regime of multiple near-degeneracies, particularly when $g_1$ and $g_2$ make comparable contributions, the delocalization in Fock space leads to a substantial increase in $\text{PR}_{s}$, giving rise to the deviation from this correspondence. Consequently, the ENET alone does not fully characterize the charging behavior, thereby raising a natural question: what governs the stable stored energy in the multiple-near-degeneracy regime?

\begin{figure}[htbp]
\centering
\includegraphics[width=0.49\textwidth]{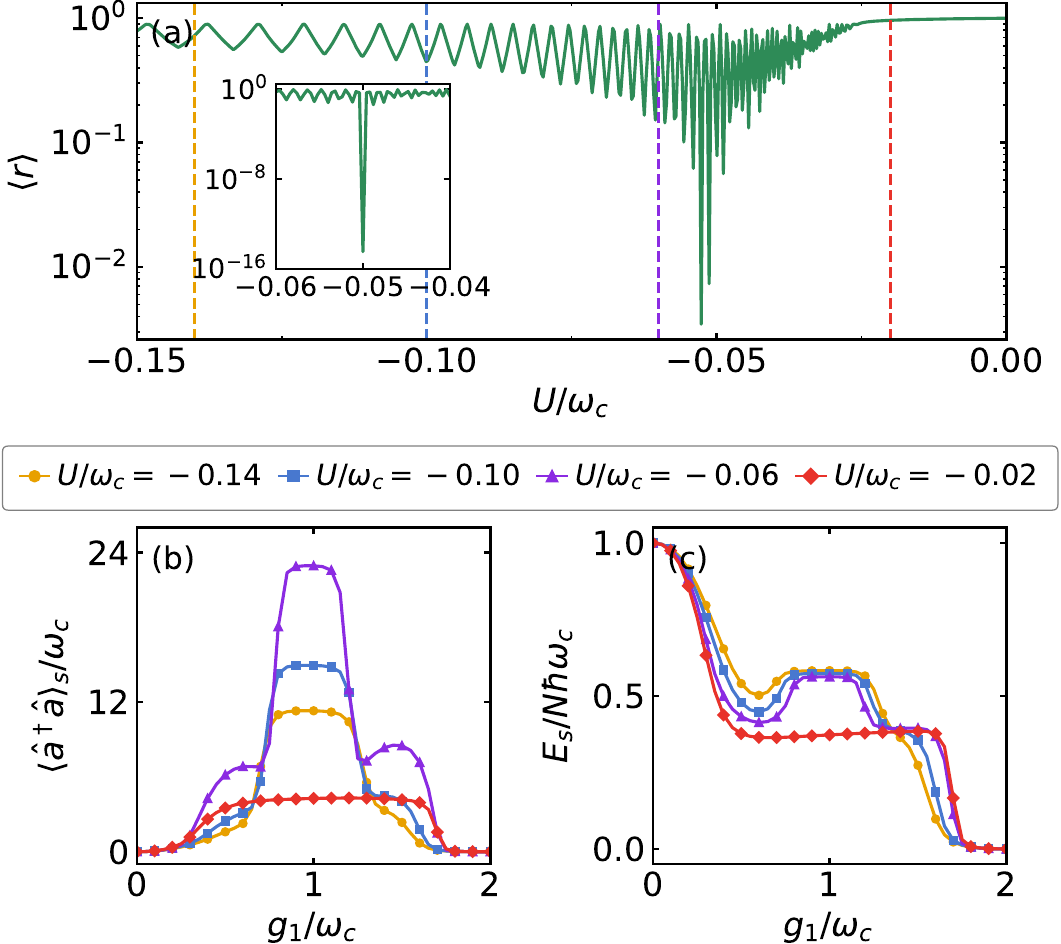}
\caption{\label{fig.4}(a) The dependence of the average level spacing ratio $ \langle r \rangle$ on Kerr nonlinearity strengths $U$ (in units of $\omega_{c}$). Four vertical dashed lines indicate specific Kerr nonlinearity strengths: $U = -0.14$ (orange circles), $-0.10$ (blue squares), $-0.06$ (purple triangles), and $-0.02$ (red diamonds), corresponding to the colors in the legends of panels (b) and (c). The inset shows an enlarged view of the region with the largest number of pairwise degeneracies. (b) The steady-state mean photon number $\langle \hat{a}^\dagger \hat{a} \rangle_{s}$ and (c) the stable stored energy $E_{s}$ (in units of $N\hbar\omega_{c}$) as functions of the coupling strength $g_1$ (in units of $\omega_{c}$) for different Kerr nonlinearity strengths $U$. The photon-number truncation is $N_{\rm ph}=40$.}
\end{figure}

\subsection{Stored energy and first- and second-order DPTs} \label{subsection2}
In this subsection, we address this question by examining the first- and second-order DPTs and their relation to the stable stored energy. The signatures of the first-order DPTs are closely tied to the multiple near-degeneracies in the cavity spectrum. 
We study the average level spacing ratio $  \langle r \rangle$ to provide a more complete characterization of the multiple-near-degeneracy regime. After arranging the cavity eigenenergies in ascending order $\epsilon_0 \leq \epsilon_1 \leq \cdots$, we define the consecutive level spacings as $s_n=\epsilon_{n}-\epsilon_{n-1}$, and the 
corresponding spacing ratio as
\begin{figure*}[htbp]
\centering
\includegraphics[width=0.9\textwidth]{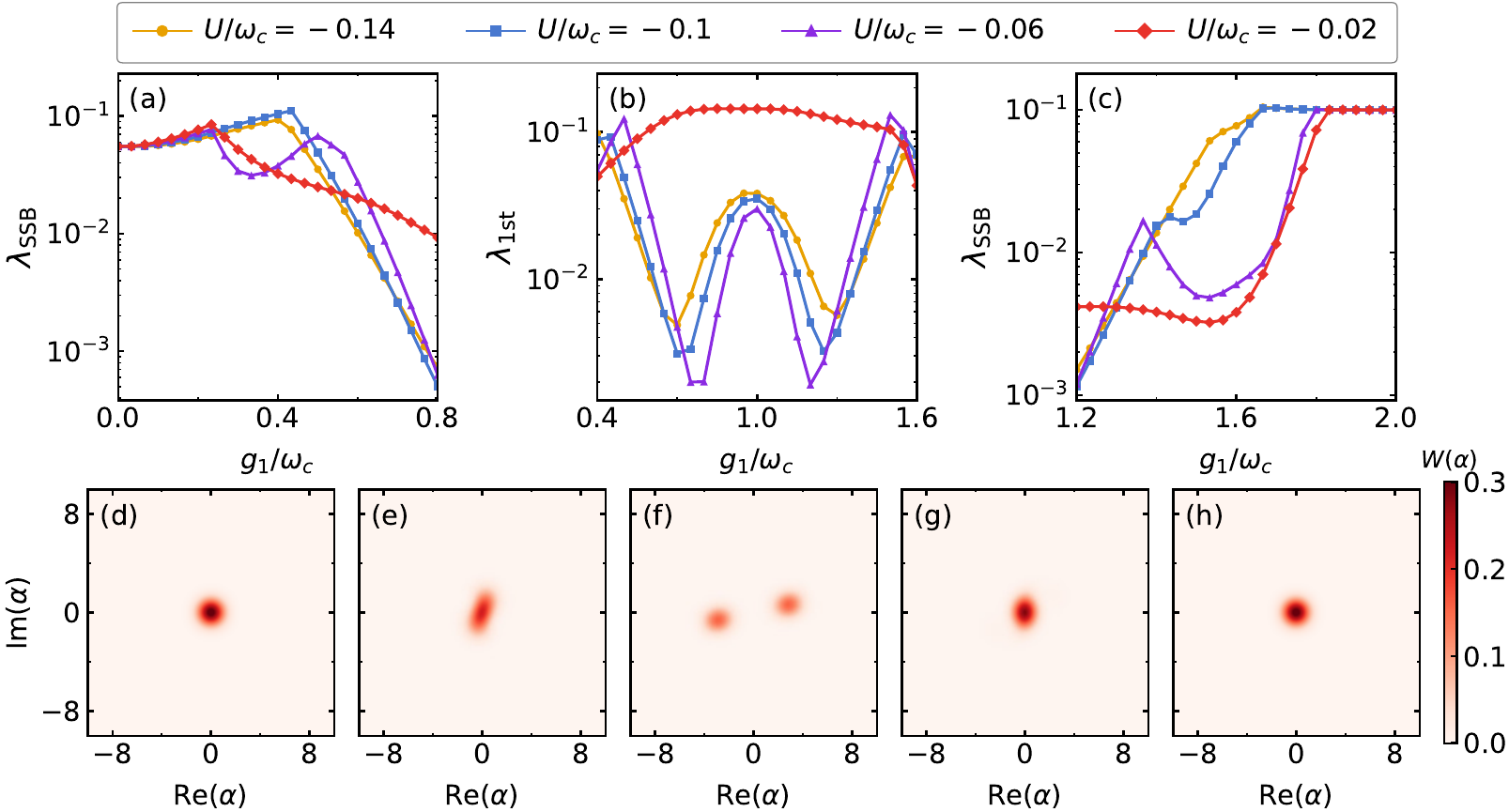}
\caption{\label{fig.5}The Liouvillian gap $\lambda_{\rm SSB}$ and $\lambda_{\rm 1st}$ as a function of $g_1$ (in units of $\omega_{c}$) for different Kerr nonlinearity strengths  $U$ (in units of $\omega_{c}$) . The Wigner function corresponding to the steady-state of the cavity at $U = 0.02$ for different cavity-atoms coupling strength $g_1$. The values of $g_1$ are as follows: (d) $g_1=0$, (e) $g_1=0.2$, (f) $g_1=1$, (g) $g_1=1.7$ and (h) $g_1=2$, respectively.}
\end{figure*}

\begin{equation}\label{r_n}
r_n = \frac{\min(s_{n-1}, s_n)}{\max(s_{n-1}, s_n)}.
\end{equation}
The average level spacing ratio $ \langle r \rangle$ provides a dimensionless measure of the local uniformity of the cavity spectrum. When one of two consecutive level spacings becomes much smaller than the other, as occurs near a level degeneracy, $ r_n $ approaches zero. Therefore, the proliferation of near-degenerate level pairs leads to a strong suppression of $ \langle r \rangle$. By contrast, in the absence of Kerr nonlinearity, the cavity reduces to a harmonic resonator with equally spaced energy levels, for which $ r_n =1 $ and hence $ \langle r \rangle=1$.
We examine the behavior of the average level spacing ratio $ \langle r \rangle$, the stable stored energy $E_{s}$, and the order parameter $\langle \hat{a}^\dagger \hat{a} \rangle_{s}$ near the multiple-near-degeneracy regime, as displayed in Fig.~\ref{fig.4}.

For the cavity spectrum $E_n=n\omega_c+Un(n-1)/2$, exact degeneracies between distinct cavity levels occur at discrete values of the Kerr nonlinearity. For two different levels $n$ and $m$, the condition $E_n=E_m$ is satisfied when
\begin{equation}
\label{eq:U_nm}
U=U_{n,m}\equiv-\frac{2\omega_c}{n+m-1}.
\end{equation} 
Under a finite photon-number truncation, the allowed values of $n$ and $m$ are finite, so that these exact degeneracies occur only at a discrete and nonuniform set of values of $U$. Whenever $U$ approaches one of these degeneracy points, one or more level spacings become vanishingly small, leading to a sharp suppression of $\langle r\rangle$. Away from these points, $\langle r\rangle$ recovers, giving rise to the pronounced oscillatory structure in Fig.~\ref{fig.4}(a). For the present truncation $N_{\rm ph}=40$, the largest number of pairwise degeneracies occurs around $U\simeq-0.05$, accounting for the particularly deep minimum of $\langle r\rangle$ in this region. As $U$ increases toward zero ($U>-0.05$), the level spacings become uniform and multiple near-degeneracies rapidly disappear. For $U<-0.05$, the number of near-degenerate level pairs also decreases, but more gradually as $U$ becomes increasingly negative.

Figure~\ref{fig.4}(b) shows the order parameter $\langle \hat a^\dagger \hat a\rangle_{s}$ as a function of $g_1$ for several values of the Kerr nonlinearity $U$. When one coupling channel dominates, the cavity changes from the vacuum state to a photon-occupied state on the $g_2$-dominated side and subsequently returns to the vacuum state as the $g_1$ channel becomes dominant. These two continuous changes of the order parameter represent finite-size precursors of second-order DPTs. To better characterize the second-order DPTs, we show the first and second derivatives of $\langle \hat{a}^\dagger \hat{a} \rangle_{s}$ with respect to $g_1$ [presented in Fig.~\ref{fig.8} of Appendix \ref{appendix2}]. In the regime where the two couplings become comparable, the mean photon number develops two abrupt jumps, which serve as finite-size signatures of first-order DPTs. Figures~\ref{fig.4}(a) and (b) exhibit a connection between multiple near-degeneracies and the first-order-DPT signatures. As the number of near-degenerate level pairs decreases, the abrupt jumps in the mean photon number are progressively suppressed and eventually evolve into smooth crossovers. At $U=-0.02$, the abrupt jumps have completely disappeared.

The evolution of the DPT signatures is accompanied by corresponding changes in the stable stored energy. We calculate the stable stored energy as a function of $g_1$ for the same values of $U$ in Fig.~\ref{fig.4}(c). In regimes where one coupling channel dominates, $E_{s}$ exhibits rapid changes near the transition points associated with the second-order DPTs. For $U=-0.02$, $E_{s}$ changes smoothly, consistent with the absence of abrupt jumps in the mean photon number shown in Fig.~\ref{fig.4}(b). In the presence of multiple near-degeneracies, $E_{s}$ is closely correlated with both first- and second-order DPTs. As the number of near-degenerate level pairs decreases, the stored-energy behavior gradually becomes associated with ENET.

We further characterize the first- and second-order DPTs by analyzing the Liouvillian gaps. In the thermodynamic limit, the closing of the Liouvillian gap leads to a divergent relaxation time. In a finite-size system, the gap generally remains finite but can become strongly suppressed near DPTs, resulting in a pronounced increase in the relaxation time, i.e., critical slowing down \cite{RevModPhys.49.435}.  By diagonalizing the Liouvillian, we obtain its eigenvalues $\mu_{j}$ and eigenoperators $\rho_j$. In particular, the dynamics of any state can be recast as
\begin{equation}\label{rhot}
\rho(t) = \rho_{\mathrm{s}} + \sum_{j} c_j e^{\mu_{j}t} \rho_j,
\end{equation}
where the coefficients $c_j$ depend only on the initial state of the system.
Owing to the weak $\mathbb{Z}_2$ symmetry, the Liouvillian superoperator $\mathcal{L}$ has a block-diagonal structure in operator space and decomposes into two independent blocks, denoted by $\mathcal{L}_1$ and $\mathcal{L}_2$ \cite{PhysRevA.98.042118}. Critical slowing down in the two symmetry sectors is associated with different types of DPTs. We denote by $\mu_{\rm 1st}$ the nonzero eigenvalue of $\mathcal{L}_1$ whose real part lies closest to zero, and $\lambda_{\mathrm{1st}} = -\operatorname{Re}(\mu_{\mathrm{1st}})$ characterizes the critical slowing down associated with the first-order DPT. Similarly, $\mu_{\rm SSB}$ denotes the eigenvalue of $\mathcal{L}_2$ whose real part lies closest to zero, and $\lambda_{\rm{SSB}} = -\operatorname{Re}(\mu_{\rm{SSB}})$ characterizes the critical slowing down signaling the second-order DPT and spontaneous symmetry breaking (SSB) in the thermodynamic limit \cite{Beaulieu2025}.

Figure~\ref{fig.5} illustrates the Liouvillian gaps and the steady-state Wigner function associated with the DPTs. $\lambda_{\rm SSB}$ exhibits asymmetric behavior on the two sides of the second-order transition points, as shown in Figs.~\ref{fig.5}(a) and (c). On the vacuum phase side [see Figs.~\ref{fig.5}(d) and (h)], $\lambda_{\rm SSB}$ varies smoothly and remains of the order of the dissipation rate $\lambda_{\rm SSB}\sim\mathcal{O}(\kappa)$. The system therefore relaxes to the steady-state on an ordinary dissipative time scale, and the steady-state respects the symmetry constraint $\rho_{s}=\hat{\Pi}\rho_{s}\hat{\Pi}^{\dagger}$ \cite{PhysRevA.89.022118}. In the vicinity of the second-order transition points, the steady-state of the cavity exhibits squeezing below the vacuum as shown in Figs.~\ref{fig.5}(e) and (g). Between the two second-order transition points, the Wigner function develops a pronounced bimodal structure, as illustrated in Fig.~\ref{fig.5}(f). For a finite-size system, the unique steady-state preserves the $\mathbb Z_2$ symmetry and can be viewed as a mixture of two symmetry-related metastable states, giving rise to the bimodal Wigner function. Correspondingly, $\lambda_{\rm SSB}$ becomes strongly suppressed and $\lambda_{\rm SSB}\ll\kappa$, leading to a dramatic enhancement of the relaxation time. In the thermodynamic limit, $\lambda_{\rm SSB}\to0$, rendering the symmetry-breaking mode stationary and allowing two degenerate symmetry-broken steady-states to emerge. The suppression of $\lambda_{\rm SSB}$ therefore provides a quantitative measure of critical slowing down associated with the second-order DPT. We observe a clear scaling behavior supporting the emergence of genuine SSB in the thermodynamic limit [see also Fig.~\ref{fig.9}(a) of Appendix \ref{appendix2}].

For the first-order DPT, Fig.~\ref{fig.5}(b) shows that $\lambda_{\rm 1st}$ develops two clear minima within the relevant parameter regime. These minima coincide with the abrupt jumps of the mean photon number shown in Fig.~\ref{fig.4}(b), while the finite-size scaling shows that their values approach zero in the thermodynamic limit [see also Fig.~\ref{fig.9}(b) of Appendix \ref{appendix2}]. Between the two minima, $\lambda_{\rm 1st}$ increases again, indicating a recovery of the relaxation rate away from the transition points. In contrast to the second-order DPT, the first-order DPT does not involve SSB, but is characterized by a discontinuous transition in the thermodynamic limit. Near the first-order transition points, the cavity steady-state exhibits coexistence between high- and low-photon-occupation phases, as reflected in the Wigner function [see also Fig.~\ref{fig.10} of Appendix \ref{appendix2}]. For $U=-0.02$, by contrast, $\lambda_{\rm 1st}$ no longer exhibits pronounced minima, consistent with the disappearance of the abrupt jumps in $\langle\hat a^\dagger\hat a\rangle_{\rm ss}$ in Fig.~\ref{fig.4}(b).

\subsection{Reinforcement learning for charging optimization} \label{subsection3}
The long relaxation time required to achieve stable stored energy limits the charging performance on finite time scales. Rather than
relying on this slow relaxation process, we seek a time-dependent protocol that drives the battery rapidly toward a high-energy state. 
\setlength{\textfloatsep}{12pt}
\begin{figure}[htbp]
\centering
\includegraphics[width=0.48\textwidth]{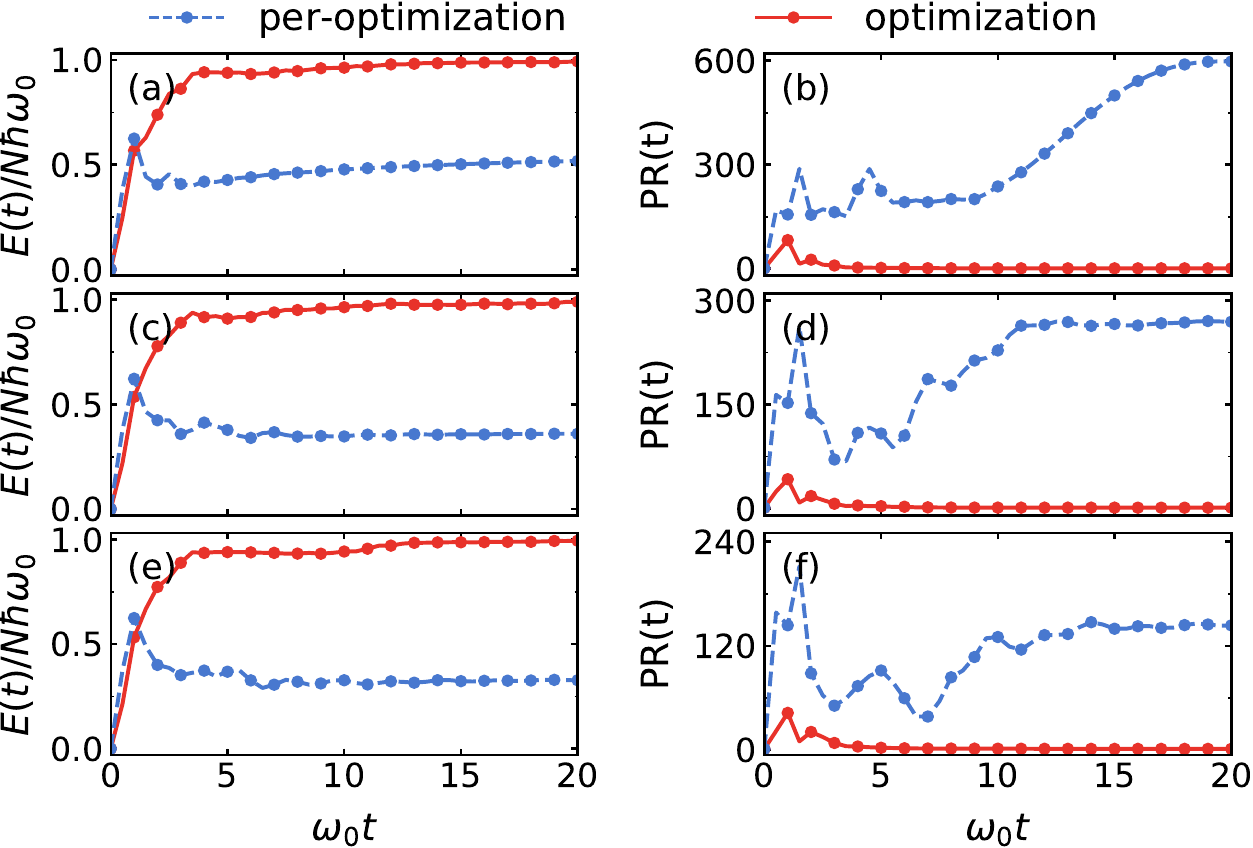}
\caption{\label{fig.6}(a), (c), and (e) The time evolution of stored energy before and after optimization under different Kerr nonlinearity strengths  $U$ (in units of $\omega_{c}$). Panels (b), (d), and (f) depict the dynamics of participation ratio before and after optimization for different Kerr nonlinearity strengths  $U$. The values of $U$ are as follows: (a) and (b) $U=-0.1$, (c) and (d) $U=0$, (e) and (f) $U=0.1$, respectively. All optimizations are performed for $\tilde{g}=1$ (in units of $\omega_{c}$) using nearly the same hyperparameters. In the pre-optimized scheme, we set $g_1=g_2=\tilde{g}$. In the optimized scheme, we set $g_1,g_2 \in[-\tilde{g},\tilde{g}]$.}
\end{figure}
RL provides an efficient framework for this optimization, while we elucidate the physical mechanism underlying the optimized protocol by examining the dynamical features associated with the ergodic and nonergodic phases.

Machine learning techniques, particularly RL, have demonstrated great potential in identifying optimal charging protocols for quantum batteries \cite{PhysRevLett.133.243602,Sun_2025}. We employ the soft actor-critic algorithm (SAC) \cite{pmlr-v80-haarnoja18b,haarnoja2019softactorcriticalgorithmsapplications} to discover the optimal time-dependent charging protocol for the battery. Specifically, two time-dependent control parameters, namely the rotating coupling $g_1(t)$ and the counter-rotating coupling $g_2(t)$, are simultaneously optimized. Discretizing charging time in steps of duration $\Delta t$ during which the controls are constant, the RL method determines the values of $g_1(t)$ and $g_2(t)$ (see Appendix \ref{appendix3} for details). Under different Kerr nonlinearity strengths, we seek to maximize the stored energy of the battery within a relatively short time. We then compare this charging protocol to the commonly employed charging protocol with time-independent coupling parameters, which corresponds to setting  the coupling terms to constants for  $t \in [0, \tau]$ .

Figures~\ref{fig.6}(a), (c), and (e) show the time evolution of the stored energy before and after optimization for different Kerr nonlinearity strengths. Compared with the pre-optimized protocol, the optimized protocol enables the battery to approach its maximum stored energy within a substantially shorter charging time and maintains it over an extended time interval. 
To gain further insight into the mechanism underlying this performance enhancement, we analyze the participation-ratio dynamics under different Kerr nonlinearity strengths presented in Figs.~\ref{fig.6}(b), (d), and (f). In closed quantum systems, participation-ratio dynamics has been employed to characterize the dynamical features associated with ergodic and nonergodic phases \cite{PhysRevA.107.043706}. We extend this approach to open quantum systems by replacing the steady-state projection in Eq. (\ref{PR}) with the projection of the time-evolved state $\rho(t)$ onto the even parity subspace, i.e., $\rho(t) = \sum_{\alpha}^{N_1} \psi_{\alpha}(t) |\alpha\rangle\rangle$. The initial state of the system lies in the nonergodic phase. 
Before optimization, the stored energy exhibits a lower energy level, while the PR fluctuates transiently and grows to considerably larger values. Under the optimized protocol, the stored energy rapidly increases and maintains a high energy level throughout the subsequent evolution, whereas PR is suppressed to a persistently low level. The comparison of $\rm{PR(t)}$ suggests that the RL protocol steers the system toward a more nonergodic dynamical regime, characterized by a strongly suppressed $\rm{PR(t)}$, which is accompanied by enhanced stored energy.

\section{CONCLUSIONS}\label{section4}
 We have investigated the charging performance and PT properties of an anisotropic Dicke quantum battery in the presence of Kerr nonlinearity and photon loss. We have shown that the competition between the rotating and counter-rotating couplings gives rise to distinct \mbox{low-,} \mbox{intermediate-,} and high-energy-storage regimes, while the Kerr nonlinearity provides an additional control parameter for tuning these regimes. Away from the parameter region with multiple near-degeneracies in the cavity spectrum, the stable stored energy is closely correlated with the ENET: the nonergodic phase is associated with extremal stored energy, whereas the ergodic phase corresponds to intermediate stored energy. Within the multiple-near-degeneracy region, this correspondence no longer holds strictly, and the stored energy is instead governed by first- and second-order DPTs. To characterize these DPTs, we have employed Liouvillian spectral theory to quantify the critical slowing down associated with the first- and second-order DPTs, and used the Wigner function to examine the properties of the cavity field near transition points. The finite-size scaling of the relevant Liouvillian gaps further supports the emergence of genuine criticality and SSB in the thermodynamic limit. 
 Finally, we have optimized the charging protocol via RL based on the SAC, which enables the battery to attain a high stored energy within a substantially shorter time and to maintain it over an extended interval. By analyzing the participation-ratio dynamics, we indicate that the performance enhancement originates from the protocol steering the system toward a more nonergodic dynamical regime. Our results offer a feasible pathway for manipulating and optimizing quantum battery performance based on intrinsic PT behaviors.
\setlength{\textfloatsep}{\originaltextfloatsep}
\section*{Acknowledgments}
The work is supported by the National Natural Science Foundation of China (Grant No. 12475026) and the Natural Science Foundation of Gansu Province (No. 25JRRA799).
\appendix
\section{CONVERGENCE OF THE PHOTONIC FOCK SPACE TRUNCATION} \label{appendix1}
\begin{figure}[htbp]
\centering
\includegraphics[width=0.42\textwidth]{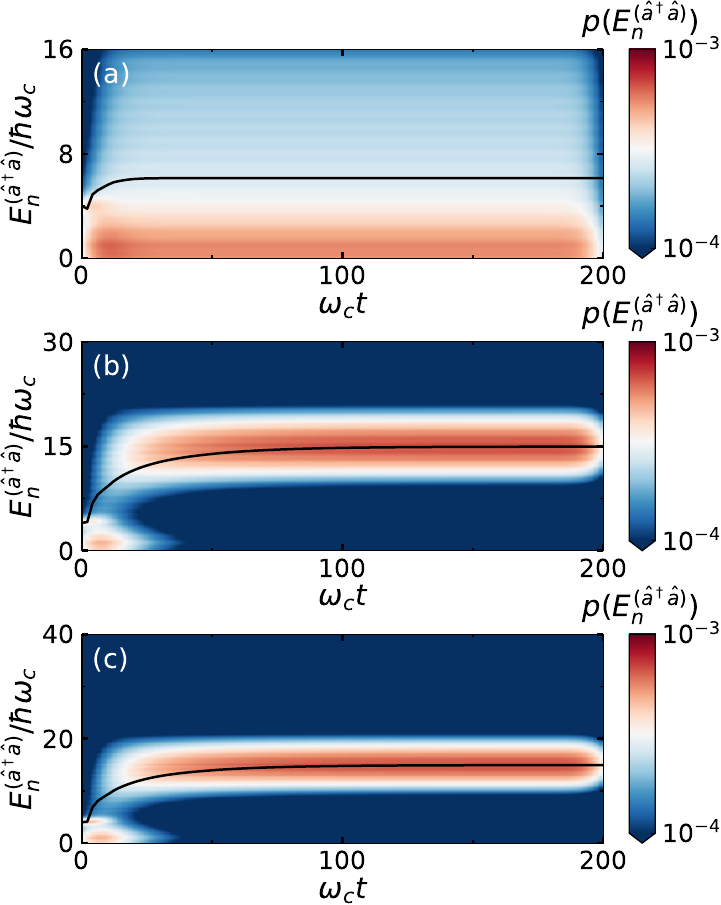}
\caption{\label{fig.7}Time evolution of the probability distribution over the spectrum of $\hat{a}^\dagger\hat{a}$ for $U=-0.1$ (in units of $\omega_{c}$), with photon-number cutoffs (a) $N_{\rm ph}=16$, (b) $N_{\rm ph}=30$, and (c) $N_{\rm ph}=40$. The black solid lines denote the mean photon number and the two coupling strengths $g_1$ and $g_2$  are equal.}
\end{figure}

In Fig.~\ref{fig.7}, we show the time evolution of the probability distribution for several truncation dimensions in the regime where the cavity spectrum exhibits multiple near-degeneracies. For the Dicke quantum battery, the commonly used cutoff $N_{\mathrm{ph}}=4N$ is generally sufficient to obtain converged dynamical observables 
\cite{PhysRevLett.120.117702}. In the present case, however, this cutoff becomes insufficient because the multiple near-degeneracies strongly enhance the occupation of high-photon-number states. As shown in Fig.~\ref{fig.7}(a), for $N_{\rm ph}=4N=16$, the distribution retains appreciable weight at the truncation boundary, indicating that the accessible Fock space is too small to faithfully capture the dynamics. Correspondingly, the long-time mean photon number is limited to $\langle \hat a^\dagger \hat a\rangle \approx 6$. Upon increasing the cutoff to $N_{\rm ph}=30$, as shown in Fig.~\ref{fig.7}(b), the population becomes well contained within the truncated Fock space, and the long-time mean photon number increases to $\langle \hat a^\dagger \hat a\rangle \approx 15$. A further increase to $N_{\rm ph}=40$, shown in Fig.~\ref{fig.7}(c), leaves both the population distribution and the mean photon number
essentially unchanged, confirming that $N_{\rm ph}=30$ is already sufficient to achieve numerical convergence.
\section{THE FIRST- AND SECOND-ORDER DPTS AND A SCALING BEHAVIOR} \label{appendix2}
\begin{figure}[htbp]
\centering
\includegraphics[width=0.49\textwidth]{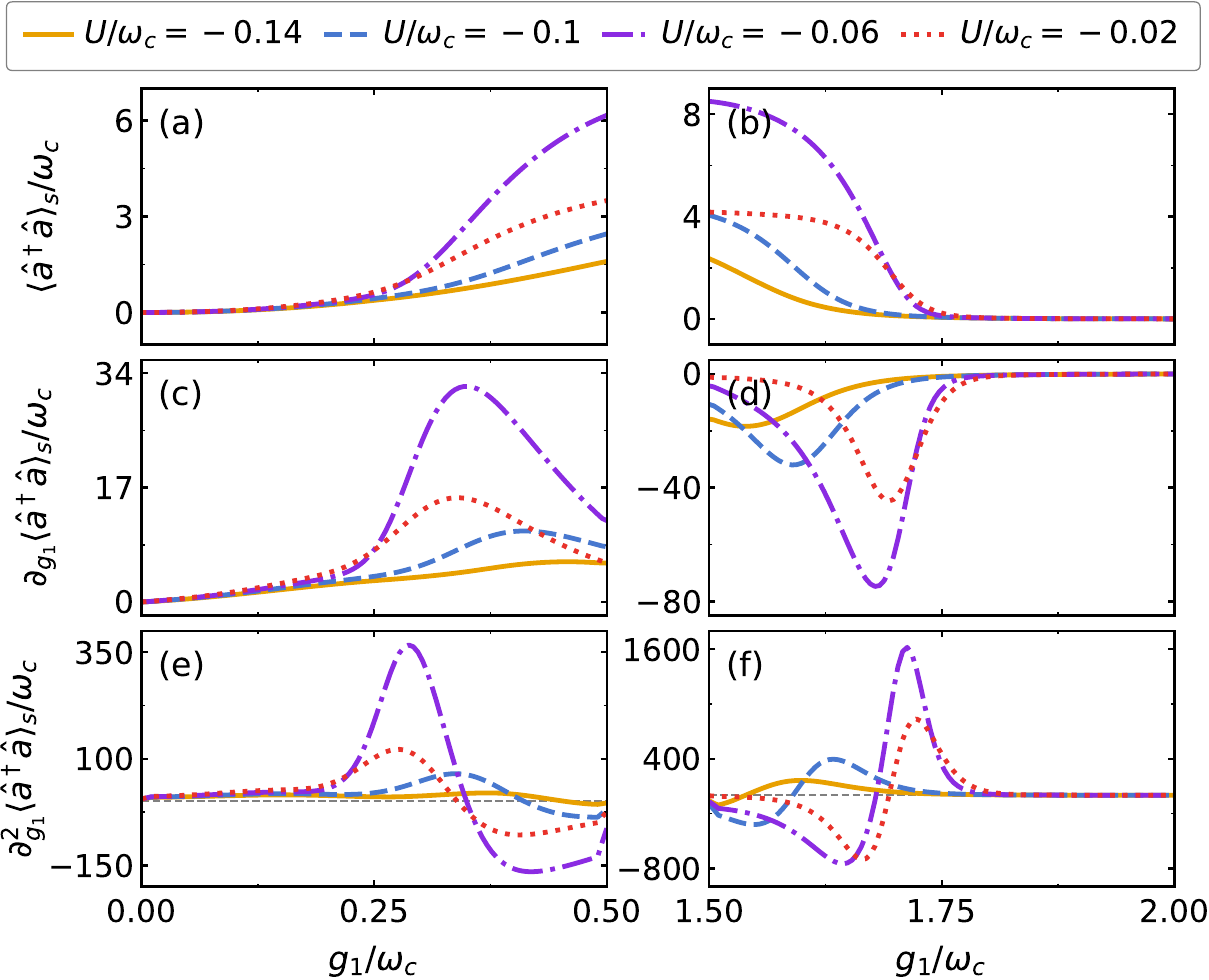}
\caption{\label{fig.8}(a), (b) The mean photon number as a function of $g_1$ (in units of $\omega_c$ ) for different Kerr nonlinearity strengths $U$ (in units of $\omega_c$). Its first (c), (d) and second derivatives (e), (f) as a function of $g_1$ for different Kerr nonlinearity strengths $U$. The thin gray horizontal lines in panels (e) and (f) indicate the zero level.}
\end{figure}

\begin{figure}[htbp]
\centering
\includegraphics[width=0.48\textwidth]{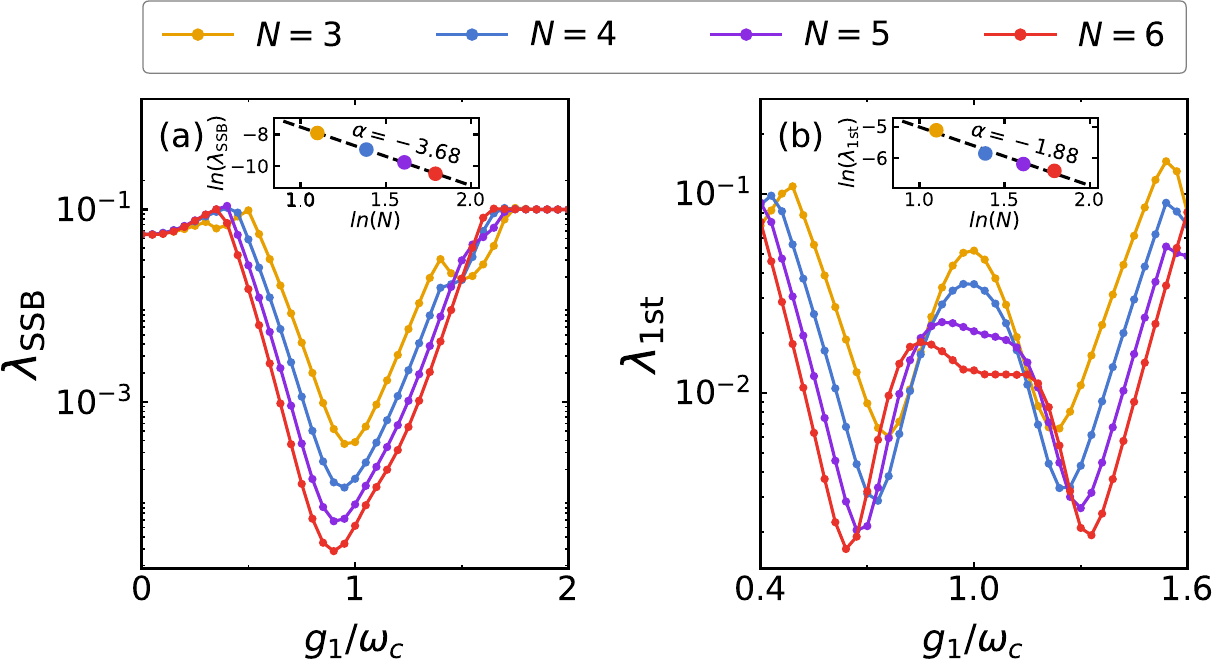}
\caption{\label{fig.9}The Liouvillian gaps $\lambda_{\mathrm{SSB}}$  and $\lambda_{\mathrm{1st}}$ as a function of $g_1$ (in units of $\omega_c$ ) for different scaling parameter $N$. The inset of (a) shows the minimum of $\lambda_{\mathrm{SSB}}$ as a function of the scaling parameter $N$. The inset of (b) shows the minimum value of $\lambda_{\mathrm{1st}}$ near the second first-order critical point as a function of the scaling parameter $N$. The black line shows the fit of the functions $\lambda_{\mathrm{1st}} \propto N^{\alpha}$ and $\lambda_{\mathrm{SSB}} \propto N^{\alpha} $.}
\end{figure}

\begin{figure}[htbp]
\centering
\includegraphics[width=0.43\textwidth,clip]{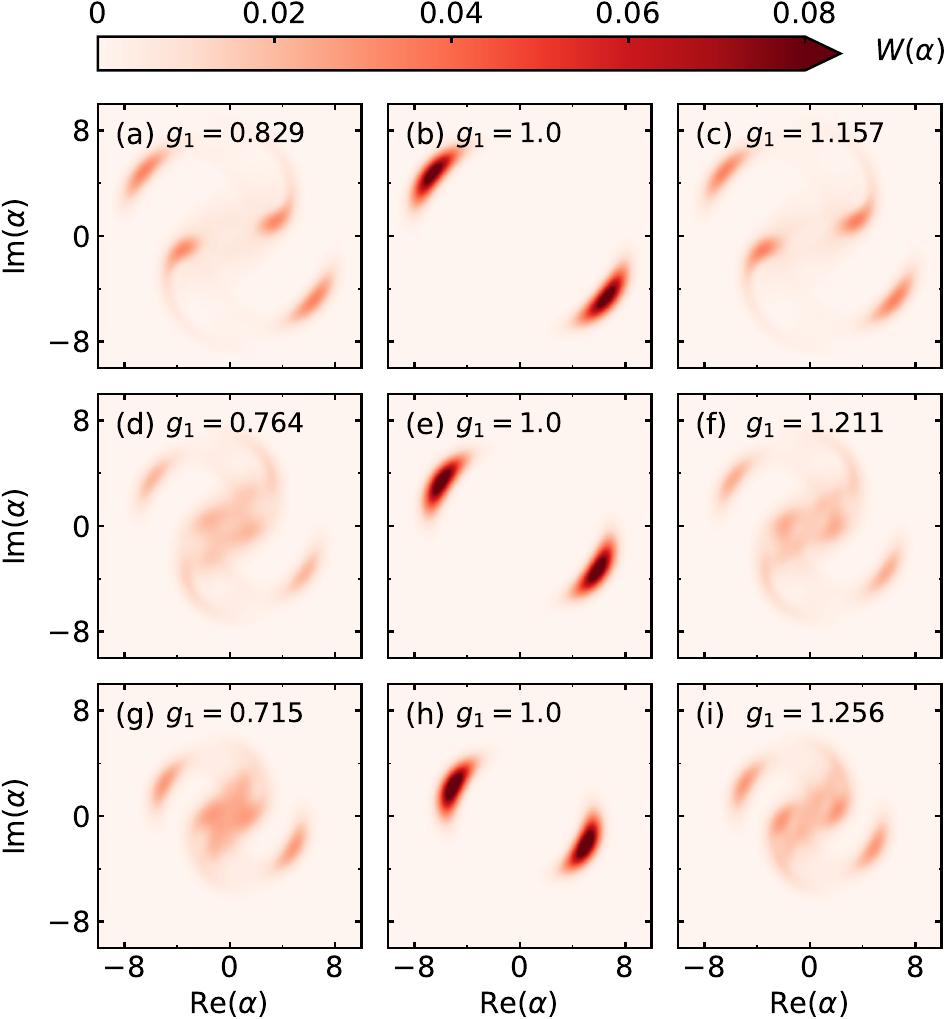}
\caption{\label{fig.10}The Wigner function corresponding to the vicinity of the first-order critical points for different Kerr nonlinearity strengths $U$ (in units of $\omega_c$). The values of $U$ are as follows: (a)-(c) $U=-0.04$, (d)-(f) $U=-0.06$, and (g)-(i) $U=-0.1$.}
\end{figure}

To further characterize the second-order DPTs, we examine the first and second derivatives of the photon number with respect to cavity-atoms coupling in Fig.~\ref{fig.8}.
We display the finite-size scaling of the Liouvillian gaps $\lambda_{\mathrm{SSB}}$ and $\lambda_{\mathrm{1st}}$, as shown in Fig.~\ref{fig.9}, revealing a progressive suppression of the relaxation rates as the system size $N$ increases. For the second-order DPT, the minimum of $\lambda_{\mathrm{SSB}}$ decreases with $N$ following a power-law behavior, $\lambda_{\mathrm{SSB}}^{\min}\propto N^{\alpha}$, with $\alpha=-3.68$, as shown in the inset of Fig.~\ref{fig.9}(a). This scaling behavior supports the closing of the Liouvillian gap and the emergence of SSB in the thermodynamic limit. For the first-order DPT, the minimum of $\lambda_{\mathrm{1st}}$ near each critical point likewise decreases with increasing $N$. The inset of Fig.~\ref{fig.9}(b) shows the scaling near the second critical point, yielding $\alpha=-1.88$, while a comparable exponent $\alpha=-1.73$ is obtained near the first critical point.
We investigate the cavity Wigner functions in the vicinity of the first-order DPTs, as displayed in Fig.~\ref{fig.10}. Panels (a), (d), and (g) correspond to the vicinity of the first transition point, while panels (c), (f), and (i) correspond to the vicinity of the second one. Panels (b), (e), and (h) show the regime between the two transition points. From top to bottom, the three rows correspond to different Kerr nonlinearity strengths $U$. 
The order parameter $\langle \hat a^\dagger \hat a\rangle_{s}$ exhibits two abrupt jumps in opposite directions as $g_1$ increases, as shown in Fig.~\ref{fig.4}(b). Near the first transition point, the system changes from a low-photon-occupation phase to a high-photon-occupation phase, whereas near the second transition point it changes back to the low-photon-occupation phase. Correspondingly, the Wigner functions near the two transition points [Figs.~\ref{fig.10}(a), (d), (g) and Figs.~\ref{fig.10}(c), (f), (i)] exhibit signatures of coexistence between the low- and high-photon-occupation phases. Between the two transition points, the system remains in the high-photon-occupation phase, and the Wigner function develops a pronounced bimodal structure, as shown in Figs.~\ref{fig.10}(b), (e), and (h).

\section{DETAILS ON REINFORCEMENT LEARNING FOR QUANTUM BATTERY} \label{appendix3}
We employ RL to identify an optimal charging protocol that maximizes the final energy stored. RL is formulated within the framework of a Markov decision process \cite{Sutton2018}, in which an agent learns to accomplish a task by repeatedly interacting with an environment. The charging interval
\begin{table}[htbp]
\caption{Hyperparameters used in all numerical calculations (k denotes thousand).
}
\label{tab:hyperparams}
\begin{tabular}{lc}
	\noalign{\hrule height 0.8pt}
	\rule[-1.5ex]{0pt}{4.5ex}Hyperparameter & Value \\
	\colrule \noalign{\vskip 3pt}
	Training steps & 500k \\
	Learning rate for $Q$-network and policy $LR$ & 0.001 \\
	Learning rate for entropy coefficient $LR_\alpha$ & 0.003 \\
	Discount factor $\gamma$ & 0.993 \\
	Size of replay buffer $D$ & 180k \\
	Polyak coefficient $\beta$ & 0.995 \\
	Units in first hidden layer & 512 \\
	Units in second hidden layer & 256 \\
	Initial random steps & 5k \\
	Update networks after this number of steps & 1k \\
	Initial value of the policy entropy & 0.7 \\
	Final value of the policy entropy & $-2.8$ \\
	Number of steps where policy entropy transition & 200k \\
	Batch size & 256 \\
	\noalign{\vskip 3pt}
	\noalign{\hrule height 0.8pt}
\end{tabular}
\end{table}
$[0,\tau]$ is discretized into $M$ time steps. At each step $i$, the agent selects an action $a_i$ according to a policy $\pi(a_i \mid s_i)$, given the current state $s_i$ of the environment. The environment subsequently evolves to a new state $s_{i+1}$ and returns a scalar reward $r_{i+1}$. Each episode starts from a fixed initial state $\sigma_0$ and terminates at time $\tau$. The objective of RL is to learn an optimal policy $\pi^*$ that maximizes the expected cumulative discounted return over an episode,
\begin{eqnarray}\label{g_0}
g_0 = r_1 + \gamma r_2 + \cdots + \gamma^{M-2} r_{M-1}     = \sum_{k=0}^{M-2} \gamma^k r_{k+1}
\end{eqnarray}
where $\gamma \in [0, 1]$ is the so-called ``discount factor'' which determines how much we privilege short or long-term rewards. An optimal policy is thus defined as
\begin{eqnarray}\label{pi}
\pi^* = \arg\max_{\pi} \mathbb{E}_{\pi} [g_0 \mid s_0 = \sigma_0]
\end{eqnarray}
where the expectation value $\mathbb{E}_{\pi} [\cdot]$ in Eq. (\ref{pi}) is taken with respect to the stochasticity in the choice of the actions according to the policy $\pi$, and with respect to the state evolution of the environment.

To formulate the charging optimization within this framework \cite{PhysRevLett.133.243602,Sun_2025}, we take the state $s_i$ of the environment to be $\rho(t_i)$, together with the last chosen action and the current time step. The action specifies the values of the control variables $g_1(t)$ and $g_2(t)\in [-1,1]$, which are held constant over the interval $[t_i,t_{i+1}]$. This construction naturally yields a piecewise-constant charging protocol.  A reward $r_{i+1}$ is constructed from the variation of stored energy between $t_i$  and $t_{i+1}$. The new state $s_{i+1}$ is obtained by evolving $s_{i}$ under the Lindblad master equation (\ref{Liouvillian}) for the chosen values of $g_1(t_i)$ and $g_2(t_i)$. The initial state $\sigma_0$ encodes the state $\rho(0)$ defined in the main text.

We optimize the charging protocol using the SAC algorithm. SAC belongs to the class of actor-critic methods: it uses separate neural networks to parameterize the policy function $\pi(a|s)$ and the value function $Q(s,a)$. The former plays the role of the ``actor'', which selects actions, while the latter serves as the ``critic'', which evaluates the expected return obtainable from a given state. Starting from a random policy, the algorithm alternates between two steps: policy evaluation and policy improvement. In the policy evaluation step, the value function is estimated by minimizing a loss derived from the Bellman equation \cite{Sutton2018}. In the policy improvement step, the policy is updated toward higher returns using the value function. By iterating these steps over many training episodes, the algorithm gradually 
converges to an optimal (or near-optimal) policy. In contrast to the standard formulation, SAC adds an entropy regularization term to the optimization objective and balances exploration against exploitation through an automatically tuned ``temperature'' parameter $\alpha$, i.e., a more exploratory behavior is encouraged early in training, while the policy becomes nearly deterministic toward the end of training. We thus refer to Refs. \cite{pmlr-v80-haarnoja18b,haarnoja2019softactorcriticalgorithmsapplications} for further details of the method. All hyperparameters used in our numerical calculations are provided in Table \ref{tab:hyperparams}. We ran our simulations on a desktop computer using an NVIDIA RTX A6000 as GPU.

\bibliography{reference}

\begin{thebibliography}{117}%
\makeatletter
\providecommand \@ifxundefined [1]{%
 \@ifx{#1\undefined}
}%
\providecommand \@ifnum [1]{%
 \ifnum #1\expandafter \@firstoftwo
 \else \expandafter \@secondoftwo
 \fi
}%
\providecommand \@ifx [1]{%
 \ifx #1\expandafter \@firstoftwo
 \else \expandafter \@secondoftwo
 \fi
}%
\providecommand \natexlab [1]{#1}%
\providecommand \enquote  [1]{``#1''}%
\providecommand \bibnamefont  [1]{#1}%
\providecommand \bibfnamefont [1]{#1}%
\providecommand \citenamefont [1]{#1}%
\providecommand \href@noop [0]{\@secondoftwo}%
\providecommand \href [0]{\begingroup \@sanitize@url \@href}%
\providecommand \@href[1]{\@@startlink{#1}\@@href}%
\providecommand \@@href[1]{\endgroup#1\@@endlink}%
\providecommand \@sanitize@url [0]{\catcode `\\12\catcode `\$12\catcode
  `\&12\catcode `\#12\catcode `\^12\catcode `\_12\catcode `\%12\relax}%
\providecommand \@@startlink[1]{}%
\providecommand \@@endlink[0]{}%
\providecommand \url  [0]{\begingroup\@sanitize@url \@url }%
\providecommand \@url [1]{\endgroup\@href {#1}{\urlprefix }}%
\providecommand \urlprefix  [0]{URL }%
\providecommand \Eprint [0]{\href }%
\providecommand \doibase [0]{https://doi.org/}%
\providecommand \selectlanguage [0]{\@gobble}%
\providecommand \bibinfo  [0]{\@secondoftwo}%
\providecommand \bibfield  [0]{\@secondoftwo}%
\providecommand \translation [1]{[#1]}%
\providecommand \BibitemOpen [0]{}%
\providecommand \bibitemStop [0]{}%
\providecommand \bibitemNoStop [0]{.\EOS\space}%
\providecommand \EOS [0]{\spacefactor3000\relax}%
\providecommand \BibitemShut  [1]{\csname bibitem#1\endcsname}%
\let\auto@bib@innerbib\@empty
\bibitem [{\citenamefont {Acharya}\ \emph {et~al.}(2025)\citenamefont
  {Acharya}, \citenamefont {Abanin}, \citenamefont {Aghababaie-Beni} \emph
  {et~al.}}]{Acharya2025}%
  \BibitemOpen
  \bibfield  {author} {\bibinfo {author} {\bibfnamefont {R.}~\bibnamefont
  {Acharya}}, \bibinfo {author} {\bibfnamefont {D.~A.}\ \bibnamefont {Abanin}},
  \bibinfo {author} {\bibfnamefont {L.}~\bibnamefont {Aghababaie-Beni}}, \emph
  {et~al.},\ }\bibfield  {title} {\bibinfo {title} {Quantum error correction
  below the surface code threshold},\ }\href
  {https://doi.org/10.1038/s41586-024-08449-y} {\bibfield  {journal} {\bibinfo
  {journal} {Nature}\ }\textbf {\bibinfo {volume} {638}},\ \bibinfo {pages}
  {920} (\bibinfo {year} {2025})}\BibitemShut {NoStop}%
\bibitem [{\citenamefont {Daley}\ \emph {et~al.}(2022)\citenamefont {Daley},
  \citenamefont {Bloch}, \citenamefont {Kokail}, \citenamefont {Flannigan},
  \citenamefont {Pearson}, \citenamefont {Troyer},\ and\ \citenamefont
  {Zoller}}]{Daley2022}%
  \BibitemOpen
  \bibfield  {author} {\bibinfo {author} {\bibfnamefont {A.~J.}\ \bibnamefont
  {Daley}}, \bibinfo {author} {\bibfnamefont {I.}~\bibnamefont {Bloch}},
  \bibinfo {author} {\bibfnamefont {C.}~\bibnamefont {Kokail}}, \bibinfo
  {author} {\bibfnamefont {S.}~\bibnamefont {Flannigan}}, \bibinfo {author}
  {\bibfnamefont {N.}~\bibnamefont {Pearson}}, \bibinfo {author} {\bibfnamefont
  {M.}~\bibnamefont {Troyer}},\ and\ \bibinfo {author} {\bibfnamefont
  {P.}~\bibnamefont {Zoller}},\ }\bibfield  {title} {\bibinfo {title}
  {Practical quantum advantage in quantum simulation},\ }\href
  {https://doi.org/10.1038/s41586-022-04940-6} {\bibfield  {journal} {\bibinfo
  {journal} {Nature}\ }\textbf {\bibinfo {volume} {607}},\ \bibinfo {pages}
  {667} (\bibinfo {year} {2022})}\BibitemShut {NoStop}%
\bibitem [{\citenamefont {Chen}\ \emph {et~al.}(2021)\citenamefont {Chen},
  \citenamefont {Zhang}, \citenamefont {Chen}, \citenamefont {Cai} \emph
  {et~al.}}]{Chen2021}%
  \BibitemOpen
  \bibfield  {author} {\bibinfo {author} {\bibfnamefont {Y.-A.}\ \bibnamefont
  {Chen}}, \bibinfo {author} {\bibfnamefont {Q.}~\bibnamefont {Zhang}},
  \bibinfo {author} {\bibfnamefont {T.-Y.}\ \bibnamefont {Chen}}, \bibinfo
  {author} {\bibfnamefont {W.-Q.}\ \bibnamefont {Cai}}, \emph {et~al.},\
  }\bibfield  {title} {\bibinfo {title} {An integrated space-to-ground quantum
  communication network over 4,600 kilometres},\ }\href
  {https://doi.org/10.1038/s41586-020-03093-8} {\bibfield  {journal} {\bibinfo
  {journal} {Nature}\ }\textbf {\bibinfo {volume} {589}},\ \bibinfo {pages}
  {214} (\bibinfo {year} {2021})}\BibitemShut {NoStop}%
\bibitem [{\citenamefont {Degen}\ \emph {et~al.}(2017)\citenamefont {Degen},
  \citenamefont {Reinhard},\ and\ \citenamefont
  {Cappellaro}}]{RevModPhys.89.035002}%
  \BibitemOpen
  \bibfield  {author} {\bibinfo {author} {\bibfnamefont {C.~L.}\ \bibnamefont
  {Degen}}, \bibinfo {author} {\bibfnamefont {F.}~\bibnamefont {Reinhard}},\
  and\ \bibinfo {author} {\bibfnamefont {P.}~\bibnamefont {Cappellaro}},\
  }\bibfield  {title} {\bibinfo {title} {Quantum sensing},\ }\href
  {https://doi.org/10.1103/RevModPhys.89.035002} {\bibfield  {journal}
  {\bibinfo  {journal} {Rev. Mod. Phys.}\ }\textbf {\bibinfo {volume} {89}},\
  \bibinfo {pages} {035002} (\bibinfo {year} {2017})}\BibitemShut {NoStop}%
\bibitem [{\citenamefont {Alicki}\ and\ \citenamefont
  {Fannes}(2013)}]{PhysRevE.87.042123}%
  \BibitemOpen
  \bibfield  {author} {\bibinfo {author} {\bibfnamefont {R.}~\bibnamefont
  {Alicki}}\ and\ \bibinfo {author} {\bibfnamefont {M.}~\bibnamefont
  {Fannes}},\ }\bibfield  {title} {\bibinfo {title} {Entanglement boost for
  extractable work from ensembles of quantum batteries},\ }\href
  {https://doi.org/10.1103/PhysRevE.87.042123} {\bibfield  {journal} {\bibinfo
  {journal} {Phys. Rev. E}\ }\textbf {\bibinfo {volume} {87}},\ \bibinfo
  {pages} {042123} (\bibinfo {year} {2013})}\BibitemShut {NoStop}%
\bibitem [{\citenamefont {Binder}\ \emph {et~al.}(2015)\citenamefont {Binder},
  \citenamefont {Vinjanampathy}, \citenamefont {Modi},\ and\ \citenamefont
  {Goold}}]{Binder_2015}%
  \BibitemOpen
  \bibfield  {author} {\bibinfo {author} {\bibfnamefont {F.~C.}\ \bibnamefont
  {Binder}}, \bibinfo {author} {\bibfnamefont {S.}~\bibnamefont
  {Vinjanampathy}}, \bibinfo {author} {\bibfnamefont {K.}~\bibnamefont
  {Modi}},\ and\ \bibinfo {author} {\bibfnamefont {J.}~\bibnamefont {Goold}},\
  }\bibfield  {title} {\bibinfo {title} {Quantacell: powerful charging of
  quantum batteries},\ }\href {https://doi.org/10.1088/1367-2630/17/7/075015}
  {\bibfield  {journal} {\bibinfo  {journal} {New J. Phys.}\ }\textbf {\bibinfo
  {volume} {17}},\ \bibinfo {pages} {075015} (\bibinfo {year}
  {2015})}\BibitemShut {NoStop}%
\bibitem [{\citenamefont {Campaioli}\ \emph {et~al.}(2024)\citenamefont
  {Campaioli}, \citenamefont {Gherardini}, \citenamefont {Quach}, \citenamefont
  {Polini},\ and\ \citenamefont {Andolina}}]{RevModPhys.96.031001}%
  \BibitemOpen
  \bibfield  {author} {\bibinfo {author} {\bibfnamefont {F.}~\bibnamefont
  {Campaioli}}, \bibinfo {author} {\bibfnamefont {S.}~\bibnamefont
  {Gherardini}}, \bibinfo {author} {\bibfnamefont {J.~Q.}\ \bibnamefont
  {Quach}}, \bibinfo {author} {\bibfnamefont {M.}~\bibnamefont {Polini}},\ and\
  \bibinfo {author} {\bibfnamefont {G.~M.}\ \bibnamefont {Andolina}},\
  }\bibfield  {title} {\bibinfo {title} {Colloquium: Quantum batteries},\
  }\href {https://doi.org/10.1103/RevModPhys.96.031001} {\bibfield  {journal}
  {\bibinfo  {journal} {Rev. Mod. Phys.}\ }\textbf {\bibinfo {volume} {96}},\
  \bibinfo {pages} {031001} (\bibinfo {year} {2024})}\BibitemShut {NoStop}%
\bibitem [{\citenamefont {Ferraro}\ \emph {et~al.}(2026)\citenamefont
  {Ferraro}, \citenamefont {Cavaliere}, \citenamefont {Genoni}, \citenamefont
  {Benenti},\ and\ \citenamefont {Sassetti}}]{Ferraro2026}%
  \BibitemOpen
  \bibfield  {author} {\bibinfo {author} {\bibfnamefont {D.}~\bibnamefont
  {Ferraro}}, \bibinfo {author} {\bibfnamefont {F.}~\bibnamefont {Cavaliere}},
  \bibinfo {author} {\bibfnamefont {M.~G.}\ \bibnamefont {Genoni}}, \bibinfo
  {author} {\bibfnamefont {G.}~\bibnamefont {Benenti}},\ and\ \bibinfo {author}
  {\bibfnamefont {M.}~\bibnamefont {Sassetti}},\ }\bibfield  {title} {\bibinfo
  {title} {Opportunities and challenges of quantum batteries},\ }\href
  {https://doi.org/10.1038/s42254-025-00906-5} {\bibfield  {journal} {\bibinfo
  {journal} {Nat. Rev. Phys.}\ }\textbf {\bibinfo {volume} {8}},\ \bibinfo
  {pages} {115} (\bibinfo {year} {2026})}\BibitemShut {NoStop}%
\bibitem [{\citenamefont {Campaioli}\ \emph {et~al.}(2017)\citenamefont
  {Campaioli}, \citenamefont {Pollock}, \citenamefont {Binder}, \citenamefont
  {C\'eleri}, \citenamefont {Goold}, \citenamefont {Vinjanampathy},\ and\
  \citenamefont {Modi}}]{PhysRevLett.118.150601}%
  \BibitemOpen
  \bibfield  {author} {\bibinfo {author} {\bibfnamefont {F.}~\bibnamefont
  {Campaioli}}, \bibinfo {author} {\bibfnamefont {F.~A.}\ \bibnamefont
  {Pollock}}, \bibinfo {author} {\bibfnamefont {F.~C.}\ \bibnamefont {Binder}},
  \bibinfo {author} {\bibfnamefont {L.}~\bibnamefont {C\'eleri}}, \bibinfo
  {author} {\bibfnamefont {J.}~\bibnamefont {Goold}}, \bibinfo {author}
  {\bibfnamefont {S.}~\bibnamefont {Vinjanampathy}},\ and\ \bibinfo {author}
  {\bibfnamefont {K.}~\bibnamefont {Modi}},\ }\bibfield  {title} {\bibinfo
  {title} {Enhancing the charging power of quantum batteries},\ }\href
  {https://doi.org/10.1103/PhysRevLett.118.150601} {\bibfield  {journal}
  {\bibinfo  {journal} {Phys. Rev. Lett.}\ }\textbf {\bibinfo {volume} {118}},\
  \bibinfo {pages} {150601} (\bibinfo {year} {2017})}\BibitemShut {NoStop}%
\bibitem [{\citenamefont {Andolina}\ \emph
  {et~al.}(2019{\natexlab{a}})\citenamefont {Andolina}, \citenamefont {Keck},
  \citenamefont {Mari}, \citenamefont {Campisi}, \citenamefont {Giovannetti},\
  and\ \citenamefont {Polini}}]{PhysRevLett.122.047702}%
  \BibitemOpen
  \bibfield  {author} {\bibinfo {author} {\bibfnamefont {G.~M.}\ \bibnamefont
  {Andolina}}, \bibinfo {author} {\bibfnamefont {M.}~\bibnamefont {Keck}},
  \bibinfo {author} {\bibfnamefont {A.}~\bibnamefont {Mari}}, \bibinfo {author}
  {\bibfnamefont {M.}~\bibnamefont {Campisi}}, \bibinfo {author} {\bibfnamefont
  {V.}~\bibnamefont {Giovannetti}},\ and\ \bibinfo {author} {\bibfnamefont
  {M.}~\bibnamefont {Polini}},\ }\bibfield  {title} {\bibinfo {title}
  {Extractable work, the role of correlations, and asymptotic freedom in
  quantum batteries},\ }\href {https://doi.org/10.1103/PhysRevLett.122.047702}
  {\bibfield  {journal} {\bibinfo  {journal} {Phys. Rev. Lett.}\ }\textbf
  {\bibinfo {volume} {122}},\ \bibinfo {pages} {047702} (\bibinfo {year}
  {2019}{\natexlab{a}})}\BibitemShut {NoStop}%
\bibitem [{\citenamefont {Andolina}\ \emph
  {et~al.}(2019{\natexlab{b}})\citenamefont {Andolina}, \citenamefont {Keck},
  \citenamefont {Mari}, \citenamefont {Giovannetti},\ and\ \citenamefont
  {Polini}}]{PhysRevB.99.205437}%
  \BibitemOpen
  \bibfield  {author} {\bibinfo {author} {\bibfnamefont {G.~M.}\ \bibnamefont
  {Andolina}}, \bibinfo {author} {\bibfnamefont {M.}~\bibnamefont {Keck}},
  \bibinfo {author} {\bibfnamefont {A.}~\bibnamefont {Mari}}, \bibinfo {author}
  {\bibfnamefont {V.}~\bibnamefont {Giovannetti}},\ and\ \bibinfo {author}
  {\bibfnamefont {M.}~\bibnamefont {Polini}},\ }\bibfield  {title} {\bibinfo
  {title} {Quantum versus classical many-body batteries},\ }\href
  {https://doi.org/10.1103/PhysRevB.99.205437} {\bibfield  {journal} {\bibinfo
  {journal} {Phys. Rev. B}\ }\textbf {\bibinfo {volume} {99}},\ \bibinfo
  {pages} {205437} (\bibinfo {year} {2019}{\natexlab{b}})}\BibitemShut
  {NoStop}%
\bibitem [{\citenamefont {Ferraro}\ \emph {et~al.}(2018)\citenamefont
  {Ferraro}, \citenamefont {Campisi}, \citenamefont {Andolina}, \citenamefont
  {Pellegrini},\ and\ \citenamefont {Polini}}]{PhysRevLett.120.117702}%
  \BibitemOpen
  \bibfield  {author} {\bibinfo {author} {\bibfnamefont {D.}~\bibnamefont
  {Ferraro}}, \bibinfo {author} {\bibfnamefont {M.}~\bibnamefont {Campisi}},
  \bibinfo {author} {\bibfnamefont {G.~M.}\ \bibnamefont {Andolina}}, \bibinfo
  {author} {\bibfnamefont {V.}~\bibnamefont {Pellegrini}},\ and\ \bibinfo
  {author} {\bibfnamefont {M.}~\bibnamefont {Polini}},\ }\bibfield  {title}
  {\bibinfo {title} {High-power collective charging of a solid-state quantum
  battery},\ }\href {https://doi.org/10.1103/PhysRevLett.120.117702} {\bibfield
   {journal} {\bibinfo  {journal} {Phys. Rev. Lett.}\ }\textbf {\bibinfo
  {volume} {120}},\ \bibinfo {pages} {117702} (\bibinfo {year}
  {2018})}\BibitemShut {NoStop}%
\bibitem [{\citenamefont {Farina}\ \emph {et~al.}(2019)\citenamefont {Farina},
  \citenamefont {Andolina}, \citenamefont {Mari}, \citenamefont {Polini},\ and\
  \citenamefont {Giovannetti}}]{PhysRevB.99.035421}%
  \BibitemOpen
  \bibfield  {author} {\bibinfo {author} {\bibfnamefont {D.}~\bibnamefont
  {Farina}}, \bibinfo {author} {\bibfnamefont {G.~M.}\ \bibnamefont
  {Andolina}}, \bibinfo {author} {\bibfnamefont {A.}~\bibnamefont {Mari}},
  \bibinfo {author} {\bibfnamefont {M.}~\bibnamefont {Polini}},\ and\ \bibinfo
  {author} {\bibfnamefont {V.}~\bibnamefont {Giovannetti}},\ }\bibfield
  {title} {\bibinfo {title} {Charger-mediated energy transfer for quantum
  batteries: An open-system approach},\ }\href
  {https://doi.org/10.1103/PhysRevB.99.035421} {\bibfield  {journal} {\bibinfo
  {journal} {Phys. Rev. B}\ }\textbf {\bibinfo {volume} {99}},\ \bibinfo
  {pages} {035421} (\bibinfo {year} {2019})}\BibitemShut {NoStop}%
\bibitem [{\citenamefont {Zhang}\ \emph {et~al.}(2019)\citenamefont {Zhang},
  \citenamefont {Yang}, \citenamefont {Fu},\ and\ \citenamefont
  {Wang}}]{PhysRevE.99.052106}%
  \BibitemOpen
  \bibfield  {author} {\bibinfo {author} {\bibfnamefont {Y.-Y.}\ \bibnamefont
  {Zhang}}, \bibinfo {author} {\bibfnamefont {T.-R.}\ \bibnamefont {Yang}},
  \bibinfo {author} {\bibfnamefont {L.}~\bibnamefont {Fu}},\ and\ \bibinfo
  {author} {\bibfnamefont {X.-G.}\ \bibnamefont {Wang}},\ }\bibfield  {title}
  {\bibinfo {title} {Powerful harmonic charging in a quantum battery},\ }\href
  {https://doi.org/10.1103/PhysRevE.99.052106} {\bibfield  {journal} {\bibinfo
  {journal} {Phys. Rev. E}\ }\textbf {\bibinfo {volume} {99}},\ \bibinfo
  {pages} {052106} (\bibinfo {year} {2019})}\BibitemShut {NoStop}%
\bibitem [{\citenamefont {Quach}\ and\ \citenamefont
  {Munro}(2020)}]{PhysRevApplied.14.024092}%
  \BibitemOpen
  \bibfield  {author} {\bibinfo {author} {\bibfnamefont {J.~Q.}\ \bibnamefont
  {Quach}}\ and\ \bibinfo {author} {\bibfnamefont {W.~J.}\ \bibnamefont
  {Munro}},\ }\bibfield  {title} {\bibinfo {title} {Using dark states to charge
  and stabilize open quantum batteries},\ }\href
  {https://doi.org/10.1103/PhysRevApplied.14.024092} {\bibfield  {journal}
  {\bibinfo  {journal} {Phys. Rev. Appl.}\ }\textbf {\bibinfo {volume} {14}},\
  \bibinfo {pages} {024092} (\bibinfo {year} {2020})}\BibitemShut {NoStop}%
\bibitem [{\citenamefont {Juli{\`a}-Farr{\'e}}\ \emph
  {et~al.}(2020)\citenamefont {Juli{\`a}-Farr{\'e}}, \citenamefont {Salamon},
  \citenamefont {Riera}, \citenamefont {Bera},\ and\ \citenamefont
  {Lewenstein}}]{PhysRevResearch.2.023113}%
  \BibitemOpen
  \bibfield  {author} {\bibinfo {author} {\bibfnamefont {S.}~\bibnamefont
  {Juli{\`a}-Farr{\'e}}}, \bibinfo {author} {\bibfnamefont {T.}~\bibnamefont
  {Salamon}}, \bibinfo {author} {\bibfnamefont {A.}~\bibnamefont {Riera}},
  \bibinfo {author} {\bibfnamefont {M.~N.}\ \bibnamefont {Bera}},\ and\
  \bibinfo {author} {\bibfnamefont {M.}~\bibnamefont {Lewenstein}},\ }\bibfield
   {title} {\bibinfo {title} {Bounds on the capacity and power of quantum
  batteries},\ }\href {https://doi.org/10.1103/PhysRevResearch.2.023113}
  {\bibfield  {journal} {\bibinfo  {journal} {Phys. Rev. Res.}\ }\textbf
  {\bibinfo {volume} {2}},\ \bibinfo {pages} {023113} (\bibinfo {year}
  {2020})}\BibitemShut {NoStop}%
\bibitem [{\citenamefont {Gyhm}\ \emph {et~al.}(2022)\citenamefont {Gyhm},
  \citenamefont {{\ifmmode \check{S}\else \v{S}\fi{}}afr{\'a}nek},\ and\
  \citenamefont {Rosa}}]{PhysRevLett.128.140501}%
  \BibitemOpen
  \bibfield  {author} {\bibinfo {author} {\bibfnamefont {J.-Y.}\ \bibnamefont
  {Gyhm}}, \bibinfo {author} {\bibfnamefont {D.}~\bibnamefont {{\ifmmode
  \check{S}\else \v{S}\fi{}}afr{\'a}nek}},\ and\ \bibinfo {author}
  {\bibfnamefont {D.}~\bibnamefont {Rosa}},\ }\bibfield  {title} {\bibinfo
  {title} {Quantum charging advantage cannot be extensive without global
  operations},\ }\href {https://doi.org/10.1103/PhysRevLett.128.140501}
  {\bibfield  {journal} {\bibinfo  {journal} {Phys. Rev. Lett.}\ }\textbf
  {\bibinfo {volume} {128}},\ \bibinfo {pages} {140501} (\bibinfo {year}
  {2022})}\BibitemShut {NoStop}%
\bibitem [{\citenamefont {Dou}\ \emph {et~al.}(2022{\natexlab{a}})\citenamefont
  {Dou}, \citenamefont {Zhou},\ and\ \citenamefont
  {Sun}}]{PhysRevA.106.032212}%
  \BibitemOpen
  \bibfield  {author} {\bibinfo {author} {\bibfnamefont {F.-Q.}\ \bibnamefont
  {Dou}}, \bibinfo {author} {\bibfnamefont {H.}~\bibnamefont {Zhou}},\ and\
  \bibinfo {author} {\bibfnamefont {J.-A.}\ \bibnamefont {Sun}},\ }\bibfield
  {title} {\bibinfo {title} {Cavity {Heisenberg}-spin-chain quantum battery},\
  }\href {https://doi.org/10.1103/PhysRevA.106.032212} {\bibfield  {journal}
  {\bibinfo  {journal} {Phys. Rev. A}\ }\textbf {\bibinfo {volume} {106}},\
  \bibinfo {pages} {032212} (\bibinfo {year} {2022}{\natexlab{a}})}\BibitemShut
  {NoStop}%
\bibitem [{\citenamefont {Dou}\ and\ \citenamefont
  {Yang}(2023)}]{PhysRevA.107.023725}%
  \BibitemOpen
  \bibfield  {author} {\bibinfo {author} {\bibfnamefont {F.-Q.}\ \bibnamefont
  {Dou}}\ and\ \bibinfo {author} {\bibfnamefont {F.-M.}\ \bibnamefont {Yang}},\
  }\bibfield  {title} {\bibinfo {title} {Superconducting transmon
  qubit-resonator quantum battery},\ }\href
  {https://doi.org/10.1103/PhysRevA.107.023725} {\bibfield  {journal} {\bibinfo
   {journal} {Phys. Rev. A}\ }\textbf {\bibinfo {volume} {107}},\ \bibinfo
  {pages} {023725} (\bibinfo {year} {2023})}\BibitemShut {NoStop}%
\bibitem [{\citenamefont {Song}\ \emph
  {et~al.}(2024{\natexlab{a}})\citenamefont {Song}, \citenamefont {Liu},
  \citenamefont {Zhou}, \citenamefont {Yang},\ and\ \citenamefont
  {An}}]{PhysRevLett.132.090401}%
  \BibitemOpen
  \bibfield  {author} {\bibinfo {author} {\bibfnamefont {W.-L.}\ \bibnamefont
  {Song}}, \bibinfo {author} {\bibfnamefont {H.-B.}\ \bibnamefont {Liu}},
  \bibinfo {author} {\bibfnamefont {B.}~\bibnamefont {Zhou}}, \bibinfo {author}
  {\bibfnamefont {W.-L.}\ \bibnamefont {Yang}},\ and\ \bibinfo {author}
  {\bibfnamefont {J.-H.}\ \bibnamefont {An}},\ }\bibfield  {title} {\bibinfo
  {title} {Remote charging and degradation suppression for the quantum
  battery},\ }\href {https://doi.org/10.1103/PhysRevLett.132.090401} {\bibfield
   {journal} {\bibinfo  {journal} {Phys. Rev. Lett.}\ }\textbf {\bibinfo
  {volume} {132}},\ \bibinfo {pages} {090401} (\bibinfo {year}
  {2024}{\natexlab{a}})}\BibitemShut {NoStop}%
\bibitem [{\citenamefont {Andolina}\ \emph {et~al.}(2025)\citenamefont
  {Andolina}, \citenamefont {Stanzione}, \citenamefont {Giovannetti},\ and\
  \citenamefont {Polini}}]{kzvn-dj7v}%
  \BibitemOpen
  \bibfield  {author} {\bibinfo {author} {\bibfnamefont {G.~M.}\ \bibnamefont
  {Andolina}}, \bibinfo {author} {\bibfnamefont {V.}~\bibnamefont {Stanzione}},
  \bibinfo {author} {\bibfnamefont {V.}~\bibnamefont {Giovannetti}},\ and\
  \bibinfo {author} {\bibfnamefont {M.}~\bibnamefont {Polini}},\ }\bibfield
  {title} {\bibinfo {title} {Genuine quantum advantage in anharmonic bosonic
  quantum batteries},\ }\href {https://doi.org/10.1103/kzvn-dj7v} {\bibfield
  {journal} {\bibinfo  {journal} {Phys. Rev. Lett.}\ }\textbf {\bibinfo
  {volume} {134}},\ \bibinfo {pages} {240403} (\bibinfo {year}
  {2025})}\BibitemShut {NoStop}%
\bibitem [{\citenamefont {Yao}\ and\ \citenamefont {Shao}(2025)}]{vqnk-kzqg}%
  \BibitemOpen
  \bibfield  {author} {\bibinfo {author} {\bibfnamefont {Y.}~\bibnamefont
  {Yao}}\ and\ \bibinfo {author} {\bibfnamefont {X.~Q.}\ \bibnamefont {Shao}},\
  }\bibfield  {title} {\bibinfo {title} {Reservoir-assisted quantum battery
  charging at finite temperatures},\ }\href {https://doi.org/10.1103/vqnk-kzqg}
  {\bibfield  {journal} {\bibinfo  {journal} {Phys. Rev. A}\ }\textbf {\bibinfo
  {volume} {111}},\ \bibinfo {pages} {062616} (\bibinfo {year}
  {2025})}\BibitemShut {NoStop}%
\bibitem [{\citenamefont {Monsel}\ \emph {et~al.}(2020)\citenamefont {Monsel},
  \citenamefont {Fellous-Asiani}, \citenamefont {Huard},\ and\ \citenamefont
  {Auff\`eves}}]{PhysRevLett.124.130601}%
  \BibitemOpen
  \bibfield  {author} {\bibinfo {author} {\bibfnamefont {J.}~\bibnamefont
  {Monsel}}, \bibinfo {author} {\bibfnamefont {M.}~\bibnamefont
  {Fellous-Asiani}}, \bibinfo {author} {\bibfnamefont {B.}~\bibnamefont
  {Huard}},\ and\ \bibinfo {author} {\bibfnamefont {A.}~\bibnamefont
  {Auff\`eves}},\ }\bibfield  {title} {\bibinfo {title} {The energetic cost of
  work extraction},\ }\href {https://doi.org/10.1103/PhysRevLett.124.130601}
  {\bibfield  {journal} {\bibinfo  {journal} {Phys. Rev. Lett.}\ }\textbf
  {\bibinfo {volume} {124}},\ \bibinfo {pages} {130601} (\bibinfo {year}
  {2020})}\BibitemShut {NoStop}%
\bibitem [{\citenamefont {Crescente}\ \emph {et~al.}(2020)\citenamefont
  {Crescente}, \citenamefont {Carrega}, \citenamefont {Sassetti},\ and\
  \citenamefont {Ferraro}}]{PhysRevB.102.245407}%
  \BibitemOpen
  \bibfield  {author} {\bibinfo {author} {\bibfnamefont {A.}~\bibnamefont
  {Crescente}}, \bibinfo {author} {\bibfnamefont {M.}~\bibnamefont {Carrega}},
  \bibinfo {author} {\bibfnamefont {M.}~\bibnamefont {Sassetti}},\ and\
  \bibinfo {author} {\bibfnamefont {D.}~\bibnamefont {Ferraro}},\ }\bibfield
  {title} {\bibinfo {title} {Ultrafast charging in a two-photon {Dicke} quantum
  battery},\ }\href {https://doi.org/10.1103/PhysRevB.102.245407} {\bibfield
  {journal} {\bibinfo  {journal} {Phys. Rev. B}\ }\textbf {\bibinfo {volume}
  {102}},\ \bibinfo {pages} {245407} (\bibinfo {year} {2020})}\BibitemShut
  {NoStop}%
\bibitem [{\citenamefont {Rossini}\ \emph {et~al.}(2020)\citenamefont
  {Rossini}, \citenamefont {Andolina}, \citenamefont {Rosa}, \citenamefont
  {Carrega},\ and\ \citenamefont {Polini}}]{PhysRevLett.125.236402}%
  \BibitemOpen
  \bibfield  {author} {\bibinfo {author} {\bibfnamefont {D.}~\bibnamefont
  {Rossini}}, \bibinfo {author} {\bibfnamefont {G.~M.}\ \bibnamefont
  {Andolina}}, \bibinfo {author} {\bibfnamefont {D.}~\bibnamefont {Rosa}},
  \bibinfo {author} {\bibfnamefont {M.}~\bibnamefont {Carrega}},\ and\ \bibinfo
  {author} {\bibfnamefont {M.}~\bibnamefont {Polini}},\ }\bibfield  {title}
  {\bibinfo {title} {Quantum advantage in the charging process of
  {Sachdev-Ye-Kitaev} batteries},\ }\href
  {https://doi.org/10.1103/PhysRevLett.125.236402} {\bibfield  {journal}
  {\bibinfo  {journal} {Phys. Rev. Lett.}\ }\textbf {\bibinfo {volume} {125}},\
  \bibinfo {pages} {236402} (\bibinfo {year} {2020})}\BibitemShut {NoStop}%
\bibitem [{\citenamefont {Lu}\ \emph {et~al.}(2021)\citenamefont {Lu},
  \citenamefont {Chen}, \citenamefont {Kuang},\ and\ \citenamefont
  {Wang}}]{PhysRevA.104.043706}%
  \BibitemOpen
  \bibfield  {author} {\bibinfo {author} {\bibfnamefont {W.-J.}\ \bibnamefont
  {Lu}}, \bibinfo {author} {\bibfnamefont {J.}~\bibnamefont {Chen}}, \bibinfo
  {author} {\bibfnamefont {L.-M.}\ \bibnamefont {Kuang}},\ and\ \bibinfo
  {author} {\bibfnamefont {X.-G.}\ \bibnamefont {Wang}},\ }\bibfield  {title}
  {\bibinfo {title} {Optimal state for a tavis-cummings quantum battery via the
  bethe ansatz method},\ }\href {https://doi.org/10.1103/PhysRevA.104.043706}
  {\bibfield  {journal} {\bibinfo  {journal} {Phys. Rev. A}\ }\textbf {\bibinfo
  {volume} {104}},\ \bibinfo {pages} {043706} (\bibinfo {year}
  {2021})}\BibitemShut {NoStop}%
\bibitem [{\citenamefont {Dou}\ \emph {et~al.}(2021)\citenamefont {Dou},
  \citenamefont {Wang},\ and\ \citenamefont {Sun}}]{Dou2021}%
  \BibitemOpen
  \bibfield  {author} {\bibinfo {author} {\bibfnamefont {F.-Q.}\ \bibnamefont
  {Dou}}, \bibinfo {author} {\bibfnamefont {Y.-J.}\ \bibnamefont {Wang}},\ and\
  \bibinfo {author} {\bibfnamefont {J.-A.}\ \bibnamefont {Sun}},\ }\bibfield
  {title} {\bibinfo {title} {Highly efficient charging and discharging of
  three-level quantum batteries through shortcuts to adiabaticity},\ }\href
  {https://doi.org/10.1007/s11467-021-1130-5} {\bibfield  {journal} {\bibinfo
  {journal} {Front. Phys.}\ }\textbf {\bibinfo {volume} {17}},\ \bibinfo
  {pages} {31503} (\bibinfo {year} {2021})}\BibitemShut {NoStop}%
\bibitem [{\citenamefont {Zhu}\ \emph {et~al.}(2023)\citenamefont {Zhu},
  \citenamefont {Chen}, \citenamefont {Hasegawa},\ and\ \citenamefont
  {Xue}}]{PhysRevLett.131.240401}%
  \BibitemOpen
  \bibfield  {author} {\bibinfo {author} {\bibfnamefont {G.}~\bibnamefont
  {Zhu}}, \bibinfo {author} {\bibfnamefont {Y.}~\bibnamefont {Chen}}, \bibinfo
  {author} {\bibfnamefont {Y.}~\bibnamefont {Hasegawa}},\ and\ \bibinfo
  {author} {\bibfnamefont {P.}~\bibnamefont {Xue}},\ }\bibfield  {title}
  {\bibinfo {title} {Charging quantum batteries via indefinite causal order:
  Theory and experiment},\ }\href
  {https://doi.org/10.1103/PhysRevLett.131.240401} {\bibfield  {journal}
  {\bibinfo  {journal} {Phys. Rev. Lett.}\ }\textbf {\bibinfo {volume} {131}},\
  \bibinfo {pages} {240401} (\bibinfo {year} {2023})}\BibitemShut {NoStop}%
\bibitem [{\citenamefont {Yan}\ and\ \citenamefont
  {Jing}(2023)}]{PhysRevApplied.19.064069}%
  \BibitemOpen
  \bibfield  {author} {\bibinfo {author} {\bibfnamefont {J.-S.}\ \bibnamefont
  {Yan}}\ and\ \bibinfo {author} {\bibfnamefont {J.}~\bibnamefont {Jing}},\
  }\bibfield  {title} {\bibinfo {title} {Charging by quantum measurement},\
  }\href {https://doi.org/10.1103/PhysRevApplied.19.064069} {\bibfield
  {journal} {\bibinfo  {journal} {Phys. Rev. Appl.}\ }\textbf {\bibinfo
  {volume} {19}},\ \bibinfo {pages} {064069} (\bibinfo {year}
  {2023})}\BibitemShut {NoStop}%
\bibitem [{\citenamefont {Ukhtary}\ \emph {et~al.}(2023)\citenamefont
  {Ukhtary}, \citenamefont {Nugraha}, \citenamefont {Cahaya}, \citenamefont
  {Rusydi},\ and\ \citenamefont {Majidi}}]{10.1063/5.0156618}%
  \BibitemOpen
  \bibfield  {author} {\bibinfo {author} {\bibfnamefont {M.~S.}\ \bibnamefont
  {Ukhtary}}, \bibinfo {author} {\bibfnamefont {A.~R.~T.}\ \bibnamefont
  {Nugraha}}, \bibinfo {author} {\bibfnamefont {A.~B.}\ \bibnamefont {Cahaya}},
  \bibinfo {author} {\bibfnamefont {A.}~\bibnamefont {Rusydi}},\ and\ \bibinfo
  {author} {\bibfnamefont {M.~A.}\ \bibnamefont {Majidi}},\ }\bibfield  {title}
  {\bibinfo {title} {High-performance {Kerr} quantum battery},\ }\href
  {https://doi.org/10.1063/5.0156618} {\bibfield  {journal} {\bibinfo
  {journal} {Applied Physics Letters}\ }\textbf {\bibinfo {volume} {123}},\
  \bibinfo {pages} {034001} (\bibinfo {year} {2023})}\BibitemShut {NoStop}%
\bibitem [{\citenamefont {Ahmadi}\ \emph {et~al.}(2024)\citenamefont {Ahmadi},
  \citenamefont {Mazurek}, \citenamefont {Horodecki},\ and\ \citenamefont
  {Barzanjeh}}]{PhysRevLett.132.210402}%
  \BibitemOpen
  \bibfield  {author} {\bibinfo {author} {\bibfnamefont {B.}~\bibnamefont
  {Ahmadi}}, \bibinfo {author} {\bibfnamefont {P.}~\bibnamefont {Mazurek}},
  \bibinfo {author} {\bibfnamefont {P.}~\bibnamefont {Horodecki}},\ and\
  \bibinfo {author} {\bibfnamefont {S.}~\bibnamefont {Barzanjeh}},\ }\bibfield
  {title} {\bibinfo {title} {Nonreciprocal quantum batteries},\ }\href
  {https://doi.org/10.1103/PhysRevLett.132.210402} {\bibfield  {journal}
  {\bibinfo  {journal} {Phys. Rev. Lett.}\ }\textbf {\bibinfo {volume} {132}},\
  \bibinfo {pages} {210402} (\bibinfo {year} {2024})}\BibitemShut {NoStop}%
\bibitem [{\citenamefont {Lu}\ \emph {et~al.}(2025)\citenamefont {Lu},
  \citenamefont {Tian}, \citenamefont {L\"u},\ and\ \citenamefont
  {Shang}}]{PhysRevLett.134.180401}%
  \BibitemOpen
  \bibfield  {author} {\bibinfo {author} {\bibfnamefont {Z.-G.}\ \bibnamefont
  {Lu}}, \bibinfo {author} {\bibfnamefont {G.}~\bibnamefont {Tian}}, \bibinfo
  {author} {\bibfnamefont {X.-Y.}\ \bibnamefont {L\"u}},\ and\ \bibinfo
  {author} {\bibfnamefont {C.}~\bibnamefont {Shang}},\ }\bibfield  {title}
  {\bibinfo {title} {Topological quantum batteries},\ }\href
  {https://doi.org/10.1103/PhysRevLett.134.180401} {\bibfield  {journal}
  {\bibinfo  {journal} {Phys. Rev. Lett.}\ }\textbf {\bibinfo {volume} {134}},\
  \bibinfo {pages} {180401} (\bibinfo {year} {2025})}\BibitemShut {NoStop}%
\bibitem [{\citenamefont {Pokhrel}\ and\ \citenamefont
  {Gea-Banacloche}(2025)}]{PhysRevLett.134.130401}%
  \BibitemOpen
  \bibfield  {author} {\bibinfo {author} {\bibfnamefont {S.}~\bibnamefont
  {Pokhrel}}\ and\ \bibinfo {author} {\bibfnamefont {J.}~\bibnamefont
  {Gea-Banacloche}},\ }\bibfield  {title} {\bibinfo {title} {Large collective
  power enhancement in dissipative charging of a quantum battery},\ }\href
  {https://doi.org/10.1103/PhysRevLett.134.130401} {\bibfield  {journal}
  {\bibinfo  {journal} {Phys. Rev. Lett.}\ }\textbf {\bibinfo {volume} {134}},\
  \bibinfo {pages} {130401} (\bibinfo {year} {2025})}\BibitemShut {NoStop}%
\bibitem [{\citenamefont {Zhang}\ \emph
  {et~al.}(2025{\natexlab{a}})\citenamefont {Zhang}, \citenamefont {Ma},
  \citenamefont {Yu}, \citenamefont {Jin},\ and\ \citenamefont
  {Chen}}]{3tm5-vsqw}%
  \BibitemOpen
  \bibfield  {author} {\bibinfo {author} {\bibfnamefont {D.-Y.}\ \bibnamefont
  {Zhang}}, \bibinfo {author} {\bibfnamefont {S.-Q.}\ \bibnamefont {Ma}},
  \bibinfo {author} {\bibfnamefont {Y.-B.}\ \bibnamefont {Yu}}, \bibinfo
  {author} {\bibfnamefont {G.-R.}\ \bibnamefont {Jin}},\ and\ \bibinfo {author}
  {\bibfnamefont {A.-X.}\ \bibnamefont {Chen}},\ }\bibfield  {title} {\bibinfo
  {title} {Unique quantum advantages by manipulation of atoms in quantum
  batteries},\ }\href {https://doi.org/10.1103/3tm5-vsqw} {\bibfield  {journal}
  {\bibinfo  {journal} {Phys. Rev. A}\ }\textbf {\bibinfo {volume} {112}},\
  \bibinfo {pages} {022615} (\bibinfo {year} {2025}{\natexlab{a}})}\BibitemShut
  {NoStop}%
\bibitem [{\citenamefont {Francica}\ \emph {et~al.}(2020)\citenamefont
  {Francica}, \citenamefont {Binder}, \citenamefont {Guarnieri}, \citenamefont
  {Mitchison}, \citenamefont {Goold},\ and\ \citenamefont
  {Plastina}}]{PhysRevLett.125.180603}%
  \BibitemOpen
  \bibfield  {author} {\bibinfo {author} {\bibfnamefont {G.}~\bibnamefont
  {Francica}}, \bibinfo {author} {\bibfnamefont {F.~C.}\ \bibnamefont
  {Binder}}, \bibinfo {author} {\bibfnamefont {G.}~\bibnamefont {Guarnieri}},
  \bibinfo {author} {\bibfnamefont {M.~T.}\ \bibnamefont {Mitchison}}, \bibinfo
  {author} {\bibfnamefont {J.}~\bibnamefont {Goold}},\ and\ \bibinfo {author}
  {\bibfnamefont {F.}~\bibnamefont {Plastina}},\ }\bibfield  {title} {\bibinfo
  {title} {Quantum coherence and ergotropy},\ }\href
  {https://doi.org/10.1103/PhysRevLett.125.180603} {\bibfield  {journal}
  {\bibinfo  {journal} {Phys. Rev. Lett.}\ }\textbf {\bibinfo {volume} {125}},\
  \bibinfo {pages} {180603} (\bibinfo {year} {2020})}\BibitemShut {NoStop}%
\bibitem [{\citenamefont {Kamin}\ \emph {et~al.}(2020)\citenamefont {Kamin},
  \citenamefont {Tabesh}, \citenamefont {Salimi},\ and\ \citenamefont
  {Santos}}]{PhysRevE.102.052109}%
  \BibitemOpen
  \bibfield  {author} {\bibinfo {author} {\bibfnamefont {F.~H.}\ \bibnamefont
  {Kamin}}, \bibinfo {author} {\bibfnamefont {F.~T.}\ \bibnamefont {Tabesh}},
  \bibinfo {author} {\bibfnamefont {S.}~\bibnamefont {Salimi}},\ and\ \bibinfo
  {author} {\bibfnamefont {A.~C.}\ \bibnamefont {Santos}},\ }\bibfield  {title}
  {\bibinfo {title} {Entanglement, coherence, and charging process of quantum
  batteries},\ }\href {https://doi.org/10.1103/PhysRevE.102.052109} {\bibfield
  {journal} {\bibinfo  {journal} {Phys. Rev. E}\ }\textbf {\bibinfo {volume}
  {102}},\ \bibinfo {pages} {052109} (\bibinfo {year} {2020})}\BibitemShut
  {NoStop}%
\bibitem [{\citenamefont {Shi}\ \emph {et~al.}(2022)\citenamefont {Shi},
  \citenamefont {Ding}, \citenamefont {Wan}, \citenamefont {Wang},\ and\
  \citenamefont {Yang}}]{PhysRevLett.129.130602}%
  \BibitemOpen
  \bibfield  {author} {\bibinfo {author} {\bibfnamefont {H.-L.}\ \bibnamefont
  {Shi}}, \bibinfo {author} {\bibfnamefont {S.}~\bibnamefont {Ding}}, \bibinfo
  {author} {\bibfnamefont {Q.-K.}\ \bibnamefont {Wan}}, \bibinfo {author}
  {\bibfnamefont {X.-H.}\ \bibnamefont {Wang}},\ and\ \bibinfo {author}
  {\bibfnamefont {W.-L.}\ \bibnamefont {Yang}},\ }\bibfield  {title} {\bibinfo
  {title} {Entanglement, coherence, and extractable work in quantum
  batteries},\ }\href {https://doi.org/10.1103/PhysRevLett.129.130602}
  {\bibfield  {journal} {\bibinfo  {journal} {Phys. Rev. Lett.}\ }\textbf
  {\bibinfo {volume} {129}},\ \bibinfo {pages} {130602} (\bibinfo {year}
  {2022})}\BibitemShut {NoStop}%
\bibitem [{\citenamefont {Xu}\ \emph {et~al.}(2022)\citenamefont {Xu},
  \citenamefont {Li}, \citenamefont {Li}, \citenamefont {Zhu}, \citenamefont
  {Zhang},\ and\ \citenamefont {Liu}}]{PhysRevA.106.012425}%
  \BibitemOpen
  \bibfield  {author} {\bibinfo {author} {\bibfnamefont {K.}~\bibnamefont
  {Xu}}, \bibinfo {author} {\bibfnamefont {H.-G.}\ \bibnamefont {Li}}, \bibinfo
  {author} {\bibfnamefont {Z.-G.}\ \bibnamefont {Li}}, \bibinfo {author}
  {\bibfnamefont {H.-J.}\ \bibnamefont {Zhu}}, \bibinfo {author} {\bibfnamefont
  {G.-F.}\ \bibnamefont {Zhang}},\ and\ \bibinfo {author} {\bibfnamefont
  {W.-M.}\ \bibnamefont {Liu}},\ }\bibfield  {title} {\bibinfo {title}
  {Charging performance of quantum batteries in a double-layer environment},\
  }\href {https://doi.org/10.1103/PhysRevA.106.012425} {\bibfield  {journal}
  {\bibinfo  {journal} {Phys. Rev. A}\ }\textbf {\bibinfo {volume} {106}},\
  \bibinfo {pages} {012425} (\bibinfo {year} {2022})}\BibitemShut {NoStop}%
\bibitem [{\citenamefont {Yang}\ \emph {et~al.}(2023)\citenamefont {Yang},
  \citenamefont {Yang}, \citenamefont {Alimuddin}, \citenamefont {Salvia},
  \citenamefont {Fei}, \citenamefont {Zhao}, \citenamefont {Nimmrichter},\ and\
  \citenamefont {Luo}}]{PhysRevLett.131.030402}%
  \BibitemOpen
  \bibfield  {author} {\bibinfo {author} {\bibfnamefont {X.}~\bibnamefont
  {Yang}}, \bibinfo {author} {\bibfnamefont {Y.-H.}\ \bibnamefont {Yang}},
  \bibinfo {author} {\bibfnamefont {M.}~\bibnamefont {Alimuddin}}, \bibinfo
  {author} {\bibfnamefont {R.}~\bibnamefont {Salvia}}, \bibinfo {author}
  {\bibfnamefont {S.-M.}\ \bibnamefont {Fei}}, \bibinfo {author} {\bibfnamefont
  {L.-M.}\ \bibnamefont {Zhao}}, \bibinfo {author} {\bibfnamefont
  {S.}~\bibnamefont {Nimmrichter}},\ and\ \bibinfo {author} {\bibfnamefont
  {M.-X.}\ \bibnamefont {Luo}},\ }\bibfield  {title} {\bibinfo {title} {Battery
  capacity of energy-storing quantum systems},\ }\href
  {https://doi.org/10.1103/PhysRevLett.131.030402} {\bibfield  {journal}
  {\bibinfo  {journal} {Phys. Rev. Lett.}\ }\textbf {\bibinfo {volume} {131}},\
  \bibinfo {pages} {030402} (\bibinfo {year} {2023})}\BibitemShut {NoStop}%
\bibitem [{\citenamefont {Zhang}\ \emph {et~al.}(2024)\citenamefont {Zhang},
  \citenamefont {Yang},\ and\ \citenamefont {Fei}}]{PhysRevA.109.042424}%
  \BibitemOpen
  \bibfield  {author} {\bibinfo {author} {\bibfnamefont {T.}~\bibnamefont
  {Zhang}}, \bibinfo {author} {\bibfnamefont {H.}~\bibnamefont {Yang}},\ and\
  \bibinfo {author} {\bibfnamefont {S.-M.}\ \bibnamefont {Fei}},\ }\bibfield
  {title} {\bibinfo {title} {Local-projective-measurement-enhanced quantum
  battery capacity},\ }\href {https://doi.org/10.1103/PhysRevA.109.042424}
  {\bibfield  {journal} {\bibinfo  {journal} {Phys. Rev. A}\ }\textbf {\bibinfo
  {volume} {109}},\ \bibinfo {pages} {042424} (\bibinfo {year}
  {2024})}\BibitemShut {NoStop}%
\bibitem [{\citenamefont {Song}\ \emph
  {et~al.}(2024{\natexlab{b}})\citenamefont {Song}, \citenamefont {Song},
  \citenamefont {Ye},\ and\ \citenamefont {Wang}}]{PhysRevE.109.064103}%
  \BibitemOpen
  \bibfield  {author} {\bibinfo {author} {\bibfnamefont {M.-L.}\ \bibnamefont
  {Song}}, \bibinfo {author} {\bibfnamefont {X.-K.}\ \bibnamefont {Song}},
  \bibinfo {author} {\bibfnamefont {L.}~\bibnamefont {Ye}},\ and\ \bibinfo
  {author} {\bibfnamefont {D.}~\bibnamefont {Wang}},\ }\bibfield  {title}
  {\bibinfo {title} {Evaluating extractable work of quantum batteries via
  entropic uncertainty relations},\ }\href
  {https://doi.org/10.1103/PhysRevE.109.064103} {\bibfield  {journal} {\bibinfo
   {journal} {Phys. Rev. E}\ }\textbf {\bibinfo {volume} {109}},\ \bibinfo
  {pages} {064103} (\bibinfo {year} {2024}{\natexlab{b}})}\BibitemShut
  {NoStop}%
\bibitem [{\citenamefont {Wang}\ and\ \citenamefont {Dou}(2026)}]{2026-0106}%
  \BibitemOpen
  \bibfield  {author} {\bibinfo {author} {\bibfnamefont {C.-J.}\ \bibnamefont
  {Wang}}\ and\ \bibinfo {author} {\bibfnamefont {F.-Q.}\ \bibnamefont {Dou}},\
  }\bibfield  {title} {\bibinfo {title} {Ergotropy in quantum batteries},\
  }\href {https://doi.org/10.1088/0256-307X/43/6/060602} {\bibfield  {journal}
  {\bibinfo  {journal} {Chin. Phys. Lett.}\ }\textbf {\bibinfo {volume} {43}},\
  \bibinfo {pages} {060602} (\bibinfo {year} {2026})}\BibitemShut {NoStop}%
\bibitem [{\citenamefont {Le}\ \emph {et~al.}(2018)\citenamefont {Le},
  \citenamefont {Levinsen}, \citenamefont {Modi}, \citenamefont {Parish},\ and\
  \citenamefont {Pollock}}]{PhysRevA.97.022106}%
  \BibitemOpen
  \bibfield  {author} {\bibinfo {author} {\bibfnamefont {T.~P.}\ \bibnamefont
  {Le}}, \bibinfo {author} {\bibfnamefont {J.}~\bibnamefont {Levinsen}},
  \bibinfo {author} {\bibfnamefont {K.}~\bibnamefont {Modi}}, \bibinfo {author}
  {\bibfnamefont {M.~M.}\ \bibnamefont {Parish}},\ and\ \bibinfo {author}
  {\bibfnamefont {F.~A.}\ \bibnamefont {Pollock}},\ }\bibfield  {title}
  {\bibinfo {title} {Spin-chain model of a many-body quantum battery},\ }\href
  {https://doi.org/10.1103/PhysRevA.97.022106} {\bibfield  {journal} {\bibinfo
  {journal} {Phys. Rev. A}\ }\textbf {\bibinfo {volume} {97}},\ \bibinfo
  {pages} {022106} (\bibinfo {year} {2018})}\BibitemShut {NoStop}%
\bibitem [{\citenamefont {Rossini}\ \emph {et~al.}(2019)\citenamefont
  {Rossini}, \citenamefont {Andolina},\ and\ \citenamefont
  {Polini}}]{PhysRevB.100.115142}%
  \BibitemOpen
  \bibfield  {author} {\bibinfo {author} {\bibfnamefont {D.}~\bibnamefont
  {Rossini}}, \bibinfo {author} {\bibfnamefont {G.~M.}\ \bibnamefont
  {Andolina}},\ and\ \bibinfo {author} {\bibfnamefont {M.}~\bibnamefont
  {Polini}},\ }\bibfield  {title} {\bibinfo {title} {Many-body localized
  quantum batteries},\ }\href {https://doi.org/10.1103/PhysRevB.100.115142}
  {\bibfield  {journal} {\bibinfo  {journal} {Phys. Rev. B}\ }\textbf {\bibinfo
  {volume} {100}},\ \bibinfo {pages} {115142} (\bibinfo {year}
  {2019})}\BibitemShut {NoStop}%
\bibitem [{\citenamefont {Peng}\ \emph {et~al.}(2021)\citenamefont {Peng},
  \citenamefont {He}, \citenamefont {Chesi}, \citenamefont {Lin},\ and\
  \citenamefont {Guan}}]{PhysRevA.103.052220}%
  \BibitemOpen
  \bibfield  {author} {\bibinfo {author} {\bibfnamefont {L.}~\bibnamefont
  {Peng}}, \bibinfo {author} {\bibfnamefont {W.-B.}\ \bibnamefont {He}},
  \bibinfo {author} {\bibfnamefont {S.}~\bibnamefont {Chesi}}, \bibinfo
  {author} {\bibfnamefont {H.-Q.}\ \bibnamefont {Lin}},\ and\ \bibinfo {author}
  {\bibfnamefont {X.-W.}\ \bibnamefont {Guan}},\ }\bibfield  {title} {\bibinfo
  {title} {Lower and upper bounds of quantum battery power in multiple central
  spin systems},\ }\href {https://doi.org/10.1103/PhysRevA.103.052220}
  {\bibfield  {journal} {\bibinfo  {journal} {Phys. Rev. A}\ }\textbf {\bibinfo
  {volume} {103}},\ \bibinfo {pages} {052220} (\bibinfo {year}
  {2021})}\BibitemShut {NoStop}%
\bibitem [{\citenamefont {Catalano}\ \emph {et~al.}(2024)\citenamefont
  {Catalano}, \citenamefont {Giampaolo}, \citenamefont {Morsch}, \citenamefont
  {Giovannetti},\ and\ \citenamefont {Franchini}}]{PRXQuantum.5.030319}%
  \BibitemOpen
  \bibfield  {author} {\bibinfo {author} {\bibfnamefont {A.}~\bibnamefont
  {Catalano}}, \bibinfo {author} {\bibfnamefont {S.}~\bibnamefont {Giampaolo}},
  \bibinfo {author} {\bibfnamefont {O.}~\bibnamefont {Morsch}}, \bibinfo
  {author} {\bibfnamefont {V.}~\bibnamefont {Giovannetti}},\ and\ \bibinfo
  {author} {\bibfnamefont {F.}~\bibnamefont {Franchini}},\ }\bibfield  {title}
  {\bibinfo {title} {Frustrating quantum batteries},\ }\href
  {https://doi.org/10.1103/PRXQuantum.5.030319} {\bibfield  {journal} {\bibinfo
   {journal} {PRX Quantum}\ }\textbf {\bibinfo {volume} {5}},\ \bibinfo {pages}
  {030319} (\bibinfo {year} {2024})}\BibitemShut {NoStop}%
\bibitem [{\citenamefont {Guo}\ \emph {et~al.}(2024)\citenamefont {Guo},
  \citenamefont {Yang},\ and\ \citenamefont {Dou}}]{PhysRevA.109.032201}%
  \BibitemOpen
  \bibfield  {author} {\bibinfo {author} {\bibfnamefont {W.-X.}\ \bibnamefont
  {Guo}}, \bibinfo {author} {\bibfnamefont {F.-M.}\ \bibnamefont {Yang}},\ and\
  \bibinfo {author} {\bibfnamefont {F.-Q.}\ \bibnamefont {Dou}},\ }\bibfield
  {title} {\bibinfo {title} {Analytically solvable many-body rosen-zener
  quantum battery},\ }\href {https://doi.org/10.1103/PhysRevA.109.032201}
  {\bibfield  {journal} {\bibinfo  {journal} {Phys. Rev. A}\ }\textbf {\bibinfo
  {volume} {109}},\ \bibinfo {pages} {032201} (\bibinfo {year}
  {2024})}\BibitemShut {NoStop}%
\bibitem [{\citenamefont {Ahmadi}\ \emph {et~al.}(2025)\citenamefont {Ahmadi},
  \citenamefont {Mazurek}, \citenamefont {Barzanjeh},\ and\ \citenamefont
  {Horodecki}}]{PhysRevApplied.23.024010}%
  \BibitemOpen
  \bibfield  {author} {\bibinfo {author} {\bibfnamefont {B.}~\bibnamefont
  {Ahmadi}}, \bibinfo {author} {\bibfnamefont {P.}~\bibnamefont {Mazurek}},
  \bibinfo {author} {\bibfnamefont {S.}~\bibnamefont {Barzanjeh}},\ and\
  \bibinfo {author} {\bibfnamefont {P.}~\bibnamefont {Horodecki}},\ }\bibfield
  {title} {\bibinfo {title} {Superoptimal charging of quantum batteries via
  reservoir engineering: Arbitrary energy transfer unlocked},\ }\href
  {https://doi.org/10.1103/PhysRevApplied.23.024010} {\bibfield  {journal}
  {\bibinfo  {journal} {Phys. Rev. Appl.}\ }\textbf {\bibinfo {volume} {23}},\
  \bibinfo {pages} {024010} (\bibinfo {year} {2025})}\BibitemShut {NoStop}%
\bibitem [{\citenamefont {Formicola}\ \emph {et~al.}(2025)\citenamefont
  {Formicola}, \citenamefont {Di~Bello}, \citenamefont {De~Filippis},
  \citenamefont {Cataudella}, \citenamefont {Farina},\ and\ \citenamefont
  {Perroni}}]{2z1g-rgr9}%
  \BibitemOpen
  \bibfield  {author} {\bibinfo {author} {\bibfnamefont {F.}~\bibnamefont
  {Formicola}}, \bibinfo {author} {\bibfnamefont {G.}~\bibnamefont {Di~Bello}},
  \bibinfo {author} {\bibfnamefont {G.}~\bibnamefont {De~Filippis}}, \bibinfo
  {author} {\bibfnamefont {V.}~\bibnamefont {Cataudella}}, \bibinfo {author}
  {\bibfnamefont {D.}~\bibnamefont {Farina}},\ and\ \bibinfo {author}
  {\bibfnamefont {C.~A.}\ \bibnamefont {Perroni}},\ }\bibfield  {title}
  {\bibinfo {title} {Local ergotropy dynamically witnesses many-body localized
  phases},\ }\href {https://doi.org/10.1103/2z1g-rgr9} {\bibfield  {journal}
  {\bibinfo  {journal} {Phys. Rev. Res.}\ }\textbf {\bibinfo {volume} {7}},\
  \bibinfo {pages} {043086} (\bibinfo {year} {2025})}\BibitemShut {NoStop}%
\bibitem [{\citenamefont {Quach}\ \emph {et~al.}(2022)\citenamefont {Quach},
  \citenamefont {McGhee}, \citenamefont {Ganzer}, \citenamefont {Rouse},
  \citenamefont {Lovett}, \citenamefont {Gauger}, \citenamefont {Keeling},
  \citenamefont {Cerullo}, \citenamefont {Lidzey},\ and\ \citenamefont
  {Virgili}}]{doi:10.1126/sciadv.abk3160}%
  \BibitemOpen
  \bibfield  {author} {\bibinfo {author} {\bibfnamefont {J.~Q.}\ \bibnamefont
  {Quach}}, \bibinfo {author} {\bibfnamefont {K.~E.}\ \bibnamefont {McGhee}},
  \bibinfo {author} {\bibfnamefont {L.}~\bibnamefont {Ganzer}}, \bibinfo
  {author} {\bibfnamefont {D.~M.}\ \bibnamefont {Rouse}}, \bibinfo {author}
  {\bibfnamefont {B.~W.}\ \bibnamefont {Lovett}}, \bibinfo {author}
  {\bibfnamefont {E.~M.}\ \bibnamefont {Gauger}}, \bibinfo {author}
  {\bibfnamefont {J.}~\bibnamefont {Keeling}}, \bibinfo {author} {\bibfnamefont
  {G.}~\bibnamefont {Cerullo}}, \bibinfo {author} {\bibfnamefont {D.~G.}\
  \bibnamefont {Lidzey}},\ and\ \bibinfo {author} {\bibfnamefont
  {T.}~\bibnamefont {Virgili}},\ }\bibfield  {title} {\bibinfo {title}
  {Superabsorption in an organic microcavity: Toward a quantum battery},\
  }\href {https://doi.org/10.1126/sciadv.abk3160} {\bibfield  {journal}
  {\bibinfo  {journal} {Sci. Adv.}\ }\textbf {\bibinfo {volume} {8}},\ \bibinfo
  {pages} {eabk3160} (\bibinfo {year} {2022})}\BibitemShut {NoStop}%
\bibitem [{\citenamefont {Joshi}\ and\ \citenamefont
  {Mahesh}(2022)}]{PhysRevA.106.042601}%
  \BibitemOpen
  \bibfield  {author} {\bibinfo {author} {\bibfnamefont {J.}~\bibnamefont
  {Joshi}}\ and\ \bibinfo {author} {\bibfnamefont {T.~S.}\ \bibnamefont
  {Mahesh}},\ }\bibfield  {title} {\bibinfo {title} {Experimental investigation
  of a quantum battery using star-topology {NMR} spin systems},\ }\href
  {https://doi.org/10.1103/PhysRevA.106.042601} {\bibfield  {journal} {\bibinfo
   {journal} {Phys. Rev. A}\ }\textbf {\bibinfo {volume} {106}},\ \bibinfo
  {pages} {042601} (\bibinfo {year} {2022})}\BibitemShut {NoStop}%
\bibitem [{\citenamefont {Maillette~de Buy~Wenniger}\ \emph
  {et~al.}(2023)\citenamefont {Maillette~de Buy~Wenniger}, \citenamefont
  {Thomas}, \citenamefont {Maffei}, \citenamefont {Wein}, \citenamefont {Pont},
  \citenamefont {Belabas}, \citenamefont {Prasad}, \citenamefont {Harouri},
  \citenamefont {Lema\^{\i}tre}, \citenamefont {Sagnes}, \citenamefont
  {Somaschi}, \citenamefont {Auff\`eves},\ and\ \citenamefont
  {Senellart}}]{PhysRevLett.131.260401}%
  \BibitemOpen
  \bibfield  {author} {\bibinfo {author} {\bibfnamefont {I.}~\bibnamefont
  {Maillette~de Buy~Wenniger}}, \bibinfo {author} {\bibfnamefont {S.~E.}\
  \bibnamefont {Thomas}}, \bibinfo {author} {\bibfnamefont {M.}~\bibnamefont
  {Maffei}}, \bibinfo {author} {\bibfnamefont {S.~C.}\ \bibnamefont {Wein}},
  \bibinfo {author} {\bibfnamefont {M.}~\bibnamefont {Pont}}, \bibinfo {author}
  {\bibfnamefont {N.}~\bibnamefont {Belabas}}, \bibinfo {author} {\bibfnamefont
  {S.}~\bibnamefont {Prasad}}, \bibinfo {author} {\bibfnamefont
  {A.}~\bibnamefont {Harouri}}, \bibinfo {author} {\bibfnamefont
  {A.}~\bibnamefont {Lema\^{\i}tre}}, \bibinfo {author} {\bibfnamefont
  {I.}~\bibnamefont {Sagnes}}, \bibinfo {author} {\bibfnamefont
  {N.}~\bibnamefont {Somaschi}}, \bibinfo {author} {\bibfnamefont
  {A.}~\bibnamefont {Auff\`eves}},\ and\ \bibinfo {author} {\bibfnamefont
  {P.}~\bibnamefont {Senellart}},\ }\bibfield  {title} {\bibinfo {title}
  {Experimental analysis of energy transfers between a quantum emitter and
  light fields},\ }\href {https://doi.org/10.1103/PhysRevLett.131.260401}
  {\bibfield  {journal} {\bibinfo  {journal} {Phys. Rev. Lett.}\ }\textbf
  {\bibinfo {volume} {131}},\ \bibinfo {pages} {260401} (\bibinfo {year}
  {2023})}\BibitemShut {NoStop}%
\bibitem [{\citenamefont {Huang}\ \emph {et~al.}(2023)\citenamefont {Huang},
  \citenamefont {Wang}, \citenamefont {Xiao}, \citenamefont {Gao},
  \citenamefont {Lin},\ and\ \citenamefont {Xue}}]{PhysRevA.107.L030201}%
  \BibitemOpen
  \bibfield  {author} {\bibinfo {author} {\bibfnamefont {X.}~\bibnamefont
  {Huang}}, \bibinfo {author} {\bibfnamefont {K.}~\bibnamefont {Wang}},
  \bibinfo {author} {\bibfnamefont {L.}~\bibnamefont {Xiao}}, \bibinfo {author}
  {\bibfnamefont {L.}~\bibnamefont {Gao}}, \bibinfo {author} {\bibfnamefont
  {H.}~\bibnamefont {Lin}},\ and\ \bibinfo {author} {\bibfnamefont
  {P.}~\bibnamefont {Xue}},\ }\bibfield  {title} {\bibinfo {title}
  {Demonstration of the charging progress of quantum batteries},\ }\href
  {https://doi.org/10.1103/PhysRevA.107.L030201} {\bibfield  {journal}
  {\bibinfo  {journal} {Phys. Rev. A}\ }\textbf {\bibinfo {volume} {107}},\
  \bibinfo {pages} {L030201} (\bibinfo {year} {2023})}\BibitemShut {NoStop}%
\bibitem [{\citenamefont {Niu}\ \emph {et~al.}(2024)\citenamefont {Niu},
  \citenamefont {Wu}, \citenamefont {Wang}, \citenamefont {Rong},\ and\
  \citenamefont {Du}}]{PhysRevLett.133.180401}%
  \BibitemOpen
  \bibfield  {author} {\bibinfo {author} {\bibfnamefont {Z.}~\bibnamefont
  {Niu}}, \bibinfo {author} {\bibfnamefont {Y.}~\bibnamefont {Wu}}, \bibinfo
  {author} {\bibfnamefont {Y.}~\bibnamefont {Wang}}, \bibinfo {author}
  {\bibfnamefont {X.}~\bibnamefont {Rong}},\ and\ \bibinfo {author}
  {\bibfnamefont {J.}~\bibnamefont {Du}},\ }\bibfield  {title} {\bibinfo
  {title} {Experimental investigation of coherent ergotropy in a single spin
  system},\ }\href {https://doi.org/10.1103/PhysRevLett.133.180401} {\bibfield
  {journal} {\bibinfo  {journal} {Phys. Rev. Lett.}\ }\textbf {\bibinfo
  {volume} {133}},\ \bibinfo {pages} {180401} (\bibinfo {year}
  {2024})}\BibitemShut {NoStop}%
\bibitem [{\citenamefont {Gemme}\ \emph {et~al.}(2024)\citenamefont {Gemme},
  \citenamefont {Grossi}, \citenamefont {Vallecorsa}, \citenamefont
  {Sassetti},\ and\ \citenamefont {Ferraro}}]{PhysRevResearch.6.023091}%
  \BibitemOpen
  \bibfield  {author} {\bibinfo {author} {\bibfnamefont {G.}~\bibnamefont
  {Gemme}}, \bibinfo {author} {\bibfnamefont {M.}~\bibnamefont {Grossi}},
  \bibinfo {author} {\bibfnamefont {S.}~\bibnamefont {Vallecorsa}}, \bibinfo
  {author} {\bibfnamefont {M.}~\bibnamefont {Sassetti}},\ and\ \bibinfo
  {author} {\bibfnamefont {D.}~\bibnamefont {Ferraro}},\ }\bibfield  {title}
  {\bibinfo {title} {Qutrit quantum battery: Comparing different charging
  protocols},\ }\href {https://doi.org/10.1103/PhysRevResearch.6.023091}
  {\bibfield  {journal} {\bibinfo  {journal} {Phys. Rev. Res.}\ }\textbf
  {\bibinfo {volume} {6}},\ \bibinfo {pages} {023091} (\bibinfo {year}
  {2024})}\BibitemShut {NoStop}%
\bibitem [{\citenamefont {Zhang}\ \emph
  {et~al.}(2025{\natexlab{b}})\citenamefont {Zhang}, \citenamefont {Wang},
  \citenamefont {Chen}, \citenamefont {Cai}, \citenamefont {Qiao},
  \citenamefont {Li}, \citenamefont {Huang}, \citenamefont {Tian},
  \citenamefont {Luan}, \citenamefont {Tu}, \citenamefont {Cui}, \citenamefont
  {Yan}, \citenamefont {Zhang}, \citenamefont {Zhang}, \citenamefont {Yung},\
  and\ \citenamefont {Kim}}]{g45c-ssfx}%
  \BibitemOpen
  \bibfield  {author} {\bibinfo {author} {\bibfnamefont {J.}~\bibnamefont
  {Zhang}}, \bibinfo {author} {\bibfnamefont {P.}~\bibnamefont {Wang}},
  \bibinfo {author} {\bibfnamefont {W.}~\bibnamefont {Chen}}, \bibinfo {author}
  {\bibfnamefont {Z.}~\bibnamefont {Cai}}, \bibinfo {author} {\bibfnamefont
  {M.}~\bibnamefont {Qiao}}, \bibinfo {author} {\bibfnamefont {R.}~\bibnamefont
  {Li}}, \bibinfo {author} {\bibfnamefont {Y.}~\bibnamefont {Huang}}, \bibinfo
  {author} {\bibfnamefont {H.}~\bibnamefont {Tian}}, \bibinfo {author}
  {\bibfnamefont {C.}~\bibnamefont {Luan}}, \bibinfo {author} {\bibfnamefont
  {H.}~\bibnamefont {Tu}}, \bibinfo {author} {\bibfnamefont {K.}~\bibnamefont
  {Cui}}, \bibinfo {author} {\bibfnamefont {L.}~\bibnamefont {Yan}}, \bibinfo
  {author} {\bibfnamefont {J.}~\bibnamefont {Zhang}}, \bibinfo {author}
  {\bibfnamefont {J.}~\bibnamefont {Zhang}}, \bibinfo {author} {\bibfnamefont
  {M.}~\bibnamefont {Yung}},\ and\ \bibinfo {author} {\bibfnamefont
  {K.}~\bibnamefont {Kim}},\ }\bibfield  {title} {\bibinfo {title} {Single-ion
  information engine for charging quantum battery},\ }\href
  {https://doi.org/10.1103/g45c-ssfx} {\bibfield  {journal} {\bibinfo
  {journal} {Phys. Rev. Lett.}\ }\textbf {\bibinfo {volume} {135}},\ \bibinfo
  {pages} {140403} (\bibinfo {year} {2025}{\natexlab{b}})}\BibitemShut
  {NoStop}%
\bibitem [{\citenamefont {Tibben}\ \emph {et~al.}(2025)\citenamefont {Tibben},
  \citenamefont {Della~Gaspera}, \citenamefont {van Embden}, \citenamefont
  {Reineck}, \citenamefont {Quach}, \citenamefont {Campaioli},\ and\
  \citenamefont {G{\'o}mez}}]{bhyh-53np}%
  \BibitemOpen
  \bibfield  {author} {\bibinfo {author} {\bibfnamefont {D.~J.}\ \bibnamefont
  {Tibben}}, \bibinfo {author} {\bibfnamefont {E.}~\bibnamefont
  {Della~Gaspera}}, \bibinfo {author} {\bibfnamefont {J.}~\bibnamefont {van
  Embden}}, \bibinfo {author} {\bibfnamefont {P.}~\bibnamefont {Reineck}},
  \bibinfo {author} {\bibfnamefont {J.~Q.}\ \bibnamefont {Quach}}, \bibinfo
  {author} {\bibfnamefont {F.}~\bibnamefont {Campaioli}},\ and\ \bibinfo
  {author} {\bibfnamefont {D.~E.}\ \bibnamefont {G{\'o}mez}},\ }\bibfield
  {title} {\bibinfo {title} {Extending the self-discharge time of {Dicke}
  quantum batteries using molecular triplets},\ }\href
  {https://doi.org/10.1103/bhyh-53np} {\bibfield  {journal} {\bibinfo
  {journal} {PRX Energy}\ }\textbf {\bibinfo {volume} {4}},\ \bibinfo {pages}
  {023012} (\bibinfo {year} {2025})}\BibitemShut {NoStop}%
\bibitem [{\citenamefont {Li}\ \emph {et~al.}(2025)\citenamefont {Li},
  \citenamefont {Zhao}, \citenamefont {Shi}, \citenamefont {Chen},
  \citenamefont {Ruan}, \citenamefont {Liang}, \citenamefont {Yuan},
  \citenamefont {Song}, \citenamefont {Deng}, \citenamefont {Liu},
  \citenamefont {Li}, \citenamefont {Liu}, \citenamefont {Guo}, \citenamefont
  {Song}, \citenamefont {Xu}, \citenamefont {Fan}, \citenamefont {Xiang},\ and\
  \citenamefont {Zheng}}]{y3qx-cs3r}%
  \BibitemOpen
  \bibfield  {author} {\bibinfo {author} {\bibfnamefont {L.}~\bibnamefont
  {Li}}, \bibinfo {author} {\bibfnamefont {S.-L.}\ \bibnamefont {Zhao}},
  \bibinfo {author} {\bibfnamefont {Y.-H.}\ \bibnamefont {Shi}}, \bibinfo
  {author} {\bibfnamefont {B.-J.}\ \bibnamefont {Chen}}, \bibinfo {author}
  {\bibfnamefont {X.}~\bibnamefont {Ruan}}, \bibinfo {author} {\bibfnamefont
  {G.-H.}\ \bibnamefont {Liang}}, \bibinfo {author} {\bibfnamefont {W.-P.}\
  \bibnamefont {Yuan}}, \bibinfo {author} {\bibfnamefont {J.-C.}\ \bibnamefont
  {Song}}, \bibinfo {author} {\bibfnamefont {C.-L.}\ \bibnamefont {Deng}},
  \bibinfo {author} {\bibfnamefont {Y.}~\bibnamefont {Liu}}, \bibinfo {author}
  {\bibfnamefont {T.-M.}\ \bibnamefont {Li}}, \bibinfo {author} {\bibfnamefont
  {Z.-H.}\ \bibnamefont {Liu}}, \bibinfo {author} {\bibfnamefont {X.-Y.}\
  \bibnamefont {Guo}}, \bibinfo {author} {\bibfnamefont {X.}~\bibnamefont
  {Song}}, \bibinfo {author} {\bibfnamefont {K.}~\bibnamefont {Xu}}, \bibinfo
  {author} {\bibfnamefont {H.}~\bibnamefont {Fan}}, \bibinfo {author}
  {\bibfnamefont {Z.}~\bibnamefont {Xiang}},\ and\ \bibinfo {author}
  {\bibfnamefont {D.}~\bibnamefont {Zheng}},\ }\bibfield  {title} {\bibinfo
  {title} {Stable and efficient charging of superconducting capacitively
  shunted flux quantum batteries},\ }\href {https://doi.org/10.1103/y3qx-cs3r}
  {\bibfield  {journal} {\bibinfo  {journal} {Phys. Rev. Appl.}\ }\textbf
  {\bibinfo {volume} {24}},\ \bibinfo {pages} {054033} (\bibinfo {year}
  {2025})}\BibitemShut {NoStop}%
\bibitem [{\citenamefont {Hu}\ \emph {et~al.}(2026)\citenamefont {Hu},
  \citenamefont {Liu}, \citenamefont {Zhao}, \citenamefont {Zhong},
  \citenamefont {Zhou} \emph {et~al.}}]{sp5l-c6m8}%
  \BibitemOpen
  \bibfield  {author} {\bibinfo {author} {\bibfnamefont {C.-K.}\ \bibnamefont
  {Hu}}, \bibinfo {author} {\bibfnamefont {C.}~\bibnamefont {Liu}}, \bibinfo
  {author} {\bibfnamefont {J.}~\bibnamefont {Zhao}}, \bibinfo {author}
  {\bibfnamefont {L.}~\bibnamefont {Zhong}}, \bibinfo {author} {\bibfnamefont
  {Y.}~\bibnamefont {Zhou}}, \emph {et~al.},\ }\bibfield  {title} {\bibinfo
  {title} {Quantum charging advantage in superconducting solid-state
  batteries},\ }\href {https://doi.org/10.1103/sp5l-c6m8} {\bibfield  {journal}
  {\bibinfo  {journal} {Phys. Rev. Lett.}\ }\textbf {\bibinfo {volume} {136}},\
  \bibinfo {pages} {060401} (\bibinfo {year} {2026})}\BibitemShut {NoStop}%
\bibitem [{\citenamefont {Hymas}\ \emph {et~al.}(2026)\citenamefont {Hymas},
  \citenamefont {Muir}, \citenamefont {Tibben}, \citenamefont {van Embden},
  \citenamefont {Hirai}, \citenamefont {Dunn}, \citenamefont {G{\'o}mez},
  \citenamefont {Hutchison}, \citenamefont {Smith},\ and\ \citenamefont
  {Quach}}]{Hymas2026}%
  \BibitemOpen
  \bibfield  {author} {\bibinfo {author} {\bibfnamefont {K.}~\bibnamefont
  {Hymas}}, \bibinfo {author} {\bibfnamefont {J.~B.}\ \bibnamefont {Muir}},
  \bibinfo {author} {\bibfnamefont {D.}~\bibnamefont {Tibben}}, \bibinfo
  {author} {\bibfnamefont {J.}~\bibnamefont {van Embden}}, \bibinfo {author}
  {\bibfnamefont {T.}~\bibnamefont {Hirai}}, \bibinfo {author} {\bibfnamefont
  {C.~J.}\ \bibnamefont {Dunn}}, \bibinfo {author} {\bibfnamefont {D.~E.}\
  \bibnamefont {G{\'o}mez}}, \bibinfo {author} {\bibfnamefont {J.~A.}\
  \bibnamefont {Hutchison}}, \bibinfo {author} {\bibfnamefont {T.~A.}\
  \bibnamefont {Smith}},\ and\ \bibinfo {author} {\bibfnamefont {J.~Q.}\
  \bibnamefont {Quach}},\ }\bibfield  {title} {\bibinfo {title} {Superextensive
  electrical power from a quantum battery},\ }\href
  {https://doi.org/10.1038/s41377-026-02240-6} {\bibfield  {journal} {\bibinfo
  {journal} {Light: Science {\&} Applications}\ }\textbf {\bibinfo {volume}
  {15}},\ \bibinfo {pages} {168} (\bibinfo {year} {2026})}\BibitemShut
  {NoStop}%
\bibitem [{\citenamefont {Emary}\ and\ \citenamefont
  {Brandes}(2003)}]{PhysRevLett.90.044101}%
  \BibitemOpen
  \bibfield  {author} {\bibinfo {author} {\bibfnamefont {C.}~\bibnamefont
  {Emary}}\ and\ \bibinfo {author} {\bibfnamefont {T.}~\bibnamefont
  {Brandes}},\ }\bibfield  {title} {\bibinfo {title} {Quantum chaos triggered
  by precursors of a quantum phase transition: The {Dicke} model},\ }\href
  {https://doi.org/10.1103/PhysRevLett.90.044101} {\bibfield  {journal}
  {\bibinfo  {journal} {Phys. Rev. Lett.}\ }\textbf {\bibinfo {volume} {90}},\
  \bibinfo {pages} {044101} (\bibinfo {year} {2003})}\BibitemShut {NoStop}%
\bibitem [{\citenamefont {Baumann}\ \emph {et~al.}(2010)\citenamefont
  {Baumann}, \citenamefont {Guerlin}, \citenamefont {Brennecke},\ and\
  \citenamefont {Esslinger}}]{Baumann2010}%
  \BibitemOpen
  \bibfield  {author} {\bibinfo {author} {\bibfnamefont {K.}~\bibnamefont
  {Baumann}}, \bibinfo {author} {\bibfnamefont {C.}~\bibnamefont {Guerlin}},
  \bibinfo {author} {\bibfnamefont {F.}~\bibnamefont {Brennecke}},\ and\
  \bibinfo {author} {\bibfnamefont {T.}~\bibnamefont {Esslinger}},\ }\bibfield
  {title} {\bibinfo {title} {Dicke quantum phase transition with a superfluid
  gas in an optical cavity},\ }\href {https://doi.org/10.1038/nature09009}
  {\bibfield  {journal} {\bibinfo  {journal} {Nature}\ }\textbf {\bibinfo
  {volume} {464}},\ \bibinfo {pages} {1301} (\bibinfo {year}
  {2010})}\BibitemShut {NoStop}%
\bibitem [{\citenamefont {Bamba}\ \emph {et~al.}(2016)\citenamefont {Bamba},
  \citenamefont {Inomata},\ and\ \citenamefont
  {Nakamura}}]{PhysRevLett.117.173601}%
  \BibitemOpen
  \bibfield  {author} {\bibinfo {author} {\bibfnamefont {M.}~\bibnamefont
  {Bamba}}, \bibinfo {author} {\bibfnamefont {K.}~\bibnamefont {Inomata}},\
  and\ \bibinfo {author} {\bibfnamefont {Y.}~\bibnamefont {Nakamura}},\
  }\bibfield  {title} {\bibinfo {title} {Superradiant phase transition in a
  superconducting circuit in thermal equilibrium},\ }\href
  {https://doi.org/10.1103/PhysRevLett.117.173601} {\bibfield  {journal}
  {\bibinfo  {journal} {Phys. Rev. Lett.}\ }\textbf {\bibinfo {volume} {117}},\
  \bibinfo {pages} {173601} (\bibinfo {year} {2016})}\BibitemShut {NoStop}%
\bibitem [{\citenamefont {Zheng}\ \emph {et~al.}(2023)\citenamefont {Zheng},
  \citenamefont {Ning}, \citenamefont {Chen}, \citenamefont {L{\"u}},
  \citenamefont {Shen}, \citenamefont {Xu}, \citenamefont {Zhang},
  \citenamefont {Xu}, \citenamefont {Li}, \citenamefont {Xia}, \citenamefont
  {Wu}, \citenamefont {Yang}, \citenamefont {Miranowicz}, \citenamefont
  {Lambert}, \citenamefont {Zheng}, \citenamefont {Fan}, \citenamefont {Nori},\
  and\ \citenamefont {Zheng}}]{PhysRevLett.131.113601}%
  \BibitemOpen
  \bibfield  {author} {\bibinfo {author} {\bibfnamefont {R.-H.}\ \bibnamefont
  {Zheng}}, \bibinfo {author} {\bibfnamefont {W.}~\bibnamefont {Ning}},
  \bibinfo {author} {\bibfnamefont {Y.-H.}\ \bibnamefont {Chen}}, \bibinfo
  {author} {\bibfnamefont {J.-H.}\ \bibnamefont {L{\"u}}}, \bibinfo {author}
  {\bibfnamefont {L.-T.}\ \bibnamefont {Shen}}, \bibinfo {author}
  {\bibfnamefont {K.}~\bibnamefont {Xu}}, \bibinfo {author} {\bibfnamefont
  {Y.-R.}\ \bibnamefont {Zhang}}, \bibinfo {author} {\bibfnamefont
  {D.}~\bibnamefont {Xu}}, \bibinfo {author} {\bibfnamefont {H.}~\bibnamefont
  {Li}}, \bibinfo {author} {\bibfnamefont {Y.}~\bibnamefont {Xia}}, \bibinfo
  {author} {\bibfnamefont {F.}~\bibnamefont {Wu}}, \bibinfo {author}
  {\bibfnamefont {Z.-B.}\ \bibnamefont {Yang}}, \bibinfo {author}
  {\bibfnamefont {A.}~\bibnamefont {Miranowicz}}, \bibinfo {author}
  {\bibfnamefont {N.}~\bibnamefont {Lambert}}, \bibinfo {author} {\bibfnamefont
  {D.}~\bibnamefont {Zheng}}, \bibinfo {author} {\bibfnamefont
  {H.}~\bibnamefont {Fan}}, \bibinfo {author} {\bibfnamefont {F.}~\bibnamefont
  {Nori}},\ and\ \bibinfo {author} {\bibfnamefont {S.-B.}\ \bibnamefont
  {Zheng}},\ }\bibfield  {title} {\bibinfo {title} {Observation of a
  superradiant phase transition with emergent cat states},\ }\href
  {https://doi.org/10.1103/PhysRevLett.131.113601} {\bibfield  {journal}
  {\bibinfo  {journal} {Phys. Rev. Lett.}\ }\textbf {\bibinfo {volume} {131}},\
  \bibinfo {pages} {113601} (\bibinfo {year} {2023})}\BibitemShut {NoStop}%
\bibitem [{\citenamefont {Heyl}\ \emph {et~al.}(2013)\citenamefont {Heyl},
  \citenamefont {Polkovnikov},\ and\ \citenamefont
  {Kehrein}}]{PhysRevLett.110.135704}%
  \BibitemOpen
  \bibfield  {author} {\bibinfo {author} {\bibfnamefont {M.}~\bibnamefont
  {Heyl}}, \bibinfo {author} {\bibfnamefont {A.}~\bibnamefont {Polkovnikov}},\
  and\ \bibinfo {author} {\bibfnamefont {S.}~\bibnamefont {Kehrein}},\
  }\bibfield  {title} {\bibinfo {title} {Dynamical quantum phase transitions in
  the transverse-field {Ising} model},\ }\href
  {https://doi.org/10.1103/PhysRevLett.110.135704} {\bibfield  {journal}
  {\bibinfo  {journal} {Phys. Rev. Lett.}\ }\textbf {\bibinfo {volume} {110}},\
  \bibinfo {pages} {135704} (\bibinfo {year} {2013})}\BibitemShut {NoStop}%
\bibitem [{\citenamefont {Jurcevic}\ \emph {et~al.}(2017)\citenamefont
  {Jurcevic}, \citenamefont {Shen}, \citenamefont {Hauke}, \citenamefont
  {Maier}, \citenamefont {Brydges}, \citenamefont {Hempel}, \citenamefont
  {Lanyon}, \citenamefont {Heyl}, \citenamefont {Blatt},\ and\ \citenamefont
  {Roos}}]{PhysRevLett.119.080501}%
  \BibitemOpen
  \bibfield  {author} {\bibinfo {author} {\bibfnamefont {P.}~\bibnamefont
  {Jurcevic}}, \bibinfo {author} {\bibfnamefont {H.}~\bibnamefont {Shen}},
  \bibinfo {author} {\bibfnamefont {P.}~\bibnamefont {Hauke}}, \bibinfo
  {author} {\bibfnamefont {C.}~\bibnamefont {Maier}}, \bibinfo {author}
  {\bibfnamefont {T.}~\bibnamefont {Brydges}}, \bibinfo {author} {\bibfnamefont
  {C.}~\bibnamefont {Hempel}}, \bibinfo {author} {\bibfnamefont {B.~P.}\
  \bibnamefont {Lanyon}}, \bibinfo {author} {\bibfnamefont {M.}~\bibnamefont
  {Heyl}}, \bibinfo {author} {\bibfnamefont {R.}~\bibnamefont {Blatt}},\ and\
  \bibinfo {author} {\bibfnamefont {C.~F.}\ \bibnamefont {Roos}},\ }\bibfield
  {title} {\bibinfo {title} {Direct observation of dynamical quantum phase
  transitions in an interacting many-body system},\ }\href
  {https://doi.org/10.1103/PhysRevLett.119.080501} {\bibfield  {journal}
  {\bibinfo  {journal} {Phys. Rev. Lett.}\ }\textbf {\bibinfo {volume} {119}},\
  \bibinfo {pages} {080501} (\bibinfo {year} {2017})}\BibitemShut {NoStop}%
\bibitem [{\citenamefont {Keesling}\ \emph {et~al.}(2019)\citenamefont
  {Keesling}, \citenamefont {Omran}, \citenamefont {Levine}, \citenamefont
  {Bernien} \emph {et~al.}}]{Keesling2019}%
  \BibitemOpen
  \bibfield  {author} {\bibinfo {author} {\bibfnamefont {A.}~\bibnamefont
  {Keesling}}, \bibinfo {author} {\bibfnamefont {A.}~\bibnamefont {Omran}},
  \bibinfo {author} {\bibfnamefont {H.}~\bibnamefont {Levine}}, \bibinfo
  {author} {\bibfnamefont {H.}~\bibnamefont {Bernien}}, \emph {et~al.},\
  }\bibfield  {title} {\bibinfo {title} {Quantum kibble--zurek mechanism and
  critical dynamics on a programmable rydberg simulator},\ }\href
  {https://doi.org/10.1038/s41586-019-1070-1} {\bibfield  {journal} {\bibinfo
  {journal} {Nature}\ }\textbf {\bibinfo {volume} {568}},\ \bibinfo {pages}
  {207} (\bibinfo {year} {2019})}\BibitemShut {NoStop}%
\bibitem [{\citenamefont {Feng}\ \emph {et~al.}(2023)\citenamefont {Feng},
  \citenamefont {Katz}, \citenamefont {Haack}, \citenamefont {Maghrebi},
  \citenamefont {Gorshkov}, \citenamefont {Gong}, \citenamefont {Cetina},\ and\
  \citenamefont {Monroe}}]{Feng2023}%
  \BibitemOpen
  \bibfield  {author} {\bibinfo {author} {\bibfnamefont {L.}~\bibnamefont
  {Feng}}, \bibinfo {author} {\bibfnamefont {O.}~\bibnamefont {Katz}}, \bibinfo
  {author} {\bibfnamefont {C.}~\bibnamefont {Haack}}, \bibinfo {author}
  {\bibfnamefont {M.}~\bibnamefont {Maghrebi}}, \bibinfo {author}
  {\bibfnamefont {A.~V.}\ \bibnamefont {Gorshkov}}, \bibinfo {author}
  {\bibfnamefont {Z.}~\bibnamefont {Gong}}, \bibinfo {author} {\bibfnamefont
  {M.}~\bibnamefont {Cetina}},\ and\ \bibinfo {author} {\bibfnamefont
  {C.}~\bibnamefont {Monroe}},\ }\bibfield  {title} {\bibinfo {title}
  {Continuous symmetry breaking in a trapped-ion spin chain},\ }\href
  {https://doi.org/10.1038/s41586-023-06656-7} {\bibfield  {journal} {\bibinfo
  {journal} {Nature}\ }\textbf {\bibinfo {volume} {623}},\ \bibinfo {pages}
  {713} (\bibinfo {year} {2023})}\BibitemShut {NoStop}%
\bibitem [{\citenamefont {Campos~Venuti}\ and\ \citenamefont
  {Zanardi}(2007)}]{PhysRevLett.99.095701}%
  \BibitemOpen
  \bibfield  {author} {\bibinfo {author} {\bibfnamefont {L.}~\bibnamefont
  {Campos~Venuti}}\ and\ \bibinfo {author} {\bibfnamefont {P.}~\bibnamefont
  {Zanardi}},\ }\bibfield  {title} {\bibinfo {title} {Quantum critical scaling
  of the geometric tensors},\ }\href
  {https://doi.org/10.1103/PhysRevLett.99.095701} {\bibfield  {journal}
  {\bibinfo  {journal} {Phys. Rev. Lett.}\ }\textbf {\bibinfo {volume} {99}},\
  \bibinfo {pages} {095701} (\bibinfo {year} {2007})}\BibitemShut {NoStop}%
\bibitem [{\citenamefont {Garbe}\ \emph {et~al.}(2020)\citenamefont {Garbe},
  \citenamefont {Bina}, \citenamefont {Keller}, \citenamefont {Paris},\ and\
  \citenamefont {Felicetti}}]{PhysRevLett.124.120504}%
  \BibitemOpen
  \bibfield  {author} {\bibinfo {author} {\bibfnamefont {L.}~\bibnamefont
  {Garbe}}, \bibinfo {author} {\bibfnamefont {M.}~\bibnamefont {Bina}},
  \bibinfo {author} {\bibfnamefont {A.}~\bibnamefont {Keller}}, \bibinfo
  {author} {\bibfnamefont {M.~G.~A.}\ \bibnamefont {Paris}},\ and\ \bibinfo
  {author} {\bibfnamefont {S.}~\bibnamefont {Felicetti}},\ }\bibfield  {title}
  {\bibinfo {title} {Critical quantum metrology with a finite-component quantum
  phase transition},\ }\href {https://doi.org/10.1103/PhysRevLett.124.120504}
  {\bibfield  {journal} {\bibinfo  {journal} {Phys. Rev. Lett.}\ }\textbf
  {\bibinfo {volume} {124}},\ \bibinfo {pages} {120504} (\bibinfo {year}
  {2020})}\BibitemShut {NoStop}%
\bibitem [{\citenamefont {Ding}\ \emph {et~al.}(2022)\citenamefont {Ding},
  \citenamefont {Liu}, \citenamefont {Shi}, \citenamefont {Guo}, \citenamefont
  {M{\o}lmer},\ and\ \citenamefont {Adams}}]{Ding2022}%
  \BibitemOpen
  \bibfield  {author} {\bibinfo {author} {\bibfnamefont {D.-S.}\ \bibnamefont
  {Ding}}, \bibinfo {author} {\bibfnamefont {Z.-K.}\ \bibnamefont {Liu}},
  \bibinfo {author} {\bibfnamefont {B.-S.}\ \bibnamefont {Shi}}, \bibinfo
  {author} {\bibfnamefont {G.-C.}\ \bibnamefont {Guo}}, \bibinfo {author}
  {\bibfnamefont {K.}~\bibnamefont {M{\o}lmer}},\ and\ \bibinfo {author}
  {\bibfnamefont {C.~S.}\ \bibnamefont {Adams}},\ }\bibfield  {title} {\bibinfo
  {title} {Enhanced metrology at the critical point of a many-body rydberg
  atomic system},\ }\href {https://doi.org/10.1038/s41567-022-01777-8}
  {\bibfield  {journal} {\bibinfo  {journal} {Nature Physics}\ }\textbf
  {\bibinfo {volume} {18}},\ \bibinfo {pages} {1447} (\bibinfo {year}
  {2022})}\BibitemShut {NoStop}%
\bibitem [{\citenamefont {Ghosh}\ \emph {et~al.}(2020)\citenamefont {Ghosh},
  \citenamefont {Chanda},\ and\ \citenamefont {Sen(De)}}]{PhysRevA.101.032115}%
  \BibitemOpen
  \bibfield  {author} {\bibinfo {author} {\bibfnamefont {S.}~\bibnamefont
  {Ghosh}}, \bibinfo {author} {\bibfnamefont {T.}~\bibnamefont {Chanda}},\ and\
  \bibinfo {author} {\bibfnamefont {A.}~\bibnamefont {Sen(De)}},\ }\bibfield
  {title} {\bibinfo {title} {Enhancement in the performance of a quantum
  battery by ordered and disordered interactions},\ }\href
  {https://doi.org/10.1103/PhysRevA.101.032115} {\bibfield  {journal} {\bibinfo
   {journal} {Phys. Rev. A}\ }\textbf {\bibinfo {volume} {101}},\ \bibinfo
  {pages} {032115} (\bibinfo {year} {2020})}\BibitemShut {NoStop}%
\bibitem [{\citenamefont {Abah}\ \emph {et~al.}(2022)\citenamefont {Abah},
  \citenamefont {De~Chiara}, \citenamefont {Paternostro},\ and\ \citenamefont
  {Puebla}}]{PhysRevResearch.4.L022017}%
  \BibitemOpen
  \bibfield  {author} {\bibinfo {author} {\bibfnamefont {O.}~\bibnamefont
  {Abah}}, \bibinfo {author} {\bibfnamefont {G.}~\bibnamefont {De~Chiara}},
  \bibinfo {author} {\bibfnamefont {M.}~\bibnamefont {Paternostro}},\ and\
  \bibinfo {author} {\bibfnamefont {R.}~\bibnamefont {Puebla}},\ }\bibfield
  {title} {\bibinfo {title} {Harnessing nonadiabatic excitations promoted by a
  quantum critical point: Quantum battery and spin squeezing},\ }\href
  {https://doi.org/10.1103/PhysRevResearch.4.L022017} {\bibfield  {journal}
  {\bibinfo  {journal} {Phys. Rev. Res.}\ }\textbf {\bibinfo {volume} {4}},\
  \bibinfo {pages} {L022017} (\bibinfo {year} {2022})}\BibitemShut {NoStop}%
\bibitem [{\citenamefont {Dou}\ \emph {et~al.}(2022{\natexlab{b}})\citenamefont
  {Dou}, \citenamefont {Lu}, \citenamefont {Wang},\ and\ \citenamefont
  {Sun}}]{PhysRevB.105.115405}%
  \BibitemOpen
  \bibfield  {author} {\bibinfo {author} {\bibfnamefont {F.-Q.}\ \bibnamefont
  {Dou}}, \bibinfo {author} {\bibfnamefont {Y.-Q.}\ \bibnamefont {Lu}},
  \bibinfo {author} {\bibfnamefont {Y.-J.}\ \bibnamefont {Wang}},\ and\
  \bibinfo {author} {\bibfnamefont {J.-A.}\ \bibnamefont {Sun}},\ }\bibfield
  {title} {\bibinfo {title} {Extended {Dicke} quantum battery with interatomic
  interactions and driving field},\ }\href
  {https://doi.org/10.1103/PhysRevB.105.115405} {\bibfield  {journal} {\bibinfo
   {journal} {Phys. Rev. B}\ }\textbf {\bibinfo {volume} {105}},\ \bibinfo
  {pages} {115405} (\bibinfo {year} {2022}{\natexlab{b}})}\BibitemShut
  {NoStop}%
\bibitem [{\citenamefont {Zhao}\ \emph {et~al.}(2022)\citenamefont {Zhao},
  \citenamefont {Dou},\ and\ \citenamefont {Zhao}}]{PhysRevResearch.4.013172}%
  \BibitemOpen
  \bibfield  {author} {\bibinfo {author} {\bibfnamefont {F.}~\bibnamefont
  {Zhao}}, \bibinfo {author} {\bibfnamefont {F.-Q.}\ \bibnamefont {Dou}},\ and\
  \bibinfo {author} {\bibfnamefont {Q.}~\bibnamefont {Zhao}},\ }\bibfield
  {title} {\bibinfo {title} {Charging performance of the {Su-Schrieffer-Heeger}
  quantum battery},\ }\href {https://doi.org/10.1103/PhysRevResearch.4.013172}
  {\bibfield  {journal} {\bibinfo  {journal} {Phys. Rev. Res.}\ }\textbf
  {\bibinfo {volume} {4}},\ \bibinfo {pages} {013172} (\bibinfo {year}
  {2022})}\BibitemShut {NoStop}%
\bibitem [{\citenamefont {Yang}\ \emph {et~al.}(2024)\citenamefont {Yang},
  \citenamefont {Yang},\ and\ \citenamefont {Dou}}]{PhysRevB.109.235432}%
  \BibitemOpen
  \bibfield  {author} {\bibinfo {author} {\bibfnamefont {D.-L.}\ \bibnamefont
  {Yang}}, \bibinfo {author} {\bibfnamefont {F.-M.}\ \bibnamefont {Yang}},\
  and\ \bibinfo {author} {\bibfnamefont {F.-Q.}\ \bibnamefont {Dou}},\
  }\bibfield  {title} {\bibinfo {title} {Three-level {Dicke} quantum battery},\
  }\href {https://doi.org/10.1103/PhysRevB.109.235432} {\bibfield  {journal}
  {\bibinfo  {journal} {Phys. Rev. B}\ }\textbf {\bibinfo {volume} {109}},\
  \bibinfo {pages} {235432} (\bibinfo {year} {2024})}\BibitemShut {NoStop}%
\bibitem [{\citenamefont {Downing}\ and\ \citenamefont
  {Ukhtary}(2023)}]{Downing2023}%
  \BibitemOpen
  \bibfield  {author} {\bibinfo {author} {\bibfnamefont {C.~A.}\ \bibnamefont
  {Downing}}\ and\ \bibinfo {author} {\bibfnamefont {M.~S.}\ \bibnamefont
  {Ukhtary}},\ }\bibfield  {title} {\bibinfo {title} {A quantum battery with
  quadratic driving},\ }\href {https://doi.org/10.1038/s42005-023-01439-y}
  {\bibfield  {journal} {\bibinfo  {journal} {Communications Physics}\ }\textbf
  {\bibinfo {volume} {6}},\ \bibinfo {pages} {322} (\bibinfo {year}
  {2023})}\BibitemShut {NoStop}%
\bibitem [{\citenamefont {Kheiri}\ \emph {et~al.}(2026)\citenamefont {Kheiri},
  \citenamefont {Kheiri},\ and\ \citenamefont {Mahdavifar}}]{42rw-l31q}%
  \BibitemOpen
  \bibfield  {author} {\bibinfo {author} {\bibfnamefont {S.}~\bibnamefont
  {Kheiri}}, \bibinfo {author} {\bibfnamefont {S.}~\bibnamefont {Kheiri}},\
  and\ \bibinfo {author} {\bibfnamefont {S.}~\bibnamefont {Mahdavifar}},\
  }\bibfield  {title} {\bibinfo {title} {Dynamical phase transitions and
  criticality-enhanced charging in a topological quantum battery},\ }\href
  {https://doi.org/10.1103/42rw-l31q} {\bibfield  {journal} {\bibinfo
  {journal} {Phys. Rev. B}\ }\textbf {\bibinfo {volume} {113}},\ \bibinfo
  {pages} {205150} (\bibinfo {year} {2026})}\BibitemShut {NoStop}%
\bibitem [{\citenamefont {Grazi}\ \emph {et~al.}(2024)\citenamefont {Grazi},
  \citenamefont {Sacco~Shaikh}, \citenamefont {Sassetti}, \citenamefont
  {Traverso~Ziani},\ and\ \citenamefont {Ferraro}}]{PhysRevLett.133.197001}%
  \BibitemOpen
  \bibfield  {author} {\bibinfo {author} {\bibfnamefont {R.}~\bibnamefont
  {Grazi}}, \bibinfo {author} {\bibfnamefont {D.}~\bibnamefont {Sacco~Shaikh}},
  \bibinfo {author} {\bibfnamefont {M.}~\bibnamefont {Sassetti}}, \bibinfo
  {author} {\bibfnamefont {N.}~\bibnamefont {Traverso~Ziani}},\ and\ \bibinfo
  {author} {\bibfnamefont {D.}~\bibnamefont {Ferraro}},\ }\bibfield  {title}
  {\bibinfo {title} {Controlling energy storage crossing quantum phase
  transitions in an integrable spin quantum battery},\ }\href
  {https://doi.org/10.1103/PhysRevLett.133.197001} {\bibfield  {journal}
  {\bibinfo  {journal} {Phys. Rev. Lett.}\ }\textbf {\bibinfo {volume} {133}},\
  \bibinfo {pages} {197001} (\bibinfo {year} {2024})}\BibitemShut {NoStop}%
\bibitem [{\citenamefont {Hou}\ \emph {et~al.}(2026)\citenamefont {Hou},
  \citenamefont {Zhang},\ and\ \citenamefont {Wang}}]{Hou2026}%
  \BibitemOpen
  \bibfield  {author} {\bibinfo {author} {\bibfnamefont {L.}~\bibnamefont
  {Hou}}, \bibinfo {author} {\bibfnamefont {Y.}~\bibnamefont {Zhang}},\ and\
  \bibinfo {author} {\bibfnamefont {C.}~\bibnamefont {Wang}},\ }\bibfield
  {title} {\bibinfo {title} {Quantum battery capacity for quantum phase
  transition with {Dzyaloshinskii--Moriya} interaction},\ }\href
  {https://doi.org/10.1140/epjp/s13360-026-07486-y} {\bibfield  {journal}
  {\bibinfo  {journal} {Eur. Phys. J. Plus}\ }\textbf {\bibinfo {volume}
  {141}},\ \bibinfo {pages} {235} (\bibinfo {year} {2026})}\BibitemShut
  {NoStop}%
\bibitem [{\citenamefont {Erdman}\ \emph {et~al.}(2024)\citenamefont {Erdman},
  \citenamefont {Andolina}, \citenamefont {Giovannetti},\ and\ \citenamefont
  {No\'e}}]{PhysRevLett.133.243602}%
  \BibitemOpen
  \bibfield  {author} {\bibinfo {author} {\bibfnamefont {P.~A.}\ \bibnamefont
  {Erdman}}, \bibinfo {author} {\bibfnamefont {G.~M.}\ \bibnamefont
  {Andolina}}, \bibinfo {author} {\bibfnamefont {V.}~\bibnamefont
  {Giovannetti}},\ and\ \bibinfo {author} {\bibfnamefont {F.}~\bibnamefont
  {No\'e}},\ }\bibfield  {title} {\bibinfo {title} {Reinforcement learning
  optimization of the charging of a {Dicke} quantum battery},\ }\href
  {https://doi.org/10.1103/PhysRevLett.133.243602} {\bibfield  {journal}
  {\bibinfo  {journal} {Phys. Rev. Lett.}\ }\textbf {\bibinfo {volume} {133}},\
  \bibinfo {pages} {243602} (\bibinfo {year} {2024})}\BibitemShut {NoStop}%
\bibitem [{\citenamefont {Koutromanos}\ \emph {et~al.}(2026)\citenamefont
  {Koutromanos}, \citenamefont {Stefanatos},\ and\ \citenamefont
  {Paspalakis}}]{m1gm-f9zy}%
  \BibitemOpen
  \bibfield  {author} {\bibinfo {author} {\bibfnamefont {D.}~\bibnamefont
  {Koutromanos}}, \bibinfo {author} {\bibfnamefont {D.}~\bibnamefont
  {Stefanatos}},\ and\ \bibinfo {author} {\bibfnamefont {E.}~\bibnamefont
  {Paspalakis}},\ }\bibfield  {title} {\bibinfo {title} {Efficient charging of
  driven cavity-{Heisenberg} spin-$1/2$ chain quantum battery with deep
  learning methods},\ }\href {https://doi.org/10.1103/m1gm-f9zy} {\bibfield
  {journal} {\bibinfo  {journal} {Phys. Rev. Appl.}\ }\textbf {\bibinfo
  {volume} {25}},\ \bibinfo {pages} {054053} (\bibinfo {year}
  {2026})}\BibitemShut {NoStop}%
\bibitem [{\citenamefont {Sun}\ \emph {et~al.}(2025)\citenamefont {Sun},
  \citenamefont {Zhou},\ and\ \citenamefont {Dou}}]{Sun_2025}%
  \BibitemOpen
  \bibfield  {author} {\bibinfo {author} {\bibfnamefont {P.-Y.}\ \bibnamefont
  {Sun}}, \bibinfo {author} {\bibfnamefont {H.}~\bibnamefont {Zhou}},\ and\
  \bibinfo {author} {\bibfnamefont {F.-Q.}\ \bibnamefont {Dou}},\ }\bibfield
  {title} {\bibinfo {title} {Cavity-{Heisenberg} spin-j chain quantum battery
  and reinforcement learning optimization},\ }\href
  {https://doi.org/10.1088/1367-2630/ae2a62} {\bibfield  {journal} {\bibinfo
  {journal} {New J. Phys.}\ }\textbf {\bibinfo {volume} {27}},\ \bibinfo
  {pages} {124513} (\bibinfo {year} {2025})}\BibitemShut {NoStop}%
\bibitem [{\citenamefont {Joshi}\ \emph {et~al.}(2016)\citenamefont {Joshi},
  \citenamefont {Larson},\ and\ \citenamefont {Spiller}}]{PhysRevA.93.043818}%
  \BibitemOpen
  \bibfield  {author} {\bibinfo {author} {\bibfnamefont {C.}~\bibnamefont
  {Joshi}}, \bibinfo {author} {\bibfnamefont {J.}~\bibnamefont {Larson}},\ and\
  \bibinfo {author} {\bibfnamefont {T.~P.}\ \bibnamefont {Spiller}},\
  }\bibfield  {title} {\bibinfo {title} {Quantum state engineering in hybrid
  open quantum systems},\ }\href {https://doi.org/10.1103/PhysRevA.93.043818}
  {\bibfield  {journal} {\bibinfo  {journal} {Phys. Rev. A}\ }\textbf {\bibinfo
  {volume} {93}},\ \bibinfo {pages} {043818} (\bibinfo {year}
  {2016})}\BibitemShut {NoStop}%
\bibitem [{\citenamefont {Wang}\ \emph {et~al.}(2024)\citenamefont {Wang},
  \citenamefont {Liu}, \citenamefont {Wu}, \citenamefont {Fan},\ and\
  \citenamefont {Liu}}]{PhysRevA.110.042419}%
  \BibitemOpen
  \bibfield  {author} {\bibinfo {author} {\bibfnamefont {L.}~\bibnamefont
  {Wang}}, \bibinfo {author} {\bibfnamefont {S.-Q.}\ \bibnamefont {Liu}},
  \bibinfo {author} {\bibfnamefont {F.-l.}\ \bibnamefont {Wu}}, \bibinfo
  {author} {\bibfnamefont {H.}~\bibnamefont {Fan}},\ and\ \bibinfo {author}
  {\bibfnamefont {S.-Y.}\ \bibnamefont {Liu}},\ }\bibfield  {title} {\bibinfo
  {title} {Deep strong charging in a multiphoton anisotropic {Dicke} quantum
  battery},\ }\href {https://doi.org/10.1103/PhysRevA.110.042419} {\bibfield
  {journal} {\bibinfo  {journal} {Phys. Rev. A}\ }\textbf {\bibinfo {volume}
  {110}},\ \bibinfo {pages} {042419} (\bibinfo {year} {2024})}\BibitemShut
  {NoStop}%
\bibitem [{\citenamefont {Mondal}\ \emph {et~al.}(2026)\citenamefont {Mondal},
  \citenamefont {Santos},\ and\ \citenamefont {Sinha}}]{ymzt-2mq4}%
  \BibitemOpen
  \bibfield  {author} {\bibinfo {author} {\bibfnamefont {D.}~\bibnamefont
  {Mondal}}, \bibinfo {author} {\bibfnamefont {L.~F.}\ \bibnamefont {Santos}},\
  and\ \bibinfo {author} {\bibfnamefont {S.}~\bibnamefont {Sinha}},\ }\bibfield
   {title} {\bibinfo {title} {Transient and steady-state chaos in dissipative
  quantum systems},\ }\href {https://doi.org/10.1103/ymzt-2mq4} {\bibfield
  {journal} {\bibinfo  {journal} {Phys. Rev. Lett.}\ }\textbf {\bibinfo
  {volume} {136}},\ \bibinfo {pages} {040401} (\bibinfo {year}
  {2026})}\BibitemShut {NoStop}%
\bibitem [{\citenamefont {Buijsman}\ \emph {et~al.}(2017)\citenamefont
  {Buijsman}, \citenamefont {Gritsev},\ and\ \citenamefont
  {Sprik}}]{PhysRevLett.118.080601}%
  \BibitemOpen
  \bibfield  {author} {\bibinfo {author} {\bibfnamefont {W.}~\bibnamefont
  {Buijsman}}, \bibinfo {author} {\bibfnamefont {V.}~\bibnamefont {Gritsev}},\
  and\ \bibinfo {author} {\bibfnamefont {R.}~\bibnamefont {Sprik}},\ }\bibfield
   {title} {\bibinfo {title} {Nonergodicity in the anisotropic {Dicke} model},\
  }\href {https://doi.org/10.1103/PhysRevLett.118.080601} {\bibfield  {journal}
  {\bibinfo  {journal} {Phys. Rev. Lett.}\ }\textbf {\bibinfo {volume} {118}},\
  \bibinfo {pages} {080601} (\bibinfo {year} {2017})}\BibitemShut {NoStop}%
\bibitem [{\citenamefont {Das}\ \emph {et~al.}(2023{\natexlab{a}})\citenamefont
  {Das}, \citenamefont {Bhakuni},\ and\ \citenamefont
  {Sharma}}]{PhysRevA.107.043706}%
  \BibitemOpen
  \bibfield  {author} {\bibinfo {author} {\bibfnamefont {P.}~\bibnamefont
  {Das}}, \bibinfo {author} {\bibfnamefont {D.~S.}\ \bibnamefont {Bhakuni}},\
  and\ \bibinfo {author} {\bibfnamefont {A.}~\bibnamefont {Sharma}},\
  }\bibfield  {title} {\bibinfo {title} {Phase transitions of the anisotropic
  {Dicke} model},\ }\href {https://doi.org/10.1103/PhysRevA.107.043706}
  {\bibfield  {journal} {\bibinfo  {journal} {Phys. Rev. A}\ }\textbf {\bibinfo
  {volume} {107}},\ \bibinfo {pages} {043706} (\bibinfo {year}
  {2023}{\natexlab{a}})}\BibitemShut {NoStop}%
\bibitem [{\citenamefont {Das}\ \emph {et~al.}(2023{\natexlab{b}})\citenamefont
  {Das}, \citenamefont {Bhakuni}, \citenamefont {Santos},\ and\ \citenamefont
  {Sharma}}]{PhysRevA.108.063716}%
  \BibitemOpen
  \bibfield  {author} {\bibinfo {author} {\bibfnamefont {P.}~\bibnamefont
  {Das}}, \bibinfo {author} {\bibfnamefont {D.~S.}\ \bibnamefont {Bhakuni}},
  \bibinfo {author} {\bibfnamefont {L.~F.}\ \bibnamefont {Santos}},\ and\
  \bibinfo {author} {\bibfnamefont {A.}~\bibnamefont {Sharma}},\ }\bibfield
  {title} {\bibinfo {title} {Periodically and quasiperiodically driven
  anisotropic {Dicke} model},\ }\href
  {https://doi.org/10.1103/PhysRevA.108.063716} {\bibfield  {journal} {\bibinfo
   {journal} {Phys. Rev. A}\ }\textbf {\bibinfo {volume} {108}},\ \bibinfo
  {pages} {063716} (\bibinfo {year} {2023}{\natexlab{b}})}\BibitemShut
  {NoStop}%
\bibitem [{\citenamefont {Herrera~Romero}\ and\ \citenamefont
  {Bastarrachea-Magnani}(2024)}]{e26070574}%
  \BibitemOpen
  \bibfield  {author} {\bibinfo {author} {\bibfnamefont {R.}~\bibnamefont
  {Herrera~Romero}}\ and\ \bibinfo {author} {\bibfnamefont {M.~A.}\
  \bibnamefont {Bastarrachea-Magnani}},\ }\bibfield  {title} {\bibinfo {title}
  {Phase and amplitude modes in the anisotropic {Dicke} model with matter
  interactions},\ }\bibfield  {journal} {\bibinfo  {journal} {Entropy}\
  }\textbf {\bibinfo {volume} {26}},\ \href {https://doi.org/10.3390/e26070574}
  {10.3390/e26070574} (\bibinfo {year} {2024})\BibitemShut {NoStop}%
\bibitem [{\citenamefont {Das}\ and\ \citenamefont
  {Saha}(2026)}]{das2026detailedstudyvariousphases}%
  \BibitemOpen
  \bibfield  {author} {\bibinfo {author} {\bibfnamefont {P.}~\bibnamefont
  {Das}}\ and\ \bibinfo {author} {\bibfnamefont {S.}~\bibnamefont {Saha}},\
  }\bibfield  {title} {\bibinfo {title} {A detailed study on various phases in
  dissipative anisotropic {Dicke} model},\ }\href
  {https://arxiv.org/abs/2505.20550} {\  (\bibinfo {year} {2026})},\ \Eprint
  {https://arxiv.org/abs/2505.20550} {arXiv:2505.20550} \BibitemShut {NoStop}%
\bibitem [{\citenamefont {Chen}\ \emph {et~al.}(2024)\citenamefont {Chen},
  \citenamefont {Zhang}, \citenamefont {Chen},\ and\ \citenamefont
  {Lin}}]{PhysRevA.110.063722}%
  \BibitemOpen
  \bibfield  {author} {\bibinfo {author} {\bibfnamefont {X.-Y.}\ \bibnamefont
  {Chen}}, \bibinfo {author} {\bibfnamefont {Y.-Y.}\ \bibnamefont {Zhang}},
  \bibinfo {author} {\bibfnamefont {Q.-H.}\ \bibnamefont {Chen}},\ and\
  \bibinfo {author} {\bibfnamefont {H.-Q.}\ \bibnamefont {Lin}},\ }\bibfield
  {title} {\bibinfo {title} {Phase transitions in the anisotropic {Dicke-Stark}
  model with ${A}^{2}$ terms},\ }\href
  {https://doi.org/10.1103/PhysRevA.110.063722} {\bibfield  {journal} {\bibinfo
   {journal} {Phys. Rev. A}\ }\textbf {\bibinfo {volume} {110}},\ \bibinfo
  {pages} {063722} (\bibinfo {year} {2024})}\BibitemShut {NoStop}%
\bibitem [{\citenamefont {Baksic}\ and\ \citenamefont
  {Ciuti}(2014)}]{PhysRevLett.112.173601}%
  \BibitemOpen
  \bibfield  {author} {\bibinfo {author} {\bibfnamefont {A.}~\bibnamefont
  {Baksic}}\ and\ \bibinfo {author} {\bibfnamefont {C.}~\bibnamefont {Ciuti}},\
  }\bibfield  {title} {\bibinfo {title} {Controlling discrete and continuous
  symmetries in ``superradiant'' phase transitions with circuit qed systems},\
  }\href {https://doi.org/10.1103/PhysRevLett.112.173601} {\bibfield  {journal}
  {\bibinfo  {journal} {Phys. Rev. Lett.}\ }\textbf {\bibinfo {volume} {112}},\
  \bibinfo {pages} {173601} (\bibinfo {year} {2014})}\BibitemShut {NoStop}%
\bibitem [{\citenamefont {Forn-D{\'{\i}}az}\ \emph {et~al.}(2019)\citenamefont
  {Forn-D{\'{\i}}az}, \citenamefont {Lamata}, \citenamefont {Rico},
  \citenamefont {Kono},\ and\ \citenamefont {Solano}}]{RevModPhys.91.025005}%
  \BibitemOpen
  \bibfield  {author} {\bibinfo {author} {\bibfnamefont {P.}~\bibnamefont
  {Forn-D{\'{\i}}az}}, \bibinfo {author} {\bibfnamefont {L.}~\bibnamefont
  {Lamata}}, \bibinfo {author} {\bibfnamefont {E.}~\bibnamefont {Rico}},
  \bibinfo {author} {\bibfnamefont {J.}~\bibnamefont {Kono}},\ and\ \bibinfo
  {author} {\bibfnamefont {E.}~\bibnamefont {Solano}},\ }\bibfield  {title}
  {\bibinfo {title} {Ultrastrong coupling regimes of light-matter
  interaction},\ }\href {https://doi.org/10.1103/RevModPhys.91.025005}
  {\bibfield  {journal} {\bibinfo  {journal} {Rev. Mod. Phys.}\ }\textbf
  {\bibinfo {volume} {91}},\ \bibinfo {pages} {025005} (\bibinfo {year}
  {2019})}\BibitemShut {NoStop}%
\bibitem [{\citenamefont {Kirchmair}\ \emph {et~al.}(2013)\citenamefont
  {Kirchmair}, \citenamefont {Vlastakis}, \citenamefont {Leghtas},
  \citenamefont {Nigg}, \citenamefont {Paik}, \citenamefont {Ginossar},
  \citenamefont {Mirrahimi}, \citenamefont {Frunzio}, \citenamefont {Girvin},\
  and\ \citenamefont {Schoelkopf}}]{Kirchmair2013}%
  \BibitemOpen
  \bibfield  {author} {\bibinfo {author} {\bibfnamefont {G.}~\bibnamefont
  {Kirchmair}}, \bibinfo {author} {\bibfnamefont {B.}~\bibnamefont
  {Vlastakis}}, \bibinfo {author} {\bibfnamefont {Z.}~\bibnamefont {Leghtas}},
  \bibinfo {author} {\bibfnamefont {S.~E.}\ \bibnamefont {Nigg}}, \bibinfo
  {author} {\bibfnamefont {H.}~\bibnamefont {Paik}}, \bibinfo {author}
  {\bibfnamefont {E.}~\bibnamefont {Ginossar}}, \bibinfo {author}
  {\bibfnamefont {M.}~\bibnamefont {Mirrahimi}}, \bibinfo {author}
  {\bibfnamefont {L.}~\bibnamefont {Frunzio}}, \bibinfo {author} {\bibfnamefont
  {S.~M.}\ \bibnamefont {Girvin}},\ and\ \bibinfo {author} {\bibfnamefont
  {R.~J.}\ \bibnamefont {Schoelkopf}},\ }\bibfield  {title} {\bibinfo {title}
  {Observation of quantum state collapse and revival due to the single-photon
  {Kerr} effect},\ }\href {https://doi.org/10.1038/nature11902} {\bibfield
  {journal} {\bibinfo  {journal} {Nature}\ }\textbf {\bibinfo {volume} {495}},\
  \bibinfo {pages} {205} (\bibinfo {year} {2013})}\BibitemShut {NoStop}%
\bibitem [{\citenamefont {Tusnin}\ \emph {et~al.}(2023)\citenamefont {Tusnin},
  \citenamefont {Tikan}, \citenamefont {Komagata},\ and\ \citenamefont
  {Kippenberg}}]{Tusnin2023}%
  \BibitemOpen
  \bibfield  {author} {\bibinfo {author} {\bibfnamefont {A.}~\bibnamefont
  {Tusnin}}, \bibinfo {author} {\bibfnamefont {A.}~\bibnamefont {Tikan}},
  \bibinfo {author} {\bibfnamefont {K.}~\bibnamefont {Komagata}},\ and\
  \bibinfo {author} {\bibfnamefont {T.~J.}\ \bibnamefont {Kippenberg}},\
  }\bibfield  {title} {\bibinfo {title} {Nonlinear dynamics and {Kerr}
  frequency comb formation in lattices of coupled microresonators},\ }\href
  {https://doi.org/10.1038/s42005-023-01438-z} {\bibfield  {journal} {\bibinfo
  {journal} {Commun. Phys.}\ }\textbf {\bibinfo {volume} {6}},\ \bibinfo
  {pages} {317} (\bibinfo {year} {2023})}\BibitemShut {NoStop}%
\bibitem [{\citenamefont {Casteels}\ \emph {et~al.}(2017)\citenamefont
  {Casteels}, \citenamefont {Fazio},\ and\ \citenamefont
  {Ciuti}}]{PhysRevA.95.012128}%
  \BibitemOpen
  \bibfield  {author} {\bibinfo {author} {\bibfnamefont {W.}~\bibnamefont
  {Casteels}}, \bibinfo {author} {\bibfnamefont {R.}~\bibnamefont {Fazio}},\
  and\ \bibinfo {author} {\bibfnamefont {C.}~\bibnamefont {Ciuti}},\ }\bibfield
   {title} {\bibinfo {title} {Critical dynamical properties of a first-order
  dissipative phase transition},\ }\href
  {https://doi.org/10.1103/PhysRevA.95.012128} {\bibfield  {journal} {\bibinfo
  {journal} {Phys. Rev. A}\ }\textbf {\bibinfo {volume} {95}},\ \bibinfo
  {pages} {012128} (\bibinfo {year} {2017})}\BibitemShut {NoStop}%
\bibitem [{\citenamefont {Minganti}\ \emph {et~al.}(2018)\citenamefont
  {Minganti}, \citenamefont {Biella}, \citenamefont {Bartolo},\ and\
  \citenamefont {Ciuti}}]{PhysRevA.98.042118}%
  \BibitemOpen
  \bibfield  {author} {\bibinfo {author} {\bibfnamefont {F.}~\bibnamefont
  {Minganti}}, \bibinfo {author} {\bibfnamefont {A.}~\bibnamefont {Biella}},
  \bibinfo {author} {\bibfnamefont {N.}~\bibnamefont {Bartolo}},\ and\ \bibinfo
  {author} {\bibfnamefont {C.}~\bibnamefont {Ciuti}},\ }\bibfield  {title}
  {\bibinfo {title} {Spectral theory of {Liouvillians} for dissipative phase
  transitions},\ }\href {https://doi.org/10.1103/PhysRevA.98.042118} {\bibfield
   {journal} {\bibinfo  {journal} {Phys. Rev. A}\ }\textbf {\bibinfo {volume}
  {98}},\ \bibinfo {pages} {042118} (\bibinfo {year} {2018})}\BibitemShut
  {NoStop}%
\bibitem [{\citenamefont {Fink}\ \emph {et~al.}(2018)\citenamefont {Fink},
  \citenamefont {Schade}, \citenamefont {H{\"o}fling}, \citenamefont
  {Schneider},\ and\ \citenamefont {Imamoglu}}]{Fink2018}%
  \BibitemOpen
  \bibfield  {author} {\bibinfo {author} {\bibfnamefont {T.}~\bibnamefont
  {Fink}}, \bibinfo {author} {\bibfnamefont {A.}~\bibnamefont {Schade}},
  \bibinfo {author} {\bibfnamefont {S.}~\bibnamefont {H{\"o}fling}}, \bibinfo
  {author} {\bibfnamefont {C.}~\bibnamefont {Schneider}},\ and\ \bibinfo
  {author} {\bibfnamefont {A.}~\bibnamefont {Imamoglu}},\ }\bibfield  {title}
  {\bibinfo {title} {Signatures of a dissipative phase transition in photon
  correlation measurements},\ }\href
  {https://doi.org/10.1038/s41567-017-0020-9} {\bibfield  {journal} {\bibinfo
  {journal} {Nat. Phys.}\ }\textbf {\bibinfo {volume} {14}},\ \bibinfo {pages}
  {365} (\bibinfo {year} {2018})}\BibitemShut {NoStop}%
\bibitem [{\citenamefont {Chen}\ \emph {et~al.}(2023)\citenamefont {Chen},
  \citenamefont {Fischer}, \citenamefont {Nojiri}, \citenamefont {Renger},
  \citenamefont {Xie}, \citenamefont {Partanen}, \citenamefont {Pogorzalek},
  \citenamefont {Fedorov}, \citenamefont {Marx}, \citenamefont {Deppe},\ and\
  \citenamefont {Gross}}]{Chen2023}%
  \BibitemOpen
  \bibfield  {author} {\bibinfo {author} {\bibfnamefont {Q.-M.}\ \bibnamefont
  {Chen}}, \bibinfo {author} {\bibfnamefont {M.}~\bibnamefont {Fischer}},
  \bibinfo {author} {\bibfnamefont {Y.}~\bibnamefont {Nojiri}}, \bibinfo
  {author} {\bibfnamefont {M.}~\bibnamefont {Renger}}, \bibinfo {author}
  {\bibfnamefont {E.}~\bibnamefont {Xie}}, \bibinfo {author} {\bibfnamefont
  {M.}~\bibnamefont {Partanen}}, \bibinfo {author} {\bibfnamefont
  {S.}~\bibnamefont {Pogorzalek}}, \bibinfo {author} {\bibfnamefont {K.~G.}\
  \bibnamefont {Fedorov}}, \bibinfo {author} {\bibfnamefont {A.}~\bibnamefont
  {Marx}}, \bibinfo {author} {\bibfnamefont {F.}~\bibnamefont {Deppe}},\ and\
  \bibinfo {author} {\bibfnamefont {R.}~\bibnamefont {Gross}},\ }\bibfield
  {title} {\bibinfo {title} {Quantum behavior of the {Duffing} oscillator at
  the dissipative phase transition},\ }\href
  {https://doi.org/10.1038/s41467-023-38217-x} {\bibfield  {journal} {\bibinfo
  {journal} {Nat. Commun.}\ }\textbf {\bibinfo {volume} {14}},\ \bibinfo
  {pages} {2896} (\bibinfo {year} {2023})}\BibitemShut {NoStop}%
\bibitem [{\citenamefont {Minganti}\ \emph {et~al.}(2023)\citenamefont
  {Minganti}, \citenamefont {Savona},\ and\ \citenamefont
  {Biella}}]{Minganti_2023}%
  \BibitemOpen
  \bibfield  {author} {\bibinfo {author} {\bibfnamefont {F.}~\bibnamefont
  {Minganti}}, \bibinfo {author} {\bibfnamefont {V.}~\bibnamefont {Savona}},\
  and\ \bibinfo {author} {\bibfnamefont {A.}~\bibnamefont {Biella}},\
  }\bibfield  {title} {\bibinfo {title} {Dissipative phase transitions in
  $n$-photon driven quantum nonlinear resonators},\ }\href
  {https://doi.org/10.22331/q-2023-11-07-1170} {\bibfield  {journal} {\bibinfo
  {journal} {Quantum}\ }\textbf {\bibinfo {volume} {7}},\ \bibinfo {pages}
  {1170} (\bibinfo {year} {2023})}\BibitemShut {NoStop}%
\bibitem [{\citenamefont {Roberts}\ and\ \citenamefont
  {Clerk}(2020)}]{PhysRevX.10.021022}%
  \BibitemOpen
  \bibfield  {author} {\bibinfo {author} {\bibfnamefont {D.}~\bibnamefont
  {Roberts}}\ and\ \bibinfo {author} {\bibfnamefont {A.~A.}\ \bibnamefont
  {Clerk}},\ }\bibfield  {title} {\bibinfo {title} {Driven-dissipative quantum
  {Kerr} resonators: New exact solutions, photon blockade and quantum
  bistability},\ }\href {https://doi.org/10.1103/PhysRevX.10.021022} {\bibfield
   {journal} {\bibinfo  {journal} {Phys. Rev. X}\ }\textbf {\bibinfo {volume}
  {10}},\ \bibinfo {pages} {021022} (\bibinfo {year} {2020})}\BibitemShut
  {NoStop}%
\bibitem [{\citenamefont {Sett}\ \emph {et~al.}(2024)\citenamefont {Sett},
  \citenamefont {Hassani}, \citenamefont {Phan}, \citenamefont {Barzanjeh},
  \citenamefont {Vukics},\ and\ \citenamefont {Fink}}]{PRXQuantum.5.010327}%
  \BibitemOpen
  \bibfield  {author} {\bibinfo {author} {\bibfnamefont {R.}~\bibnamefont
  {Sett}}, \bibinfo {author} {\bibfnamefont {F.}~\bibnamefont {Hassani}},
  \bibinfo {author} {\bibfnamefont {D.}~\bibnamefont {Phan}}, \bibinfo {author}
  {\bibfnamefont {S.}~\bibnamefont {Barzanjeh}}, \bibinfo {author}
  {\bibfnamefont {A.}~\bibnamefont {Vukics}},\ and\ \bibinfo {author}
  {\bibfnamefont {J.~M.}\ \bibnamefont {Fink}},\ }\bibfield  {title} {\bibinfo
  {title} {Emergent macroscopic bistability induced by a single superconducting
  qubit},\ }\href {https://doi.org/10.1103/PRXQuantum.5.010327} {\bibfield
  {journal} {\bibinfo  {journal} {PRX Quantum}\ }\textbf {\bibinfo {volume}
  {5}},\ \bibinfo {pages} {010327} (\bibinfo {year} {2024})}\BibitemShut
  {NoStop}%
\bibitem [{\citenamefont {Beaulieu}\ \emph {et~al.}(2025)\citenamefont
  {Beaulieu}, \citenamefont {Minganti}, \citenamefont {Frasca}, \citenamefont
  {Savona}, \citenamefont {Felicetti}, \citenamefont {Di~Candia},\ and\
  \citenamefont {Scarlino}}]{Beaulieu2025}%
  \BibitemOpen
  \bibfield  {author} {\bibinfo {author} {\bibfnamefont {G.}~\bibnamefont
  {Beaulieu}}, \bibinfo {author} {\bibfnamefont {F.}~\bibnamefont {Minganti}},
  \bibinfo {author} {\bibfnamefont {S.}~\bibnamefont {Frasca}}, \bibinfo
  {author} {\bibfnamefont {V.}~\bibnamefont {Savona}}, \bibinfo {author}
  {\bibfnamefont {S.}~\bibnamefont {Felicetti}}, \bibinfo {author}
  {\bibfnamefont {R.}~\bibnamefont {Di~Candia}},\ and\ \bibinfo {author}
  {\bibfnamefont {P.}~\bibnamefont {Scarlino}},\ }\bibfield  {title} {\bibinfo
  {title} {Observation of first- and second-order dissipative phase transitions
  in a two-photon driven {Kerr} resonator},\ }\href
  {https://doi.org/10.1038/s41467-025-56830-w} {\bibfield  {journal} {\bibinfo
  {journal} {Nature Communications}\ }\textbf {\bibinfo {volume} {16}},\
  \bibinfo {pages} {1954} (\bibinfo {year} {2025})}\BibitemShut {NoStop}%
\bibitem [{\citenamefont {Imamo\ifmmode~\bar{g}\else \={g}\fi{}lu}\ \emph
  {et~al.}(1997)\citenamefont {Imamo\ifmmode~\bar{g}\else \={g}\fi{}lu},
  \citenamefont {Schmidt}, \citenamefont {Woods},\ and\ \citenamefont
  {Deutsch}}]{PhysRevLett.79.1467}%
  \BibitemOpen
  \bibfield  {author} {\bibinfo {author} {\bibfnamefont {A.}~\bibnamefont
  {Imamo\ifmmode~\bar{g}\else \={g}\fi{}lu}}, \bibinfo {author} {\bibfnamefont
  {H.}~\bibnamefont {Schmidt}}, \bibinfo {author} {\bibfnamefont
  {G.}~\bibnamefont {Woods}},\ and\ \bibinfo {author} {\bibfnamefont
  {M.}~\bibnamefont {Deutsch}},\ }\bibfield  {title} {\bibinfo {title}
  {Strongly interacting photons in a nonlinear cavity},\ }\href
  {https://doi.org/10.1103/PhysRevLett.79.1467} {\bibfield  {journal} {\bibinfo
   {journal} {Phys. Rev. Lett.}\ }\textbf {\bibinfo {volume} {79}},\ \bibinfo
  {pages} {1467} (\bibinfo {year} {1997})}\BibitemShut {NoStop}%
\bibitem [{\citenamefont {Zhu}(2010)}]{Zhu:10}%
  \BibitemOpen
  \bibfield  {author} {\bibinfo {author} {\bibfnamefont {Y.}~\bibnamefont
  {Zhu}},\ }\bibfield  {title} {\bibinfo {title} {Large {Kerr} nonlinearities
  on cavity-atom polaritons},\ }\href {https://doi.org/10.1364/OL.35.000303}
  {\bibfield  {journal} {\bibinfo  {journal} {Opt. Lett.}\ }\textbf {\bibinfo
  {volume} {35}},\ \bibinfo {pages} {303} (\bibinfo {year} {2010})}\BibitemShut
  {NoStop}%
\bibitem [{\citenamefont {Breuer}\ and\ \citenamefont
  {Petruccione}(2007)}]{10.1093/acprof:oso/9780199213900.001.0001}%
  \BibitemOpen
  \bibfield  {author} {\bibinfo {author} {\bibfnamefont {H.-P.}\ \bibnamefont
  {Breuer}}\ and\ \bibinfo {author} {\bibfnamefont {F.}~\bibnamefont
  {Petruccione}},\ }\href
  {https://doi.org/10.1093/acprof:oso/9780199213900.001.0001} {\emph {\bibinfo
  {title} {The Theory of Open Quantum Systems}}}\ (\bibinfo  {publisher}
  {Oxford University Press},\ \bibinfo {year} {2007})\BibitemShut {NoStop}%
\bibitem [{\citenamefont {Prasad}\ \emph {et~al.}(2022)\citenamefont {Prasad},
  \citenamefont {Yadalam}, \citenamefont {Aron},\ and\ \citenamefont
  {Kulkarni}}]{PhysRevA.105.L050201}%
  \BibitemOpen
  \bibfield  {author} {\bibinfo {author} {\bibfnamefont {M.}~\bibnamefont
  {Prasad}}, \bibinfo {author} {\bibfnamefont {H.~K.}\ \bibnamefont {Yadalam}},
  \bibinfo {author} {\bibfnamefont {C.}~\bibnamefont {Aron}},\ and\ \bibinfo
  {author} {\bibfnamefont {M.}~\bibnamefont {Kulkarni}},\ }\bibfield  {title}
  {\bibinfo {title} {Dissipative quantum dynamics, phase transitions, and
  non-{Hermitian} random matrices},\ }\href
  {https://doi.org/10.1103/PhysRevA.105.L050201} {\bibfield  {journal}
  {\bibinfo  {journal} {Phys. Rev. A}\ }\textbf {\bibinfo {volume} {105}},\
  \bibinfo {pages} {L050201} (\bibinfo {year} {2022})}\BibitemShut {NoStop}%
\bibitem [{\citenamefont {Lieu}\ \emph {et~al.}(2020)\citenamefont {Lieu},
  \citenamefont {Belyansky}, \citenamefont {Young}, \citenamefont {Lundgren},
  \citenamefont {Albert},\ and\ \citenamefont
  {Gorshkov}}]{PhysRevLett.125.240405}%
  \BibitemOpen
  \bibfield  {author} {\bibinfo {author} {\bibfnamefont {S.}~\bibnamefont
  {Lieu}}, \bibinfo {author} {\bibfnamefont {R.}~\bibnamefont {Belyansky}},
  \bibinfo {author} {\bibfnamefont {J.~T.}\ \bibnamefont {Young}}, \bibinfo
  {author} {\bibfnamefont {R.}~\bibnamefont {Lundgren}}, \bibinfo {author}
  {\bibfnamefont {V.~V.}\ \bibnamefont {Albert}},\ and\ \bibinfo {author}
  {\bibfnamefont {A.~V.}\ \bibnamefont {Gorshkov}},\ }\bibfield  {title}
  {\bibinfo {title} {Symmetry breaking and error correction in open quantum
  systems},\ }\href {https://doi.org/10.1103/PhysRevLett.125.240405} {\bibfield
   {journal} {\bibinfo  {journal} {Phys. Rev. Lett.}\ }\textbf {\bibinfo
  {volume} {125}},\ \bibinfo {pages} {240405} (\bibinfo {year}
  {2020})}\BibitemShut {NoStop}%
\bibitem [{\citenamefont {Johansson}\ \emph {et~al.}(2013)\citenamefont
  {Johansson}, \citenamefont {Nation},\ and\ \citenamefont
  {Nori}}]{JOHANSSON20131234}%
  \BibitemOpen
  \bibfield  {author} {\bibinfo {author} {\bibfnamefont {J.}~\bibnamefont
  {Johansson}}, \bibinfo {author} {\bibfnamefont {P.}~\bibnamefont {Nation}},\
  and\ \bibinfo {author} {\bibfnamefont {F.}~\bibnamefont {Nori}},\ }\bibfield
  {title} {\bibinfo {title} {{QuTiP} 2: A {Python} framework for the dynamics
  of open quantum systems},\ }\href
  {https://doi.org/https://doi.org/10.1016/j.cpc.2012.11.019} {\bibfield
  {journal} {\bibinfo  {journal} {Comput. Phys. Commun.}\ }\textbf {\bibinfo
  {volume} {184}},\ \bibinfo {pages} {1234} (\bibinfo {year}
  {2013})}\BibitemShut {NoStop}%
\bibitem [{\citenamefont {Atas}\ \emph {et~al.}(2013)\citenamefont {Atas},
  \citenamefont {Bogomolny}, \citenamefont {Giraud},\ and\ \citenamefont
  {Roux}}]{PhysRevLett.110.084101}%
  \BibitemOpen
  \bibfield  {author} {\bibinfo {author} {\bibfnamefont {Y.~Y.}\ \bibnamefont
  {Atas}}, \bibinfo {author} {\bibfnamefont {E.}~\bibnamefont {Bogomolny}},
  \bibinfo {author} {\bibfnamefont {O.}~\bibnamefont {Giraud}},\ and\ \bibinfo
  {author} {\bibfnamefont {G.}~\bibnamefont {Roux}},\ }\bibfield  {title}
  {\bibinfo {title} {Distribution of the ratio of consecutive level spacings in
  random matrix ensembles},\ }\href
  {https://doi.org/10.1103/PhysRevLett.110.084101} {\bibfield  {journal}
  {\bibinfo  {journal} {Phys. Rev. Lett.}\ }\textbf {\bibinfo {volume} {110}},\
  \bibinfo {pages} {084101} (\bibinfo {year} {2013})}\BibitemShut {NoStop}%
\bibitem [{\citenamefont {Mac\'e}\ \emph {et~al.}(2019)\citenamefont {Mac\'e},
  \citenamefont {Alet},\ and\ \citenamefont
  {Laflorencie}}]{PhysRevLett.123.180601}%
  \BibitemOpen
  \bibfield  {author} {\bibinfo {author} {\bibfnamefont {N.}~\bibnamefont
  {Mac\'e}}, \bibinfo {author} {\bibfnamefont {F.}~\bibnamefont {Alet}},\ and\
  \bibinfo {author} {\bibfnamefont {N.}~\bibnamefont {Laflorencie}},\
  }\bibfield  {title} {\bibinfo {title} {Multifractal scalings across the
  many-body localization transition},\ }\href
  {https://doi.org/10.1103/PhysRevLett.123.180601} {\bibfield  {journal}
  {\bibinfo  {journal} {Phys. Rev. Lett.}\ }\textbf {\bibinfo {volume} {123}},\
  \bibinfo {pages} {180601} (\bibinfo {year} {2019})}\BibitemShut {NoStop}%
\bibitem [{\citenamefont {Hohenberg}\ and\ \citenamefont
  {Halperin}(1977)}]{RevModPhys.49.435}%
  \BibitemOpen
  \bibfield  {author} {\bibinfo {author} {\bibfnamefont {P.~C.}\ \bibnamefont
  {Hohenberg}}\ and\ \bibinfo {author} {\bibfnamefont {B.~I.}\ \bibnamefont
  {Halperin}},\ }\bibfield  {title} {\bibinfo {title} {Theory of dynamic
  critical phenomena},\ }\href {https://doi.org/10.1103/RevModPhys.49.435}
  {\bibfield  {journal} {\bibinfo  {journal} {Rev. Mod. Phys.}\ }\textbf
  {\bibinfo {volume} {49}},\ \bibinfo {pages} {435} (\bibinfo {year}
  {1977})}\BibitemShut {NoStop}%
\bibitem [{\citenamefont {Albert}\ and\ \citenamefont
  {Jiang}(2014)}]{PhysRevA.89.022118}%
  \BibitemOpen
  \bibfield  {author} {\bibinfo {author} {\bibfnamefont {V.~V.}\ \bibnamefont
  {Albert}}\ and\ \bibinfo {author} {\bibfnamefont {L.}~\bibnamefont {Jiang}},\
  }\bibfield  {title} {\bibinfo {title} {Symmetries and conserved quantities in
  {Lindblad} master equations},\ }\href
  {https://doi.org/10.1103/PhysRevA.89.022118} {\bibfield  {journal} {\bibinfo
  {journal} {Phys. Rev. A}\ }\textbf {\bibinfo {volume} {89}},\ \bibinfo
  {pages} {022118} (\bibinfo {year} {2014})}\BibitemShut {NoStop}%
\bibitem [{\citenamefont {Haarnoja}\ \emph {et~al.}(2018)\citenamefont
  {Haarnoja}, \citenamefont {Zhou}, \citenamefont {Abbeel},\ and\ \citenamefont
  {Levine}}]{pmlr-v80-haarnoja18b}%
  \BibitemOpen
  \bibfield  {author} {\bibinfo {author} {\bibfnamefont {T.}~\bibnamefont
  {Haarnoja}}, \bibinfo {author} {\bibfnamefont {A.}~\bibnamefont {Zhou}},
  \bibinfo {author} {\bibfnamefont {P.}~\bibnamefont {Abbeel}},\ and\ \bibinfo
  {author} {\bibfnamefont {S.}~\bibnamefont {Levine}},\ }\bibfield  {title}
  {\bibinfo {title} {Soft {Actor-Critic}: Off-policy maximum entropy deep
  reinforcement learning with a stochastic actor},\ }in\ \href
  {https://proceedings.mlr.press/v80/haarnoja18b.html} {\emph {\bibinfo
  {booktitle} {Proceedings of the 35th International Conference on Machine
  Learning}}},\ \bibinfo {series} {Proceedings of Machine Learning Research},
  Vol.~\bibinfo {volume} {80},\ \bibinfo {editor} {edited by\ \bibinfo {editor}
  {\bibfnamefont {J.}~\bibnamefont {Dy}}\ and\ \bibinfo {editor} {\bibfnamefont
  {A.}~\bibnamefont {Krause}}}\ (\bibinfo  {publisher} {PMLR},\ \bibinfo {year}
  {2018})\ pp.\ \bibinfo {pages} {1861--1870}\BibitemShut {NoStop}%
\bibitem [{\citenamefont {Haarnoja}\ \emph {et~al.}(2019)\citenamefont
  {Haarnoja}, \citenamefont {Zhou}, \citenamefont {Hartikainen}, \citenamefont
  {Tucker}, \citenamefont {Ha}, \citenamefont {Tan}, \citenamefont {Kumar},
  \citenamefont {Zhu}, \citenamefont {Gupta}, \citenamefont {Abbeel},\ and\
  \citenamefont {Levine}}]{haarnoja2019softactorcriticalgorithmsapplications}%
  \BibitemOpen
  \bibfield  {author} {\bibinfo {author} {\bibfnamefont {T.}~\bibnamefont
  {Haarnoja}}, \bibinfo {author} {\bibfnamefont {A.}~\bibnamefont {Zhou}},
  \bibinfo {author} {\bibfnamefont {K.}~\bibnamefont {Hartikainen}}, \bibinfo
  {author} {\bibfnamefont {G.}~\bibnamefont {Tucker}}, \bibinfo {author}
  {\bibfnamefont {S.}~\bibnamefont {Ha}}, \bibinfo {author} {\bibfnamefont
  {J.}~\bibnamefont {Tan}}, \bibinfo {author} {\bibfnamefont {V.}~\bibnamefont
  {Kumar}}, \bibinfo {author} {\bibfnamefont {H.}~\bibnamefont {Zhu}}, \bibinfo
  {author} {\bibfnamefont {A.}~\bibnamefont {Gupta}}, \bibinfo {author}
  {\bibfnamefont {P.}~\bibnamefont {Abbeel}},\ and\ \bibinfo {author}
  {\bibfnamefont {S.}~\bibnamefont {Levine}},\ }\href@noop {} {\bibinfo {title}
  {Soft {Actor-Critic} algorithms and applications}} (\bibinfo {year} {2019}),\
  \Eprint {https://arxiv.org/abs/1812.05905} {arXiv:1812.05905} \BibitemShut
  {NoStop}%
\bibitem [{\citenamefont {Sutton}\ and\ \citenamefont
  {Barto}(2018)}]{Sutton2018}%
  \BibitemOpen
  \bibfield  {author} {\bibinfo {author} {\bibfnamefont {R.~S.}\ \bibnamefont
  {Sutton}}\ and\ \bibinfo {author} {\bibfnamefont {A.~G.}\ \bibnamefont
  {Barto}},\ }\href@noop {} {\emph {\bibinfo {title} {Reinforcement Learning:
  An Introduction}}},\ \bibinfo {edition} {2nd}\ ed.\ (\bibinfo  {publisher}
  {MIT Press},\ \bibinfo {address} {Cambridge, MA},\ \bibinfo {year}
  {2018})\BibitemShut {NoStop}%
\end{thebibliography}%
\end{document}